\documentclass[structabstract]{aa}
\usepackage{amsmath}
\usepackage{txfonts}
\usepackage{natbib}
\usepackage{footnote}

\usepackage{graphicx}
\usepackage{color}

\newcommand{\teff}{\rm T_{eff}}
\newcommand{\logg}{\log{g}}
\newcommand{\feh}{\rm [Fe/H]}
\newcommand{\meta}{{\rm [M/H]}}
\newcommand{\afe}{{\rm [\alpha/Fe]}}
\def\kms{{\rm km\,s^{-1}}}
\def\kpc{\,{\rm kpc}}

\def\dex{\,{\rm dex}}
\def\Gyr{\,{\rm Gyr}}
\def\url#1{{\tt#1}}

\begin{document}

%\title{ Interstellar extinction with the Gaia-ESO survey}
\title{The Gaia-ESO Survey: Tracing interstellar extinction\thanks{Based on observations collected with the FLAMES spectrograph at the VLT/UT2 telescope (Paranal Observatory, ESO, Chile), for the Gaia-ESO Large Public Survey, programme 188.B-300}}
%\author{M. Schultheis \& B.Q. Chen et al.}

\author{M. Schultheis \inst{1}
\and   G. Kordopatis \inst{2,3}
\and A. Recio-Blanco\inst{1}
\and P. de Laverny\inst{1}
\and V. Hill\inst{1}
\and G. Gilmore\inst{3}
\and  E.~J. Alfaro\inst{4}
\and M.T. Costado\inst{4}
\and T. Bensby\inst{5}
\and F. Damiani\inst{6}
\and S. Feltzing\inst{5}
\and   E. Flaccomio\inst{6}
\and C. Lardo\inst{7}
\and P. Jofre\inst{3}
\and L. Prisinzano\inst{6}
\and S. Zaggia\inst{8}
\and F. Jimenez-Esteban\inst{9,10}
\and L. Morbidelli\inst{11}
\and A.C. Lanzafame\inst{12}
\and A. Hourihane\inst{3}
\and C. Worley\inst{3}
\and P. Francois\inst{13}
}

   \institute{ Universit\'e de Nice Sophia-Antipolis, CNRS, Observatoire de C\^ote d'Azur, Laboratoire Lagrange, 06304 Nice Cedex 4, France 
 e-mail: mathias.schultheis@oca.eu
 \and
Leibniz-Institut f\"ur  Astrophysik Potsdam (AIP), An der Sternwarte 16, 14482 Potsdam, Germany
\and
 Institute of Astronomy, University of Cambridge, Madingley Road, CB3 0HA, United KIngdom
\and
Instituto de Astrof\'{i}sica de Andaluc\'{i}a-CSIC, Apdo. 3004, 18080 Granada, Spain
\and
Lund Observatory, Department of Astronomy and Theoretical Physics, Box 43, SE-221 00 Lund, Sweden
\and
INAF - Osservatorio Astronomico di Palermo, Piazza del Parlamento 1, 90134, Palermo, Italy
\and
Astrophysics Research Institute, Liverpool John Moores University, 146 Brownlow Hill, Liverpool L3 5RF, United Kingdom
\and
INAF - Padova Observatory, Vicolo dell'Osservatorio 5, 35122 Padova, Italy
\and
Centro de Astrobiolog\'{\i}a (INTA-CSIC), Departamento de Astrof\'{\i}sica, PO Box 78, E-28691, Villanueva de la Ca\~nada, Madrid, Spain
     \and
Suffolk University, Madrid Campus, C/ Valle de la Viña 3, 28003, Madrid, Spain 
\and
INAF - Osservatorio Astrofisico di Arcetri, Largo E. Fermi 5, 50125, Florence, Italy
\and
Dipartimento di Fisica e Astronomia, Sezione Astrofisica, Universit\'{a} di Catania, via S. Sofia 78, 95123, Catania, Italy
\and
GEPI, Observatoire de Paris, CNRS, Universit\'e Paris Diderot, 5 Place Jules Janssen, 92190 Meudon, France
 }

\abstract 
%Context
{Large spectroscopic surveys have enabled in the recent years the computation of three-dimensional interstellar extinction maps thanks to accurate stellar atmospheric parameters and line-of-sight distances. Such maps are complementary to 3D maps extracted from photometry, allowing a more thorough study of the dust properties.}
%Aims
 {Our goal is to use the high-resolution spectroscopic survey Gaia-ESO in order to obtain with a good distance resolution the interstellar extinction and its dependency as a function of the environment and the Galactocentric position.}
 %Methods
 { We use the stellar atmospheric parameters of more than 5000 stars, obtained from the Gaia-ESO survey second internal data release, and  combine them with optical (SDSS) and near-infrared (VISTA) photometry  as well as different sets of  theoretical stellar isochrones, in order to calculate line-of-sight extinction and distances. The extinction coefficients are then compared with the literature to discuss their dependancy on the  stellar parameters and position in the Galaxy.}
 %Results
 { Within the errors of our method, our work does not show that there is any dependence of the interstellar extinction coefficient on the atmospheric parameters of the stars.  We do not find any evidence of the variation of $E(J-H)/E(J-K)$ with the angle from the Galactic centre nor with
Galactocentric distance. This suggests that we are dealing with a uniform extinction law in the SDSS $ugriz$ bands and the near-IR $JHKs$ bands. Therefore, extinction maps using mean colour-excesses and  assuming a constant extinction coefficient can be used without introducing any systematic errors.}
%Conclusions
{}


\maketitle

\titlerunning{3D extinction map with GES}
\authorrunning{Schultheis et al.}

\section{Introduction}
 
%Interstellar dust is a significant component of the plane of the Milky Way and is closely related to the
%gas distribution which is concentrated within 100\,pc of the Galactic plane.  

Understanding the dust spatial distribution in the Milky is a crucial part of Galactic archeology. Indeed, it can reveal important features of Galaxy evolution, such as the location and intensity of past star formation episodes (\citealt{boulanger2007}). Among the literature, the
 most commonly used full-sky dust map is that of \citet{schlegel1998} obtained with COBE/DIRBE data,  later improved by \citet{schlafy2010} and  \citet{schlafy2011} by adding correction terms  mainly due to the adopted reddening law. Other 2-D extinction maps
 result from specific stellar populations. For example, red clump stars are considered to be an  ideal tracer for extinction as their mean intrinsic colour varies only slightly with metallicity therefore making them a reliable tracer of dust extinction \citep[see for example][]{gonzalez2011,gonzalez2012,nataf2013}. However, such studies require a sufficient number of  red clump stars in order to  have a good spatial coverage and hence are  limited to regions with high stellar density (close to the plane or towards the Bulge).  
On the other hand, \citet{nidever2012} mapped the extinction with the so-called Rayleigh Jeans Colour Excess method (RJCE) which is based on  a combination of near and mid-infrared photometry (e.g. H and [4.5]). RJCE determines 2D star-by-star reddening at high-resolution  ($\rm 2 \times 2\arcmin$) allowing to penetrate the  heavily obscured Galactic mid-plane. 

The recent growth of large  surveys together with the increasing volume of data has pushed forward the extinction mapping and allowed to trace its distribution in three dimensions. The first of these 3D extinction models was constructed by  \citet{drimmel2003}, by fitting the far and near-IR data from the COBE/DIRBE instrument. \citet{marshall2006} used the  2MASS colours together with the stellar population synthesis model of Besan\c{c}on  (\citealt{robin2003}) to trace extinction in 3D for the Galactic Bulge region ($ |l| < 90, |b| < 10$) with a spatial resolution of 15\arcmin. 
% \citet{schultheis2014a} applied  a similar method but used  data from the VISTA Via Lactea survey (VVV, \citealt{minniti2010}) together with an improved version of  the  Besan\c{c}on model (\citealt{robin2012}). Their angular resolution is $\rm 6 \times 6\arcmin$ with distance bins of 500\,pc.  

A number of photometric techniques have been  used to study the dust in 3D (\citealt{green2014},\citealt{schlafy2014}, \citealt{bailer2011}, \citealt{hanson2014}, \citealt{sale2012}, \citealt{sale2014}, \citealt{lallement2014}).
\citet{Chen2014} traced the stellar locus similar as done by \citet{berry2012}  method  by combining optical, 2MASS and WISE photometry and  defining the reference stellar locus as well as  fixing the extinction law.  Their  map covers about 6000. sq. degree with a small overlap to some of our GES fields (see Sect. 5)

%\edit{The universality of} the near-Infrared extinction law has been \edit{debated in the last years. For example,} \cite{gao2009} and \citet{zasowski2009} found variations of the extinction law shape as a function of the angle from the Galactic centre.  This could be  a sign of the variation of the dust grain size distribution with metallicity.  On the other hand,  \citet{wang2014} revisited the extinction law using APOGEE K-type giants and concluded a universal Near-IR extinction law from diffuse to dense regions.
%\query{Ce paragraphe tombe un peu comme un cheveu sur la soupe. Je comprend que son utilite est de donner des questions scientifiques quanta  l'utilite de l'etude de la poussiere, mais alors il faudrait en poser quelques unes supplementaires, et le deplacer soit plus haut, soit plus bas. En effet, on parle deja de APOGEE, et ce n'est que dans le paragraphe suivant qu'on dit que les releves spectroscopiques permettent egalement de faire de l'etude de la poussiere.}

 Until now, most of the interstellar extinction studies were done mainly by photometric techniques. The already available or upcoming large spectroscopic surveys (RAVE, APOGEE, Gaia-ESO , GALAH, 4MOST, etc..) provide an important quantity of  accurate stellar parameters of  different stellar populations. It is therefore possible to  probe the 3D distribution of interstellar dust using the expected colours of the targets and comparing them with the observed ones. The Gaia space mission of ESA  will provide in the following years two-dimensional maps for most of the Galaxy but also
individual extinction estimates together with  measured parallaxes (see \citealt{bailer2013} for more details).
In this paper we will use stellar properties derived from the Gaia-ESO survey \citep[GES,][]{gilmore2012} to probe the  interstellar extinction in three dimensions  as done by  \citet{schultheis2014} using APOGEE data. They compared 3D extinction models in the Galactic Bulge region with \citet{marshall2006} and  \citet{schultheis2014a} and found a steep rise in extinction in the first few kpc and a flattening of the extinction at about  4\,kpc from the Sun.
While \citet{schultheis2014} and \citet{wang2014} probed the interstellar dust properties  with APOGEE in the Galactic plane
($\rm |z| < 1\kpc$), the GES fields are  located at much higher Galactic latitudes ($\rm |b| > 20^{o}$).  The GES data are thus complementary to APOGEE   allowing to trace the dust extinction at higher Galactic height $\rm |z|$  and compare them with available 3D dust models. In Sect.\,2, we
describe the sample of stars that has been used, in Sect\,3 we present the method employed in the determination of the  extinction and the distances. In Sections 4 and 5, we compare and discuss the extinctions to existing 2D and 3D maps, and we finish in Sect. 6 with the discussion about the universality of the extinction law.

\section{The sample}

The Gaia-ESO survey (GES) is a public spectroscopic survey targeting $\sim 10^5$ stars, covering all the major components of the Milky Way, from the halo to star forming regions, with the purpose of characterising the chemistry and kinematics of these populations. It uses the FLAMES multi-object spectrograph on the VLT UT2 telescope to obtain high quality, uniformly calibrated 
spectra. 
 The GES processing flow goes from target selection, through data reduction, spectrum analysis, astrophysical parameter determination, calibration and homogenisation. A detailed description of the data processing cascade and general characterisation of the data set can be found in Gilmore et al. (2015, in prep.),
In this paper, we  analyse the second data release (DR2) GES results for $\sim$ 10 000 Milky Way stars observed with the high-resolution gratings   HR10 centred at 5488\,\AA\ (R $\sim$ 19800) and HR21 centred at 8757\,\AA\ (R $\sim$ 16200) of the GIRAFFE spectrograph. All the targets were selected from VISTA photometry, with colour cuts in the range $\rm 0.2 < (J-K) < 0.8$, and magnitude cuts between $\rm 12.5 < J < 17.5$ (c.f. \citealt{gilmore2012}). Additional
SDSS photometry is also available which we will use in this work.  Considerable effort has been invested in the determination of the stellar parameters. The stellar parameters
 have been derived from three different methods, MATISSE (\citealt{recio-blanco2006}), SME (\citealt{valenti1996}) and FERRE (\citealt{allende2006}).  This ensures a reliable determination of the derived stellar parameters, a crucial step to get accurate stellar abundances and line-of-sight distances. 
The  homogenisation of the results from the three nodes  which was verified during the GES  parameters validation process leads to the so-called  ``Recommended stellar parameters''.
 For our analysis we used those parameters, i.e. effective temperature ($\teff$),  surface gravities ($\logg$), global metallicities ($\meta$) and $\alpha$-elements ($\rm [\alpha/Fe]$).
 The relative error distributions peak at 70\,K for $\teff$, $0.10\dex$ for $\logg$, $0.08\dex$ for $\meta$ and $0.03\dex$ for $\afe$. 
 More details about the related GES parameterisation pipeline can be found in \citet{recio-blanco2014}.

\begin{figure}[!htbp]
\includegraphics[width=0.5\textwidth]{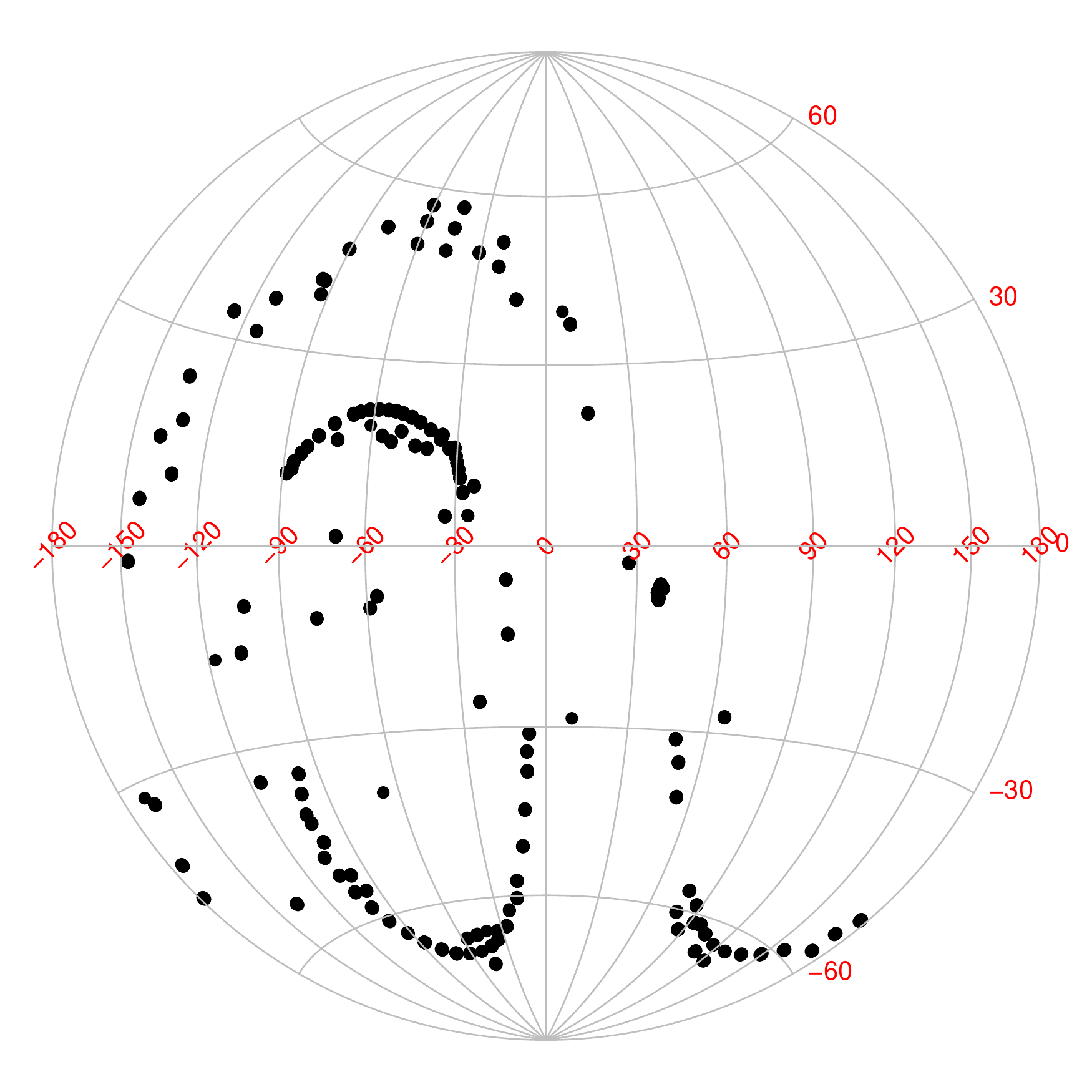}
\caption{Field centres of the GES-DR2 fields in Galactic coordinates}
\label{longlat}
\end{figure}

%We selected stars where measurements in the VVV or 2MASS bands are available and where the  spectra obtained had signal-to-noise $\rm S/N > 10$ and  the error in the radial velocity $\sigma(RV) < 1.5\,km/s$. This ensures that the error in the corresponding absolute magnitude (see Sect.~\ref{distance}) is reasonably  small. Larger errors in the radial velocites
%would result in larger errors in  the determination of the stellar parameters. An additional selection based on $\sigma_{Teff}$  and $\sigma_{log\,g}$ is included (see Sect.~\ref{distance}) in order
%to use only  stars where the internal dispersion of  $\teff$ and $\rm log\,g$ between the three different nodes is small.

To achieve the scope of this paper, we selected targets with available photometry from the VISTA Variables in the Via Lactea (VVV) survey (\citealt{minniti2010}) or 2MASS, together with reliable atmospheric parameters obtained from spectroscopy. This implied that we selected only the targets with spectra having a signal-to-noise $\rm S/N > 10$, errors in radial velocity $\sigma(RV) < 1.5\kms$, and  low internal dispersion on $\teff$ and $\logg$ between the three different nodes (see Sect.~\ref{distance}). The first two criteria ensure that the random errors of the algorithms are minimised, and the $(\sigma_{\teff}, \sigma_{\logg})$ criterium ensures a minimisation of the internal errors of the GES homogenisation pipeline.  Our sample consists thus of 5603 stars with only Near-IR photometry, while 1106 stars have ugrizJHK photometry.

Figure \ref{longlat} shows the field centres of the GES-DR2 fields in Galactic coordinates.

\section{Distance and extinction determination} \label{distance}

The routine that has been used to determine the absolute magnitudes of the stars in different photometric bands  (and from that the extinctions and the distances), is based on the one described in \citet{Kordopatis11b}, and already successfully applied in \citet{Gazzano13,Kordopatis13a,Recio-Blanco14}.
Briefly, the method projects the measured atmospheric parameters ($\hat{\theta} \equiv \teff,\logg,  \meta$) and colours on a given set of theoretical isochrones. The set of isochrones used is  defined for a given age $a$ and  a range of iron abundances $\feh$  within the $\meta$ error bars. For this paper, we assume that the $\feh$ values of the isochrones are similar enough to $\meta$, to use the approximation [Fe/H]=[M/H] while performing the projection on the isochrones. According to Recio-Blanco et al. (in prep), this is true for most of the stars in the GES iDR2.

The probability density function for a given star, $W(a,m, {\feh})$, is defined as: 
%The projection first assigns  a weight, $W(a,m)$,  to each point of the set of isochrones, \edit{defining the probability density function for a given star as:}
\begin{equation}
W(a,m, {\feh})=dm \cdot \exp \left( -\sum_i \frac{(\theta_i - \hat{\theta_i})^2 }{2\sigma^2_{\hat{\theta_i}}}\right)
\end{equation}

where $\theta_i$ is the theoretical  $\teff$, $\logg$ or $\feh$,  $\hat{\theta_i}$ and  $\sigma_{\hat{\theta_i}}$ are the measured parameters and their respective errors,  and $dm$ is the mass step between two points of the same isochrone, introduced in order to impose a uniform prior on the stellar mass. We note that \citet{Zwitter10} have shown that having such a flat prior on mass does not affect significantly  the final derived distances, when compared to the use of a combination of a more realistic mass function \citep[e.g.][]{chabrier03}  with a luminosity prior on the surveyed stars \citep[see,][ their Sect.\,2.2, for more details]{Zwitter10}.

 The expected value of the absolute magnitude $M_\tau$ in a given photometric band $\tau$   of a given star  is  then obtained by computing the weighted mean: 

\begin{equation}
M_\tau=\frac{\sum_{a,m,\feh} W(a,m, {\feh}) \cdot   M_\tau(a,m, \feh)}{\sum_{a,m,\feh} W(a,m,\feh) },
\end{equation}
where $\sum_{a,m,\feh}$ is the triple sum over the ages, masses and iron abundances. The associated variance of the expected absolute magnitude $M_\tau$ is obtained by: 

\begin{equation}
\sigma^2(M_\tau)=\frac{\sum_{a, m, \feh}   W(a,m, {\feh}) \cdot   [ M_\tau -M_\tau(a,m,\feh)] ^2}{\sum_{a,m,\feh} W(a,m,\feh) }.
\end{equation}

We have used two sets of isochrones: the Yonsei-Yale ones \citep{Demarque04} with the \citet{Lejeune98} colour tables, and the Padova ones \citep{Marigo98, Bressan12}.  The Yonsei-Yale (YY) isochrones have been computed using the provided interpolation code of YY, from which we generated a set of isochrones with a constant step in age of $1\Gyr$, starting from $1\Gyr$ to $14\Gyr$, therefore resulting to flat prior on the age of the stars. As far as the metallicities are concerned, they are within a range of $-3 < \feh < 0.8\dex$, constantly spaced by $0.1\dex$. The $\alpha-$enhancements of the isochrones have been selected in the following way: 
\begin{itemize}
\renewcommand{\labelitemi}{$\bullet$}
\item $\feh \geq 0\dex$, then $\afe=0.0\dex$
\item $-0.3 \leq \feh \leq -0.1\dex$, then $\afe=+0.1\dex$
\item $-0.6 \leq \feh \leq -0.4\dex$, then $\afe=+0.2\dex$
\item $-0.9 \leq \feh \leq -0.7\dex$, then $\afe=+0.3\dex$
\item $\feh \leq -1\dex$, then $\afe=+0.4\dex$
\end{itemize}

\begin{table}[!htbp]
\caption{Median errors in distance when adding an offset to the derived atmospheric parameters for giants ($\rm log\,{g} < 3$) and dwarfs and subgiants  ($\rm log\,{g} > 3$). The ``+'' sign specifies adding the offset, the ``-'' sign subtracting the offset}
\begin{tabular}{c c c c}
\hline \hline
  &$\teff \pm 100\,K$&$\logg \pm 0.2\dex$& $\rm \meta \pm 0.1\dex$\\
  &  \% & \% & \%\\
\hline
Giants (+)&1.1&19.7&1.8 \\
Giants (-) & 5.1&34.3&1.3\\
\hline
Dwarfs (+)&5&2.9&7.6\\
Dwarfs (-) &11&8.1&9.7\\
\hline
\end{tabular}
%\caption{Median errors in distance (in \%) when adding an offset to the derived parameters of  $\rm T_{eff} \pm 100\,K$  (first column), $\rm log\,g \pm 0.2\,dex$ (second column) and $\rm [Fe/H] \pm 0.1\,dex$ (third column).  The first row is for giants ($\rm log{g} < 3$), the second row for dwarfs ($\rm log{g} > 3$).}
\label{errors}
\end{table}

According to \citet{Carpenter01}, the $(JK)_{ESO}$ provided by  the Yonsei-Yale isochrones match very well the $(JK_s)_{2MASS}$ and $(JK_s)_{VHS}$, so no colour transformation is needed when manipulating the magnitudes from the different photometric systems.
 
\begin{figure}[!htbp]
  \includegraphics[width=0.24\textwidth]{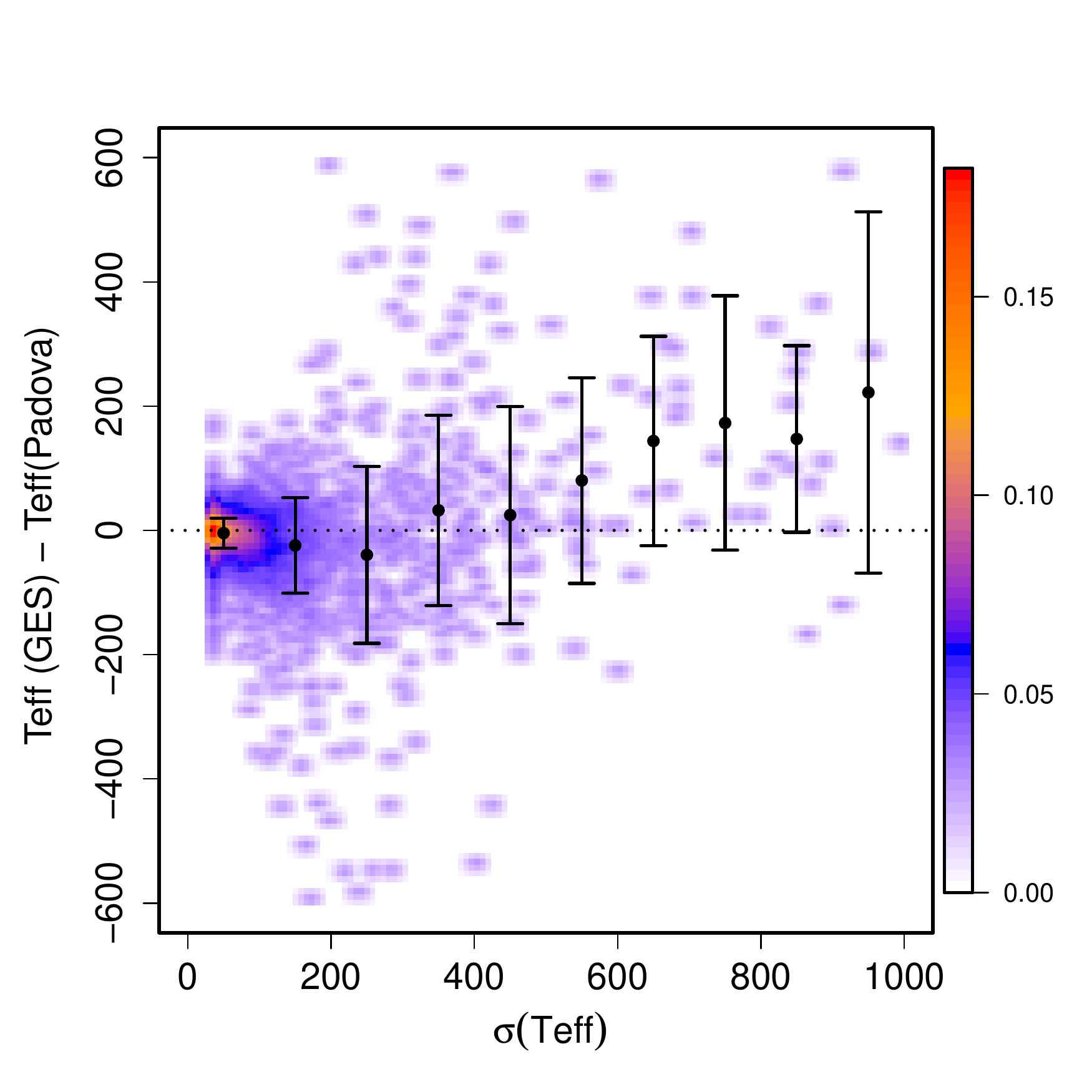}  \includegraphics[width=0.24\textwidth]{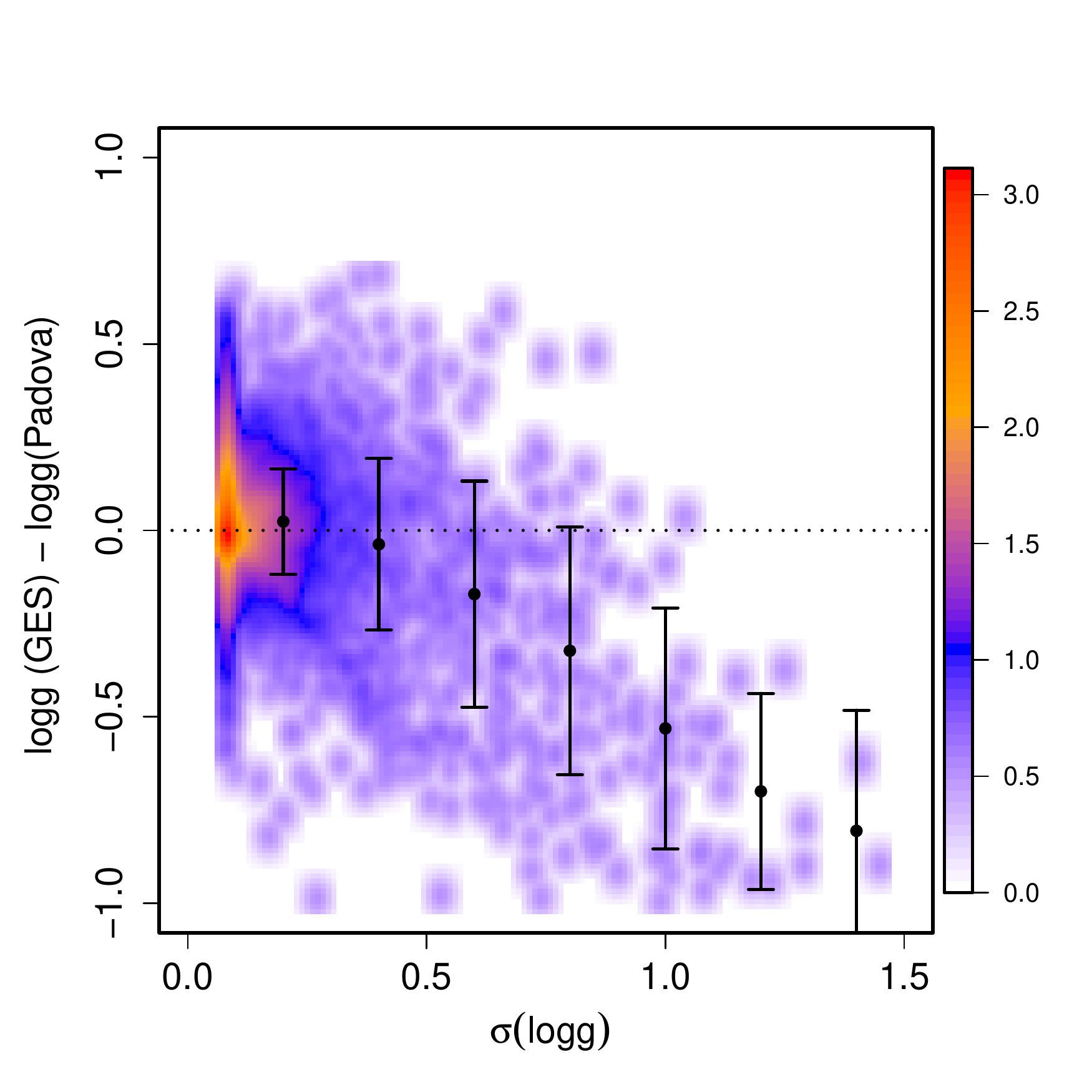} 
\caption{{\bf Left:} Density plot of the difference between $\teff$ and the corresponding temperature from the isochrones and the  dispersion of $\teff$ between the three nodes (see text). The black points indicate the mean (in 100\,K steps)  and the error bars the standard deviations. {\bf Right:}  The same but for $\logg$. The black points indicate the mean (in $0.2\dex$ steps) and the error-bars the standard deviations.}
\label{sigmaTeff}
\end{figure}

As far as the Padova isochrones are concerned, they have been downloaded using the online interpolation interface\footnote{http://stev.oapd.inaf.it/cgi-bin/cmd} which allows us to select the output photometric system (2MASS, SDSS, VISTA). The considered metallicity range is smaller than the one of YY (from $-2.2\dex$ to $+0.2\dex$ in steps of 0.1\,dex with solar-scale $\alpha-$abundances), computed with steps in age of $0.5\Gyr$.

\begin{figure*}[!htbp]
   \includegraphics[width=0.49\textwidth]{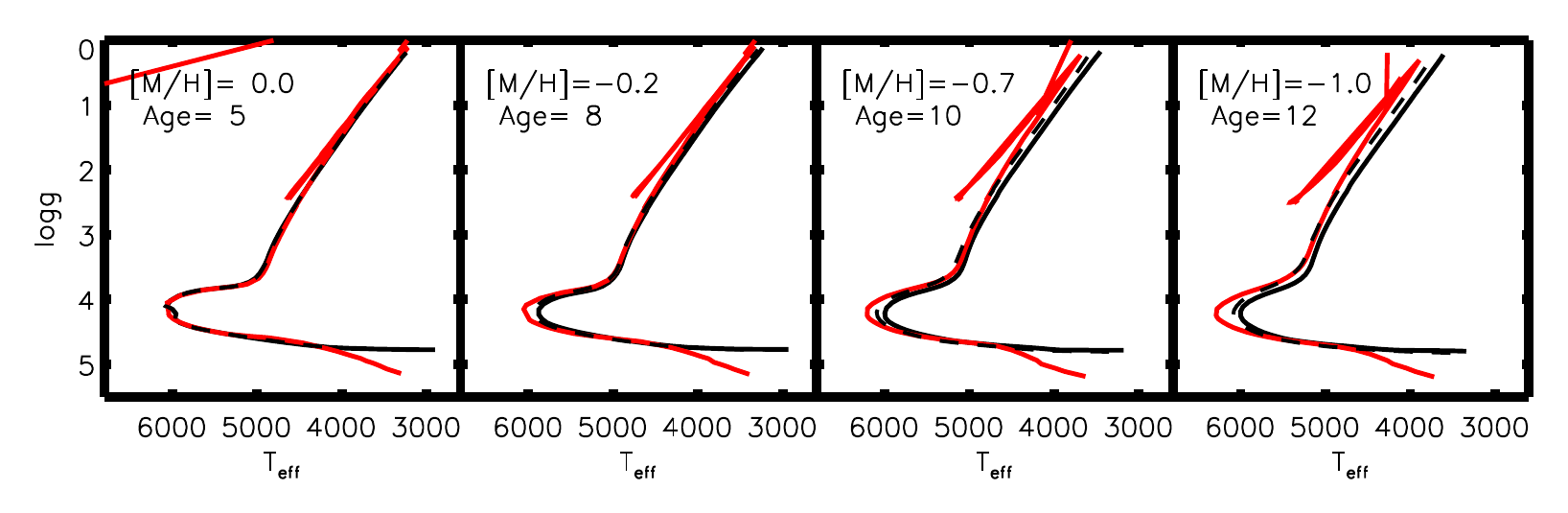} 
   \includegraphics[width=0.48\textwidth]{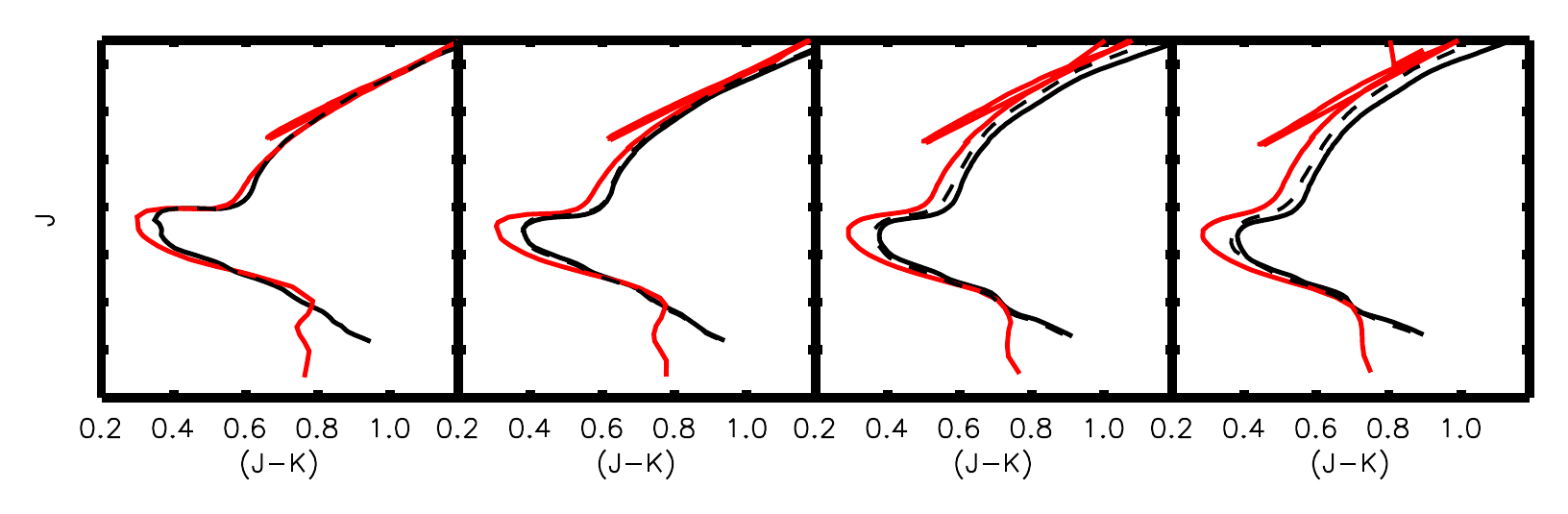} 
\caption{{\bf Left:}   Isochrones in the $\teff$ vs. $\logg$ space for different combinations of [M/H] and ages. The red line shows the Padova isochrones. The  dashed black line indicates  the Yonsei-Yale isochrones without $\rm \alpha$-elements while the plain black line is for  YY models with alpha enhancement.  {\bf Right:}  Similar as in the left panel but in the J--K vs. K colour-magnitude diagram}
%{\bf [[(1) Reverse the x-axis, so the first quadrant is on the left.  (2) There are stars shown here above the $b \le 5^\circ$ limit stated in the text!]]}}
\label{isochrones}
\end{figure*}

 Once the absolute magnitudes are computed, the colours are derived and the colour excess are deduced by subtracting the theoretical colour from the observed one in
the five SDSS and 3 VISTA filters.
 The colour excess  derived using this method is hereafter referred to as $\rm E_{\lambda_{1} - \lambda_{2}}$ with $\lambda = u,g,r,i,z,J,H,K_{s}$.  About 5\% of our stars show negative extinction, most of which  are fainter than  $\rm K_s > 14.5$. We
omitted those from our analysis. 
 The distances were calculated using  the usual relation :

\begin{equation}
\log_{10}d={\frac{K_s - M_{K_s} - A(K_s)+5}{5}}
\end{equation}
with $d$ expressed in pc, and adopting $\rm A(K_s) = 0.528 \times E(J-K_{s})$ (\citealt{nishiyama2009}) similar to what has been done  in 
\citet{schultheis2014}.  The errors in the derived distances include errors in $\teff$, $\logg$ and $\meta$ and errors on the apparent magnitude J and $\rm K_s$ as well as the extinction $\rm A_{K_s}$. For more details,  we refer to \citet{kordopatis2011}. Other than the internal errors in the stellar parameters, the absolute calibration in $\teff$, $\logg$ and $\feh$ is also crucial for the errors in our derived distances.

\begin{figure}[!htbp]
   \includegraphics[width=0.24\textwidth]{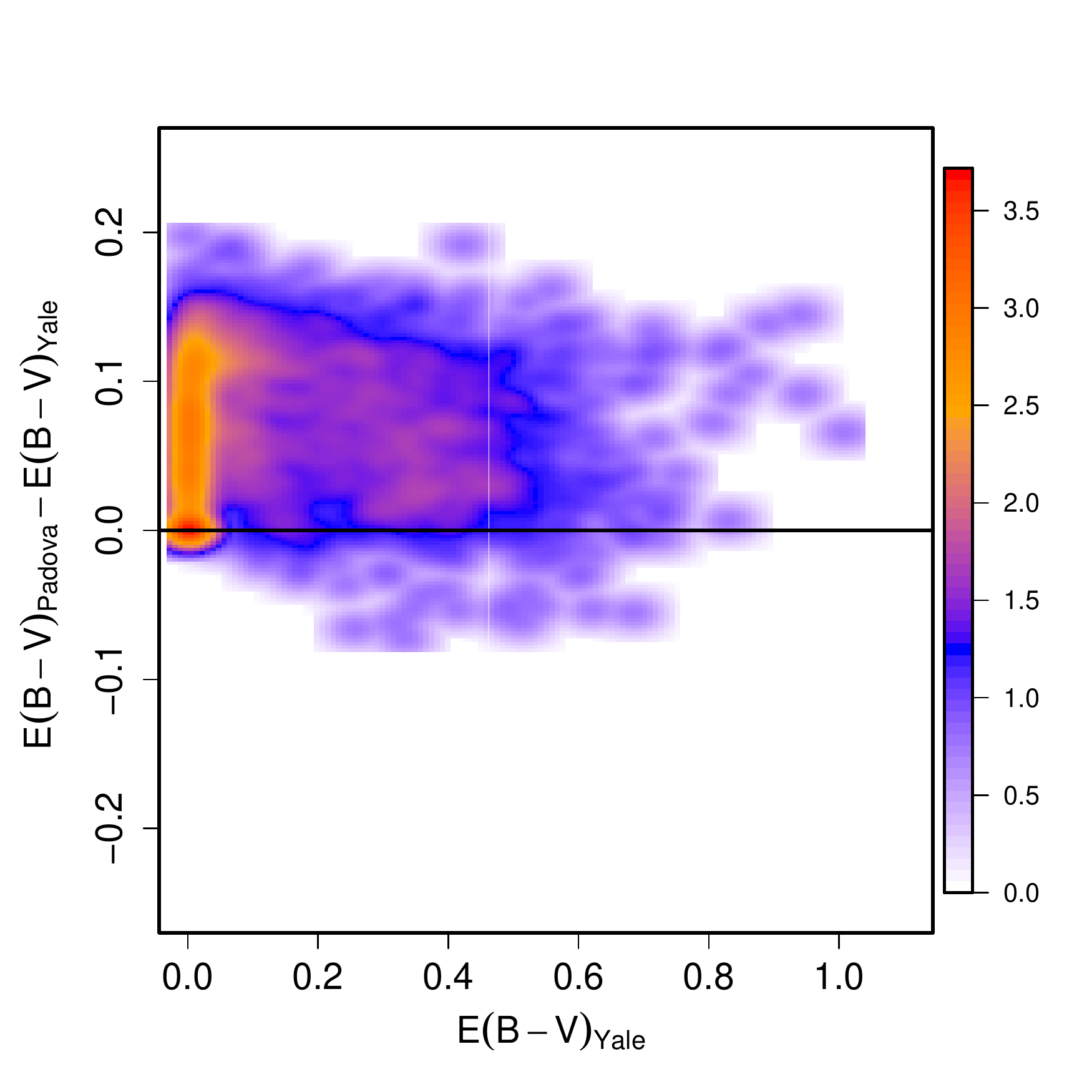}
  \includegraphics[width=0.24\textwidth]{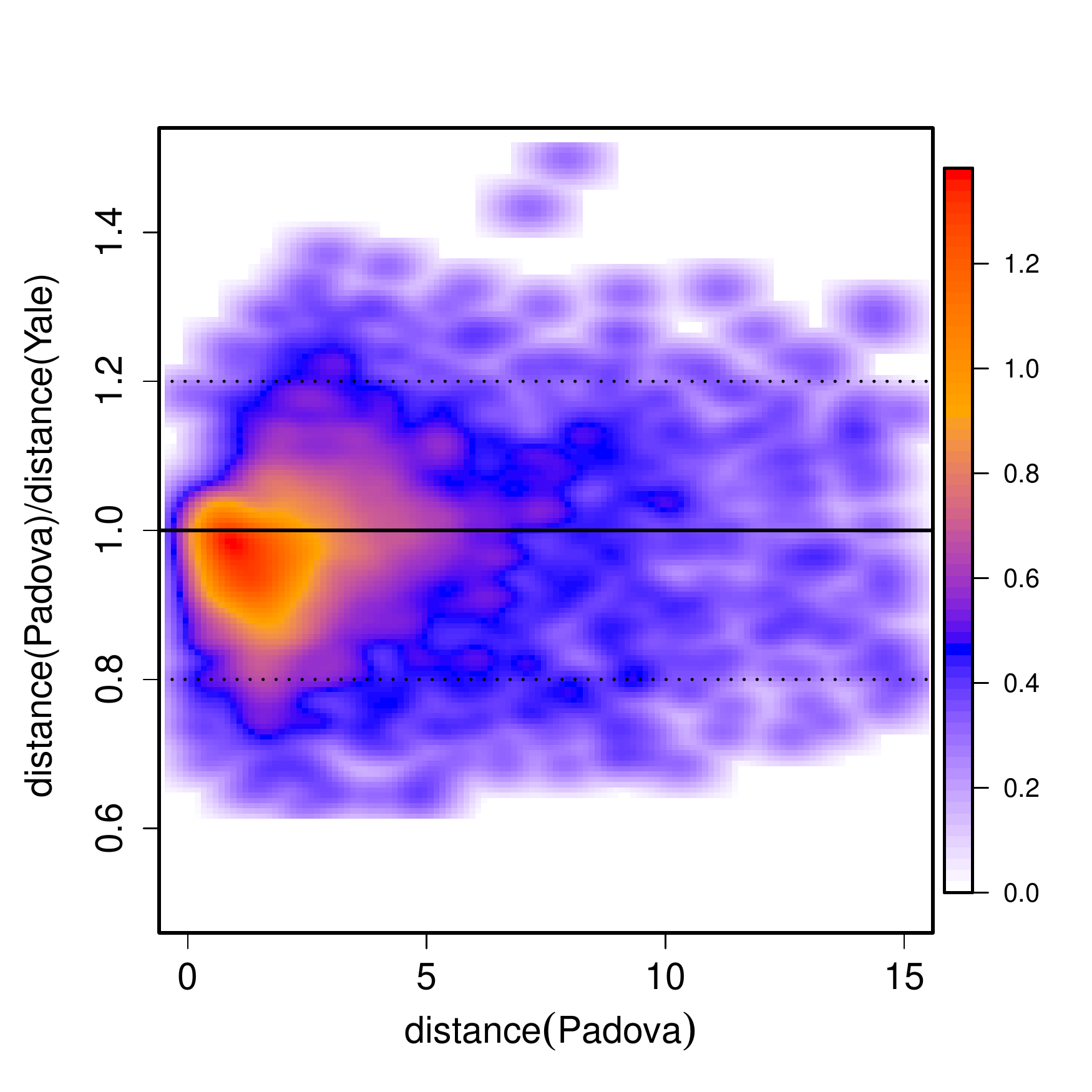}
\caption{{\bf Left:}  Density distribution of the difference in E(B-V) derived from the Padova and the Yonsei-Yale (YY) isochrones as a function of E(B-V)
from the YY isochrones. {\bf Right:} Ratio of distances derived from the Padova isochrones to the YY stellar library, as a function of distance. The dashed horizontal lines indicate $\rm \pm20\%$ difference.}
\label{stellib}
\end{figure}

\begin{figure}[!htbp]
 \includegraphics[width=0.30\textwidth]{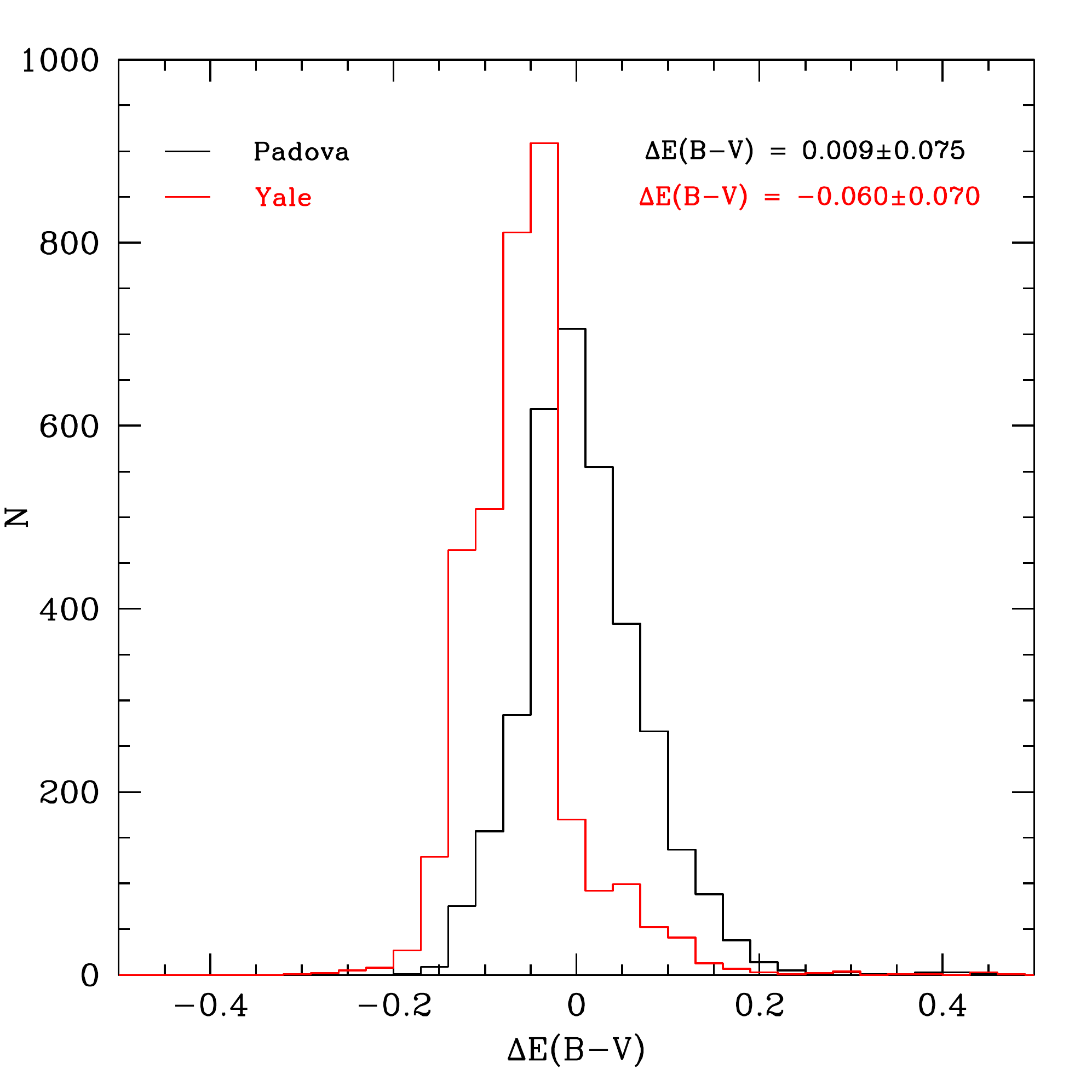}
\caption{Histogram of the differences between our derived E(B-V) and the Schlegel SFD98 value  for high galactic latitude stars ($\rm |b| > 10^{o}$).The black line are  the derived extinction values using the Padova isochrones, the red line those from the Yale isochrones.  The mean value and the r.m.s  scatter is indicated on the top left corner.}
\label{sfd98a}
\end{figure}

\begin{figure}[!htbp]

   \includegraphics[width=0.38\textwidth]{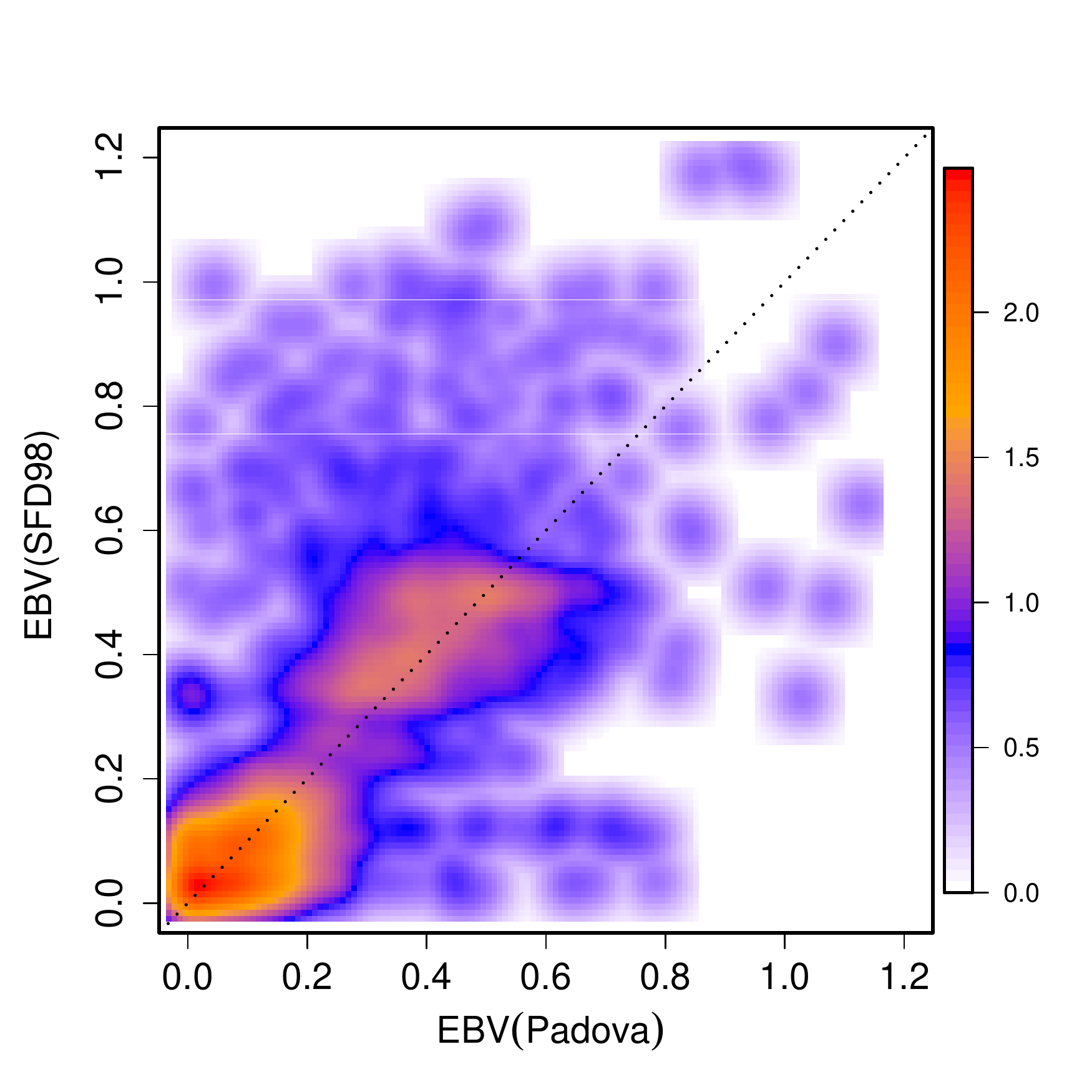} 
\caption{Density plot of E(B-V) against E(B-V) from SFD98.}
\label{sfd98}
\end{figure}

\begin{figure}[!htbp]
\includegraphics[width=0.38\textwidth]{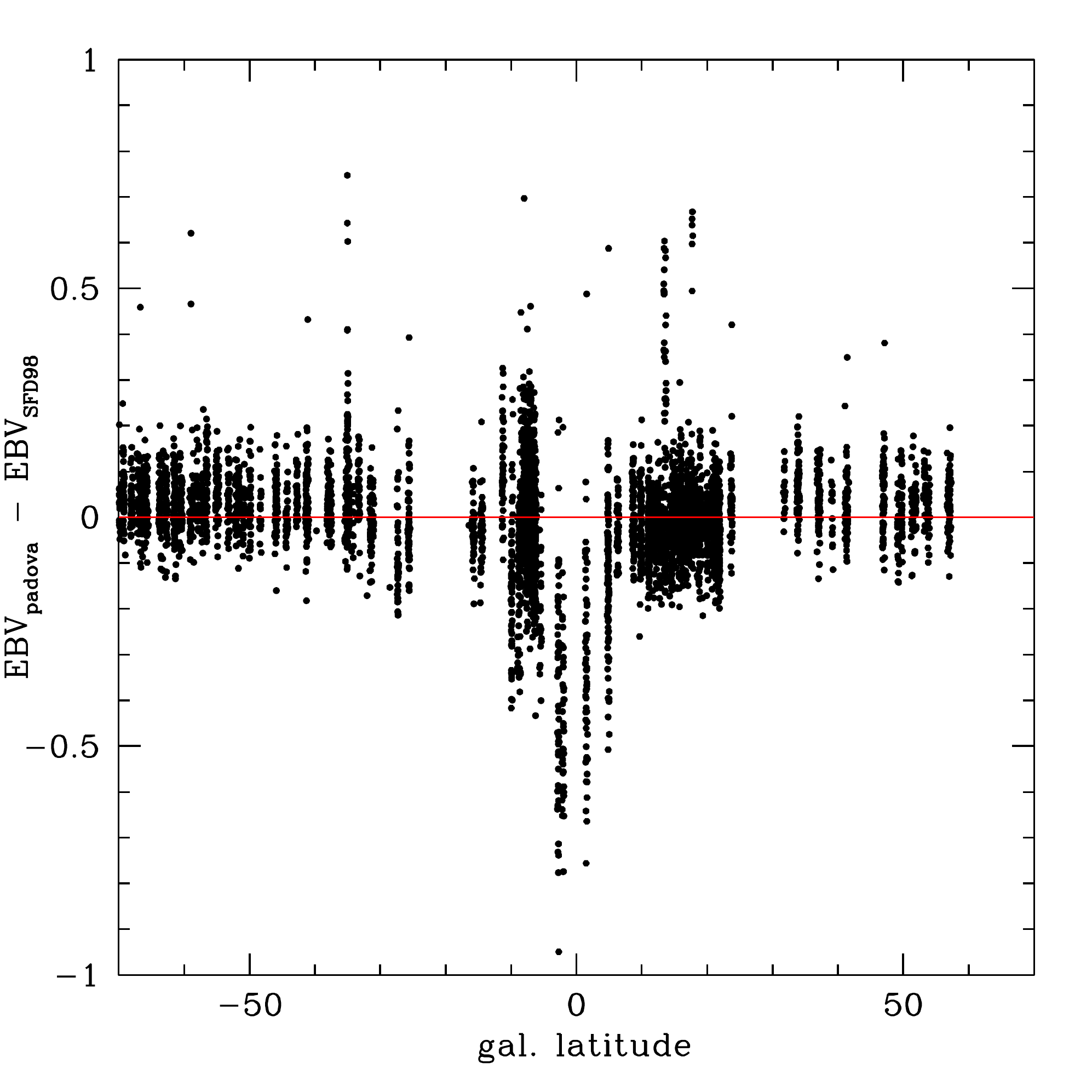}
\caption{Difference of E(B-V) to E(B-V) from SDF98 vs. Galactic latitude}
\label{hist_gal}
\end{figure}

Table~\ref{errors} shows the errors in the derived distances (in \%) if one assumes an offset in $\teff \pm 100\,K$, $\logg \pm 0.2\dex$ and $\meta \pm 0.1\dex$. 
Clearly seen is the large impact of $\logg$ for giants with up to $\sim 34\%$ error in the derived distances while temperature  and metallicity offsets play only a minor role.
 For dwarf stars and subgiants,  the effect of the surface gravity is the smallest ($\rm < 10\%$) due to the fact the main-sequences roughly overlap at all ages. On the other hand, offsets in $\teff$ have a larger effect  for dwarf  stars, because of the overall steeper slope of the main-sequences in the $J$ vs $(J-K)$ plane. We note, however, that the effect of $\teff$ offsets rarely exceeds 10\%.

Figure \ref{sigmaTeff} shows the  difference between the  GES $\teff$ and $\logg$  and  the projected value on the Padova isochrones as a function of $\rm \sigma_{\teff}$ and $\rm \sigma_{\logg}$ (the internal GES parameter dispersions resulting from the three individual nodes, see Sect.~2).
%the internal GES dispersion of the effective temperature ($\rm \sigma_{\teff}$) and gravity ($\rm \sigma_{\logg}$) resulting from the three individual GES nodes. }}
It shows that when the internal GES parameter dispersions are large, then the absolute differences between the projected $\teff$ and $\logg$ and the recommended GES ones also increase. In general, this indicates that for those few  stars, the recommended parameters do not lie on top of theoretical isochrones and that offsets are performed during the projection.
%difference gets larger for $\rm \sigma_{\teff} > 400\,K$ and $\rm \sigma_{\logg} > 0.4\dex$ and that $\teff$ is slightly overestimated while $\logg$ is underestimated. This shows the powerful tool of  $\rm \sigma_{\teff}$  and $\rm \sigma_{\logg}$  of GES  to select the most reliable stellar parameters.
 Given the trends illustrated in Fig.\,\ref{isochrones}, we will use for our analysis only the targets with  $\rm \sigma_{\teff} < 300\,K$ and $\rm \sigma_{\logg} < 0.3\dex$, resulting to a total of  5603  stars.

%\subsection{Differences in extinction and distances between stellar libraries}

\begin{figure*}[!htbp]
\begin{center}
   \includegraphics[width=0.33\textwidth]{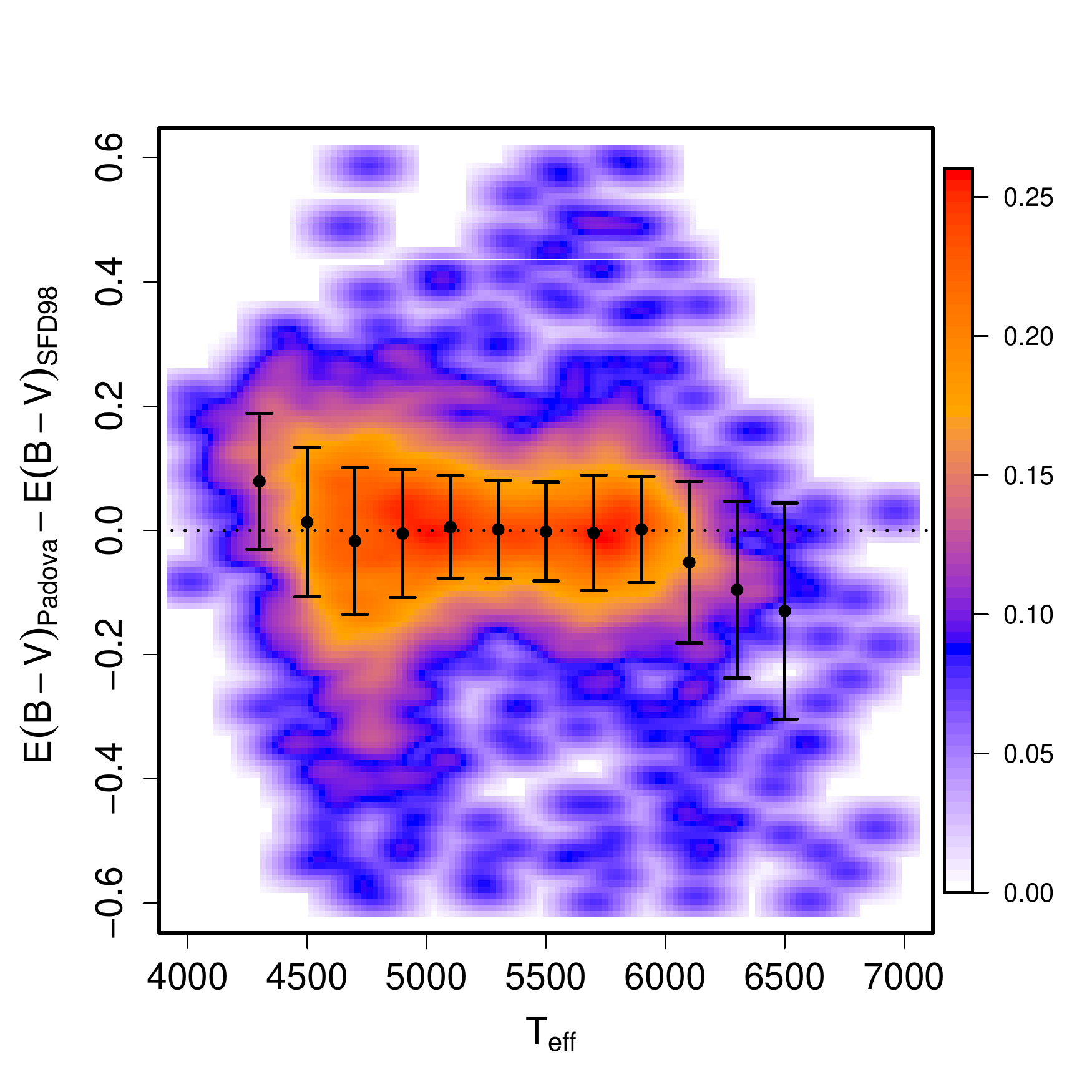}
   \includegraphics[width=0.33\textwidth]{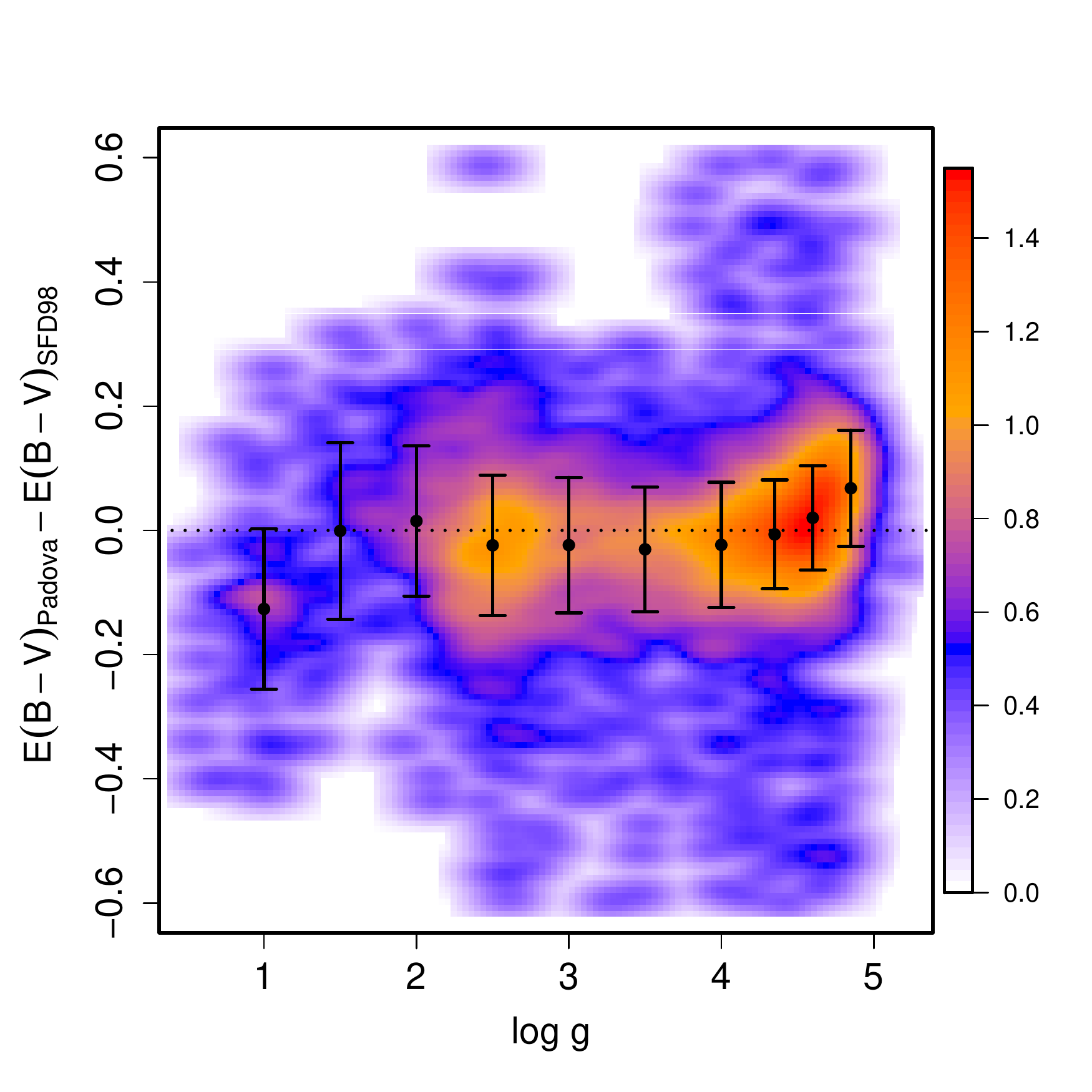}
   \includegraphics[width=0.33\textwidth]{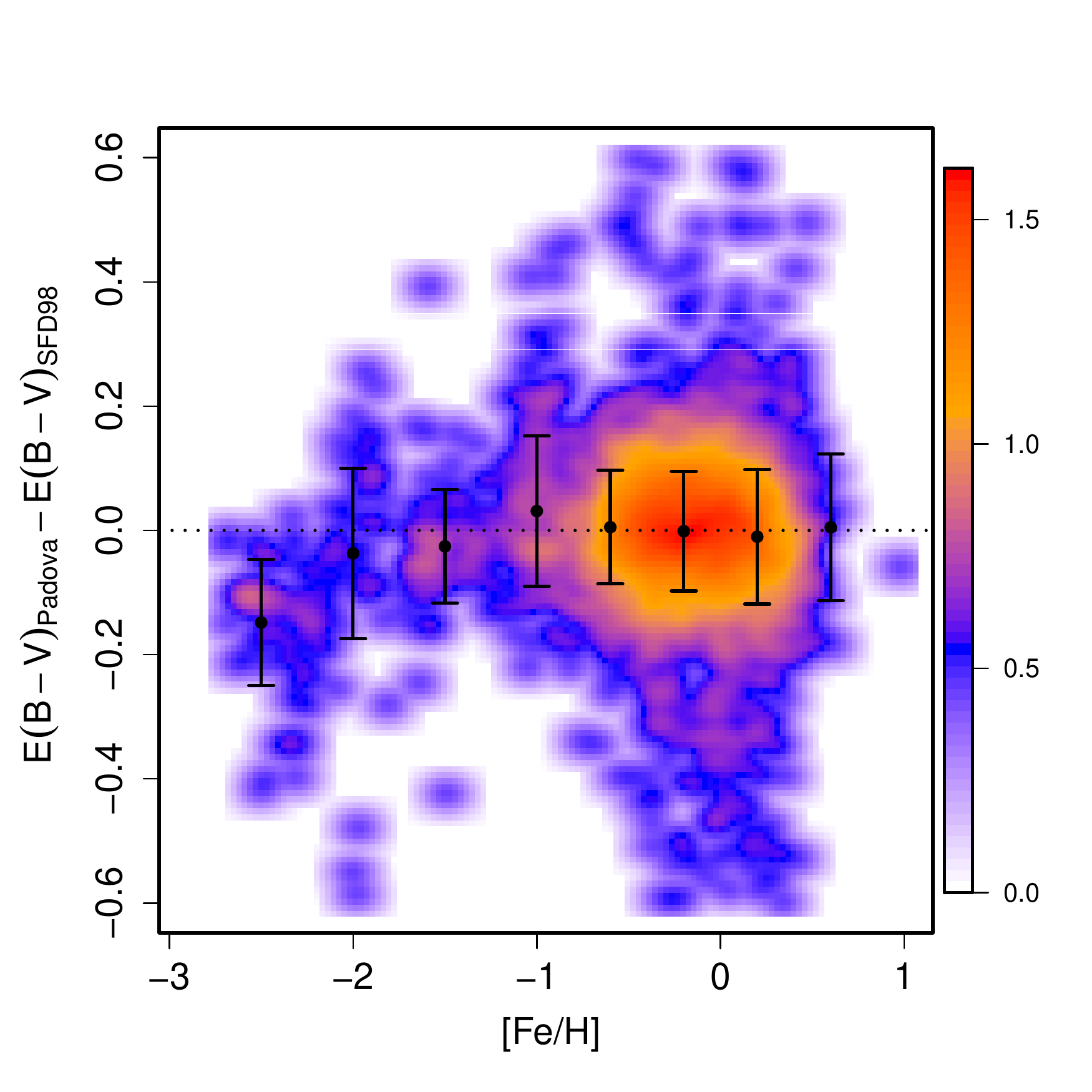}
\caption{Difference in E(B--V) to SFD98 as a function of (a) $\teff$, (b) $\logg$, and (c) [Fe/H] for high galactic latitude stars ($\rm |b| > 10^{o}$).The median value is indicated by black dots together with the standard deviation}
\vspace*{0.4cm}  
\label{distparamstellib}
\end{center}
\end{figure*}

\subsection{Effect of the  stellar isochrone libraries}

The choice of the stellar atmosphere models used to produce the isochrones affect the derived  extinctions and  distances. In what follows, we compare the extinction and distance values derived
 from the  same pipeline (see Sect. \ref{distance} and \citealt{Kordopatis11b})  using alternatively the  Padova and Yonsei-Yale models.
We recall that \citet{schultheis2014} have  shown that
there are only  small differences in the derived extinction and distances between the Padova isochrones and the corresponding Basel3.1 model
library (\citealt{lejeune1997}) for K/M giants observed by APOGEE. Here we want to demonstrate the importance of the chosen model library
for our GES sample covering a much larger $\teff$ and $\logg$ range compared to APOGEE.

Figure~\ref{isochrones} compares the Padova and Yonsei-Yale libraries for different combinations of ages and metallicities.
One can see that the differences between the two libraries increase with decreasing metallicity. These differences concern  the positions of the turn-off and the giant branch, in the sense that YY has an offset towards cooler temperatures, i.e. towards redder colours.  
However, the known age-colour (or age-metallicity) degeneracy (e.g. \citealt{Bergemann2014}, \citealt{Worthey94}) makes that the YY and the Padova models can partly overlap at a given metallicity when selecting younger YY isochrones.  Since our procedure projects the observed $\teff$, $\logg$ and [M/H] on all the ages of a set of isochrones (see above), the differences in the derived magnitudes will therefore be smaller for most types of stars than what is suggested by a simple one-to-one comparison between the isochrones. However, for the stars at the boundaries of the libraries (e.g. the hotter stars and the more metal-poor giants), the differences are expected to be the largest, due to the described offset. 
Finally, we also investigated the effect on the shape of the isochrones (and therefore on the derived absolute magnitudes) of  atmospheric models with different $\alpha-$enhancements. The dashed and plain lines in Fig.~\ref{isochrones}, obtained for the YY isochrones, indicate   only a very small effect in the  $\rm T_{eff}$ vs. log\,g  and  J vs (J--Ks) diagram.  We conclude, in agreement with other studies \citep[e.g.][]{Breddels10,Zwitter10}, that the adopted $\alpha-$enhancement level of the isochrones is not affecting significantly the final distance estimations.

Figure~\ref{stellib}  shows the comparison of the derived extinctions (left panel) and distances (right panel). It is obvious from
this comparison that the YY E(B--V) values are systematically smaller than the Padova ones. In agreement with what has been stated in the previous paragraph, this means that the intrinsic colours from
the YY models are redder than the Padova ones. The effect in E(B--V) can reach up to 0.2\,mag showing that the choice of a certain stellar library is essential for
the  extinction determination. If one transforms this into distances (see right panel of Fig.\ref{stellib}), the distances from the YY isochrones are  systematically larger, especially for 
  $\rm d < 3\kpc$.  We traced these differences as function of the stellar parameters, $\teff$, $\logg$ and $\meta$  and could identify that those arise mainly
for dwarf stars ($\logg>4$) with $\rm 5500 < \teff < 6500\,K$. This is consistent with Fig.~\ref{isochrones} 
where the Padova isochrones predict bluer J--K colours.  On the other hand, YY underestimate distances for cool giant stars with
$\rm \logg < 3$ and $\rm 4000 < \teff < 5000\,K$. Finally, the difference between Padova and YY increases for the most metal-poor stars ($\meta < -1\dex$).  All the above discrepancies indicate important differences in the stellar atmosphere models between Padova and YY. For the majority of our objects, the differences between the two stellar libraries are within 20\% (right panel of Fig.~\ref{stellib}). In the following Section~\ref{Schlegel}, we will confront
 the derived E(B--V) values with the dust map of \citealt{schlegel1998}.

\section{Comparison to the 2D extinction maps: The Schlegel map} \label{Schlegel}

Contrary to the study of \citet{schultheis2014} with APOGEE where the fields were concentrated in regions of high extinction towards the Galactic Bulge, we analyse
here with GES lower extinction fields at higher latitudes. We use the  \citet{schlegel1998} dust map,  hereafter referred to as SFD98, as it has the same sky coverage as our GES stars.  We used the conversion $\rm E(J-K_{S})/E(B-V) = 0.527$ \citep{rieke1985}.

\begin{figure*}[!htbp]
\begin{center}
\includegraphics[width=0.48\textwidth]{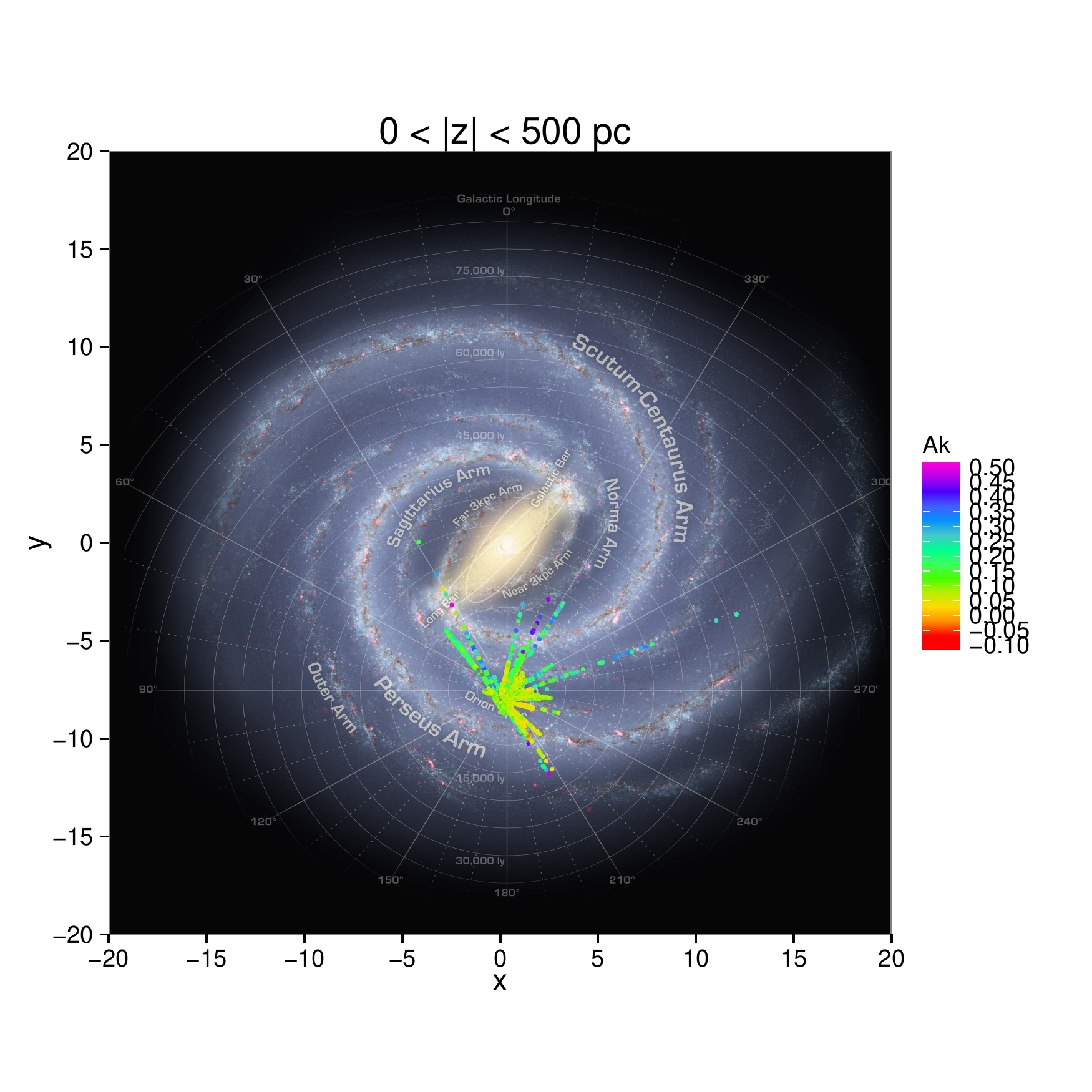}  \includegraphics[width=0.48\textwidth]{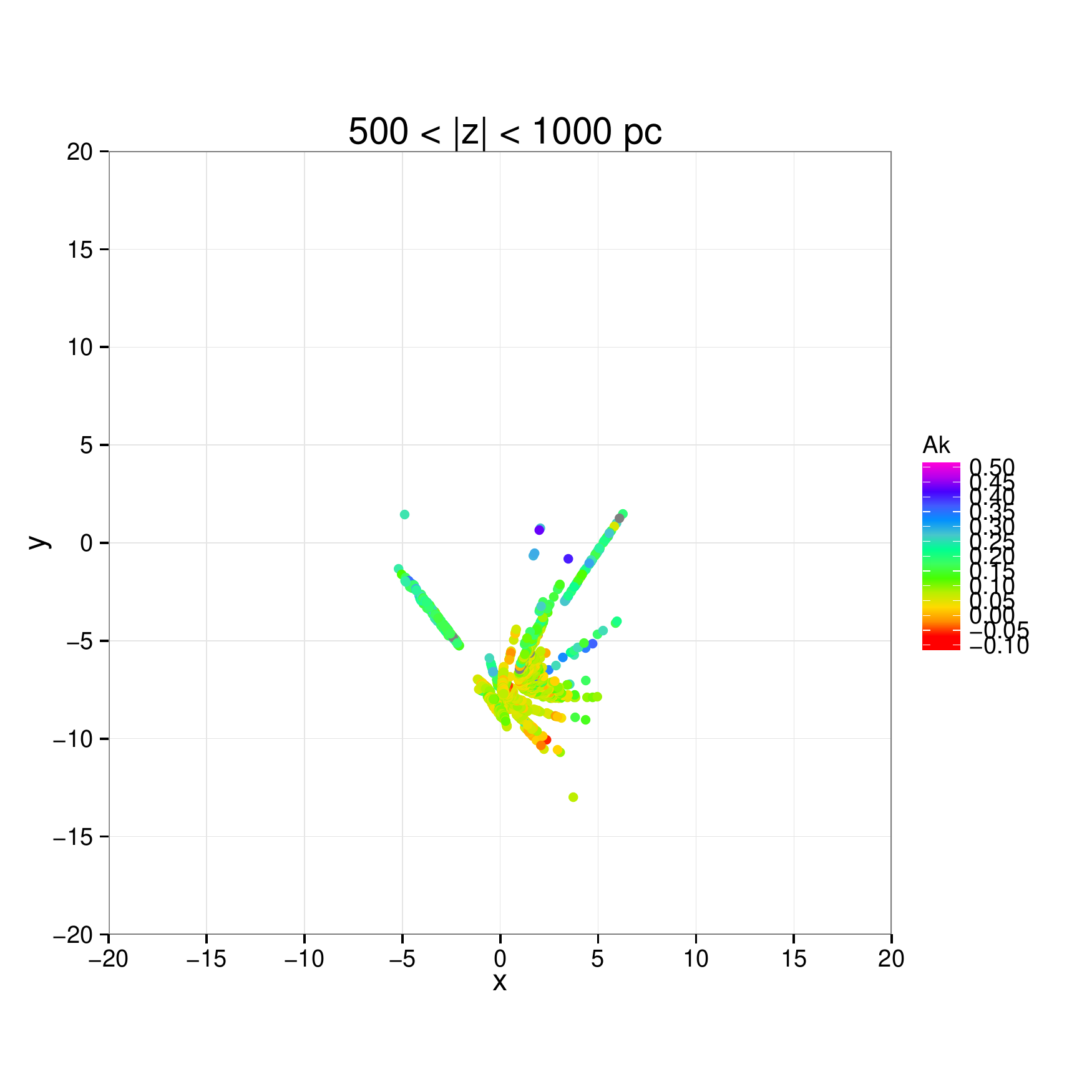}
\includegraphics[width=0.48\textwidth]{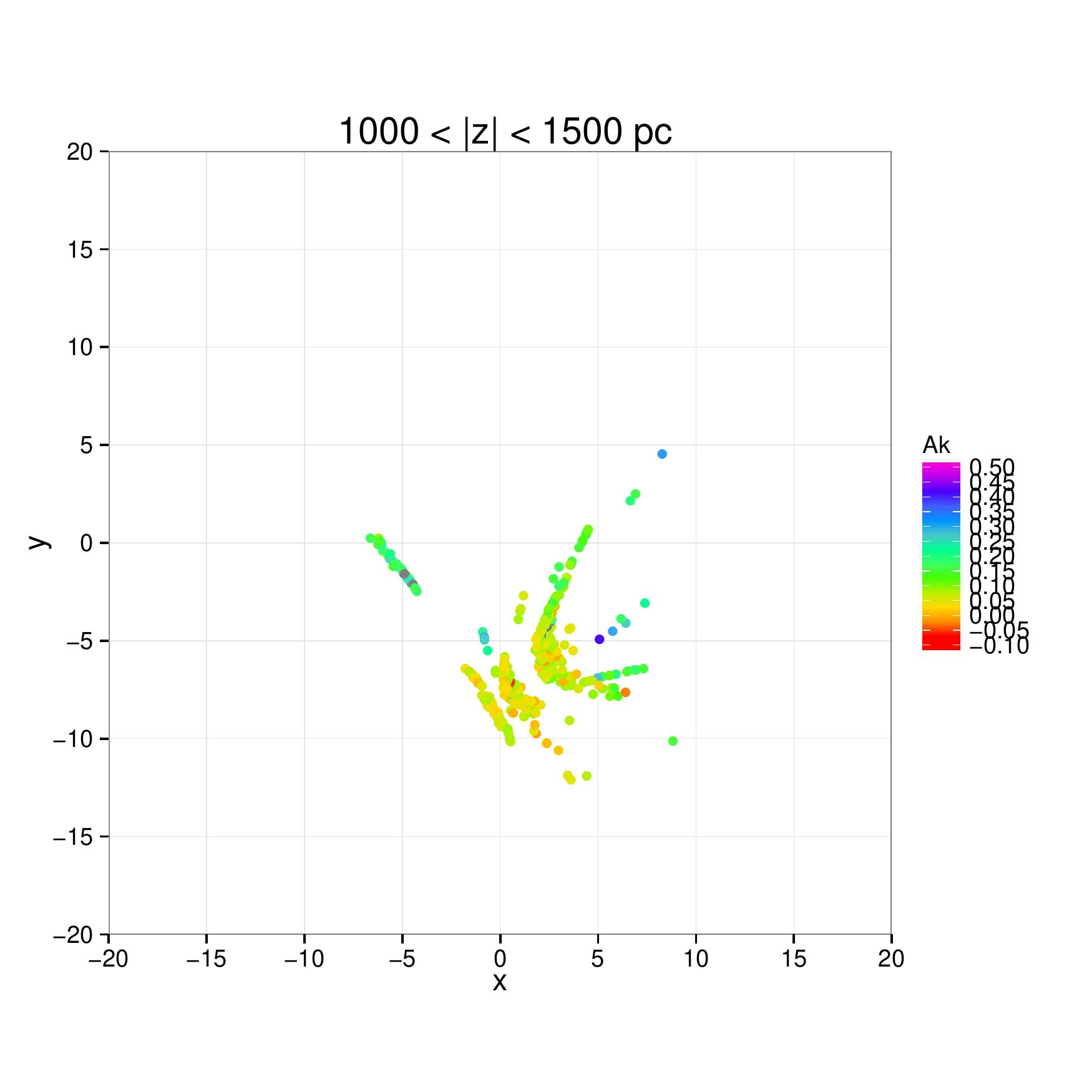}  \includegraphics[width=0.48\textwidth]{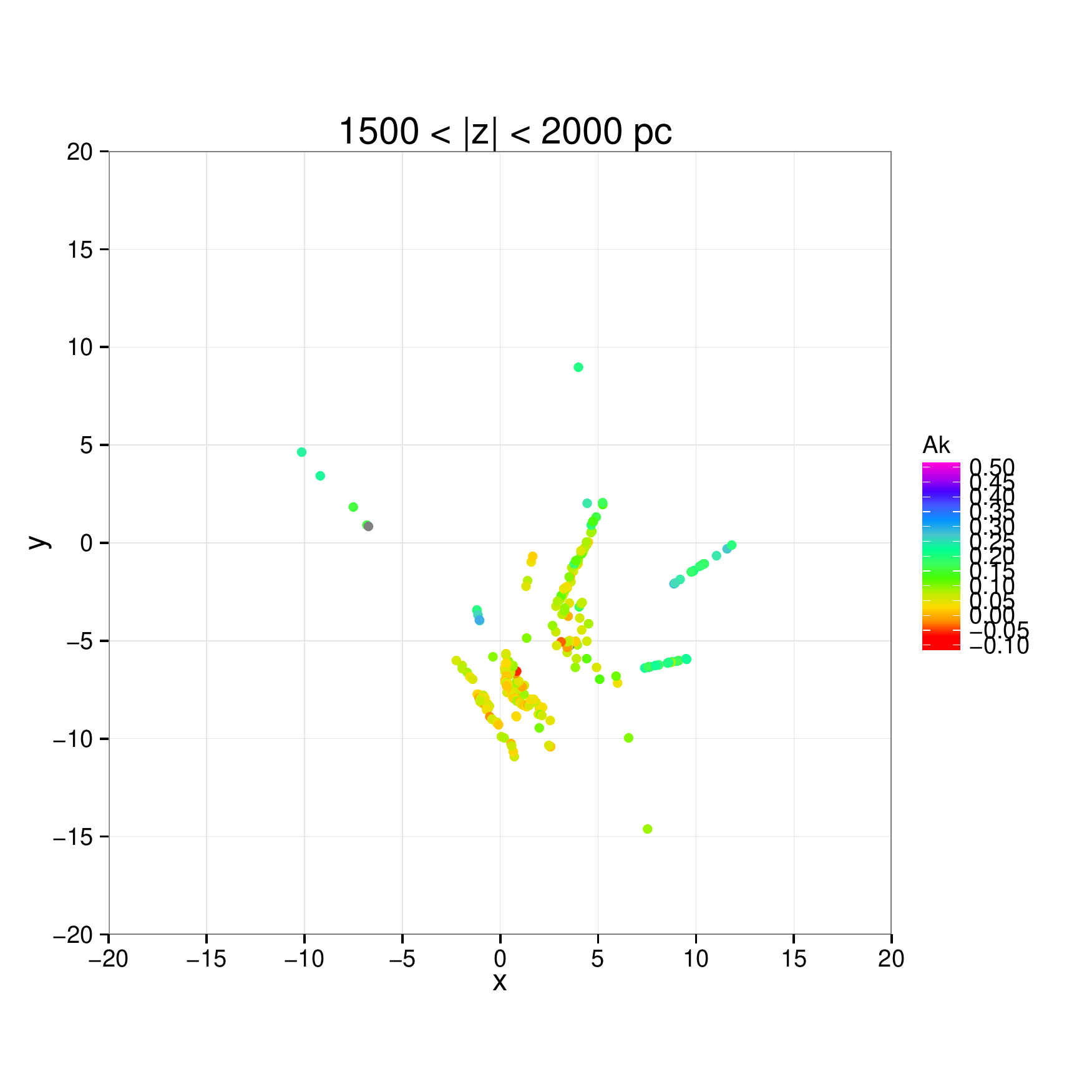}
\caption{Extinction in the (X,Y) plane for different heights above the galactic plane. On the left upper plot an illustration of our Galaxy produced by R. Hurt is superimposed. } 
\label{ext3d}
\end{center}
\end{figure*}

\begin{figure*}[!htbp]
\begin{center}
\includegraphics[angle=90,width=0.95\textwidth]{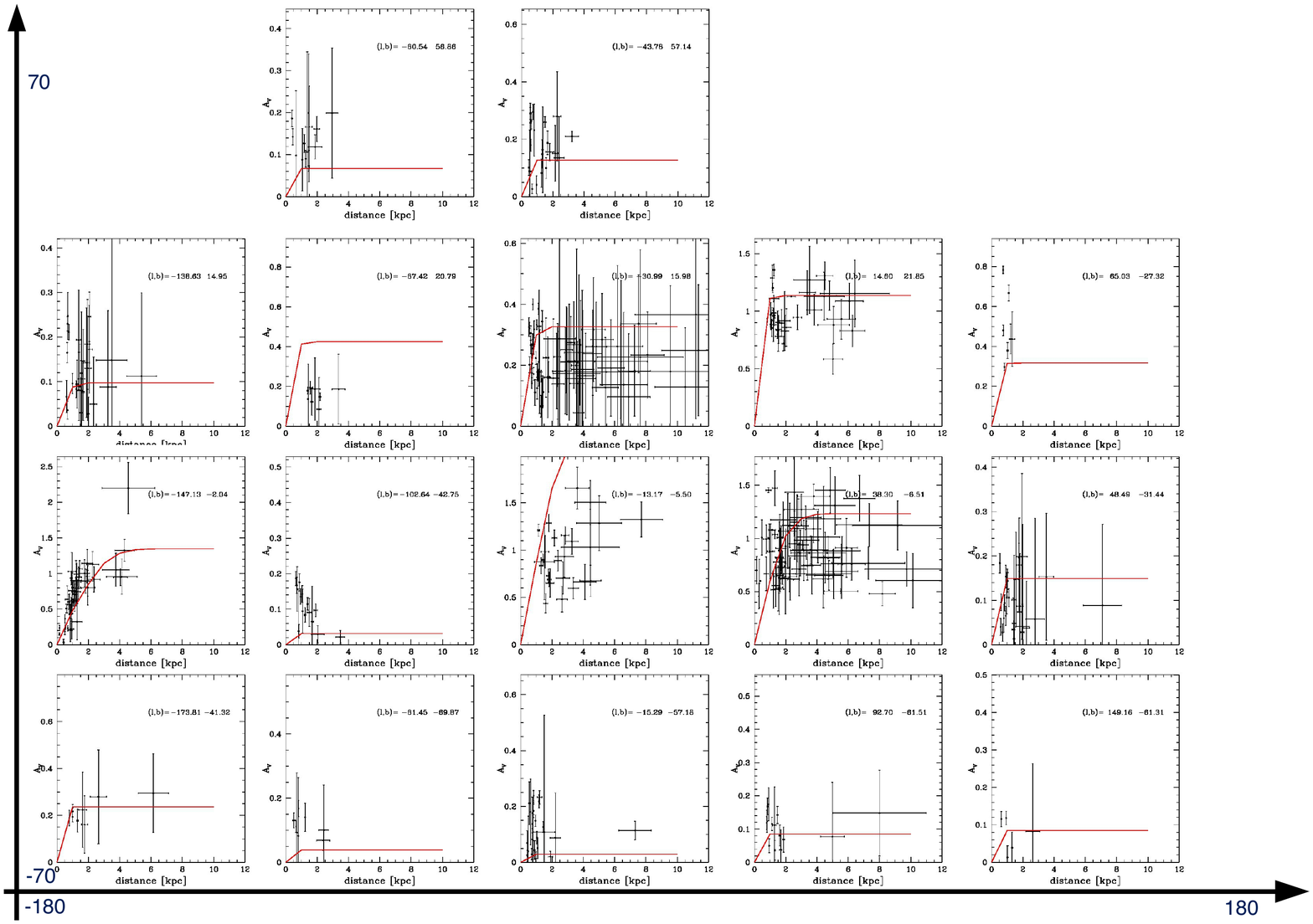}
\caption{Extinction vs distance for different lines of sight. The x-axis and y-axis give approximate the location in Galactic coordinates. The Drimmel et al. (2003) model is superimposed in red}
%\vspace*{0.4cm}  
\label{3Dextinction}
\end{center}
\end{figure*}

As already mentioned, stellar libraries can give systematically offset extinctions.  Figure~\ref{sfd98a} shows the difference in E(B--V) between  the SFD98 values and  the ones derived with the Padova isochrones (in black), and the YY isochrones (in red). We see clearly that the Padova isochrones match better the $\rm E(B-V)_{SFD98}$ with  a mean difference of $\rm 0.009 \pm 0.075$  which is the typical uncertainty of our method (derived from Eq.\,3).  If we include low galactic latitude fields with $ |b| < 10^{\circ}$,  the dispersion increases to 0.18\,mag where it is suspected that SFD98 overestimates extinction (\citealt{schlafy2014}). \citet{schlafy2011} measured the dust reddening using the colors of stars derived from stellar parameters from the SDSS. Their  uncertainty is in the order of  30 mmag  for high latitude fields and $\rm E(B-V) < 0.04$. Our  larger dispersion is due to the  lower galactic latitude fields of the GES fields. The YY isochrones systematically overestimate $\rm E(B-V)_{SFD98}$ by 0.065\,mag.  For the remaining of our 
 analysis, we decided to use the Padova isochrones, as they match better the SFD98 values.

Figure~\ref{sfd98} shows the comparison between $\rm  E(B-V)_{Padova}$ and $\rm  E(B-V)_{SFD98}$. While there is no shift
in the derived E(B-V),  a large scatter is seen, especially for $\rm E(B-V)_{SFD98}  > 0.5$ where the SFD98 values are higher than $\rm E(B-V)_{Padova}$. 
Our results are in agreement with those  of \citet{schlafy2014} where they compared their map based on  PAN-STARRS photometry with the Schlegel map. They find systematically higher E(B--V) of SFD98 for $\rm E(B-V)_{SFD98} > 0.3$.  The Planck dust map at 353\,GHz  show   a  similar behaviour  (see \citealt{schlafy2014})  indicating that for higher  E(B--V)  the  far-infrared modelling of the dust  shows some systematical  offsets.  A revision of these models is therefore needed.
\smallskip

 We now assess the existence of  biases   as a function of $\teff$, $\logg$ and $\meta$ between our derived extinction and the SFD98 map.
 Figure~\ref{distparamstellib} displays a small trend of  the differences in E(B--V) towards higher temperatures ($\teff > 6000\,K$) in the sense that SFD98 gets higher  E(B--V) with respect to $\rm E(B-V)_{Padova}$. For cooler temperatures ($\teff  < 4500\,K $) SFD98 predicts higher E(B--V) than $\rm E(B-V)_{Padova}$. While  for the range $\rm 1.5 < \logg < 4.5$ no trend is visible, giants with  $\logg < 1.5$ have a slightly overestimated extinction compared to $\rm E(B-V)_{SFD98}$. Interestingly, our E(J--K) estimation for dwarf stars with $\logg > 4.5$ and $ 4000 < \teff < 6000\,K$ is slightly offset towards higher values  with respect to $\rm E(J-K)_{SFD98}$. 

The derived extinction depends on the corresponding matched colour of the isochrone as well as of the stellar parameters, the surface gravity being a particularly sensitive parameter. Indeed, systematic offsets of 0.2 dex can significantly shift the extinctions and distances (see Sect. 3). This effects mainly giant stars while GES has mostly dwarf stars (see Fig. 2 of \citealt{recio-blanco2014}).
%The APOGEE survey for example had to apply  a systematic correction of $0.2-0.3\dex$ with respect to stars with asteroseismic stellar parameters from NASA's Kepler mission.
%A similiar study for the GES parameters is awaited for the DR3 release with the surface gravities of COROT stars.

% An  additional interesting question is if a different Galactic environment with different metallicity can bias the SFD98 map. 
% Towards low-metallicities ($\meta < -1\dex$) one can see that the SFD98 map gives  overestimated values compared to  $\rm E(B-V)_{Padova}$. This could be due to variation in the extinction law with metallicity as the dust grain size distribution depends on the metallicity. This has been demonstrated by \citet{weingartner2001} who constructed different size distributions for carbonaceous and silicate grain populations in the Milky Way, LMC and SMC which can be parameterised by the ratio of visual extinction to reddening $\rm R_{V}$. We will investigate this further in Sect.~\ref{law}.  Finally, we also investigated the effect of alpha elements, and did not notice any systematics with this parameter.

%However, the statistics of metal-poor stars in the DR2-GES release is rather  poor.

\begin{figure*}[!htbp]
\begin{center}
\includegraphics[width=0.33\textwidth]{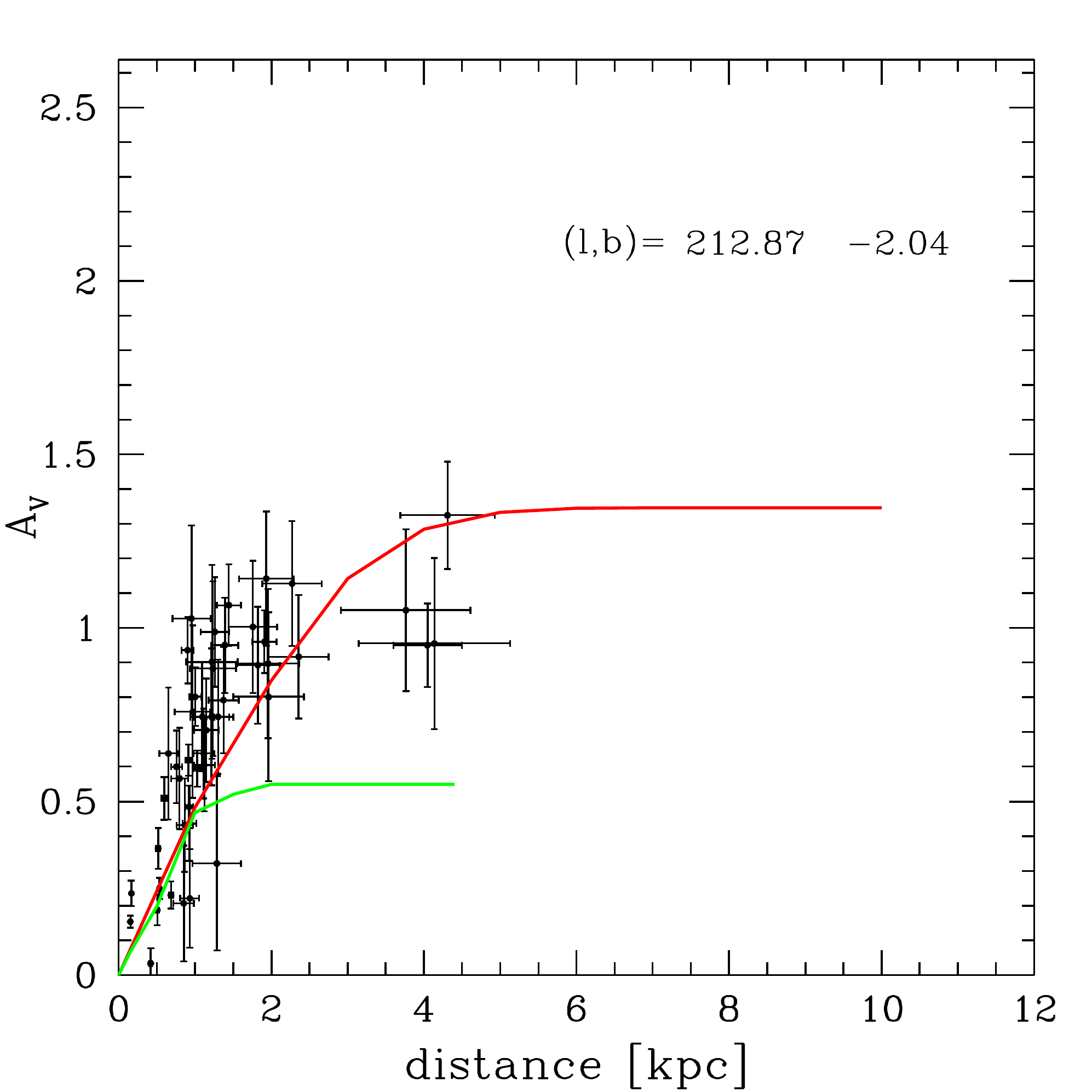} \includegraphics[width=0.33\textwidth]{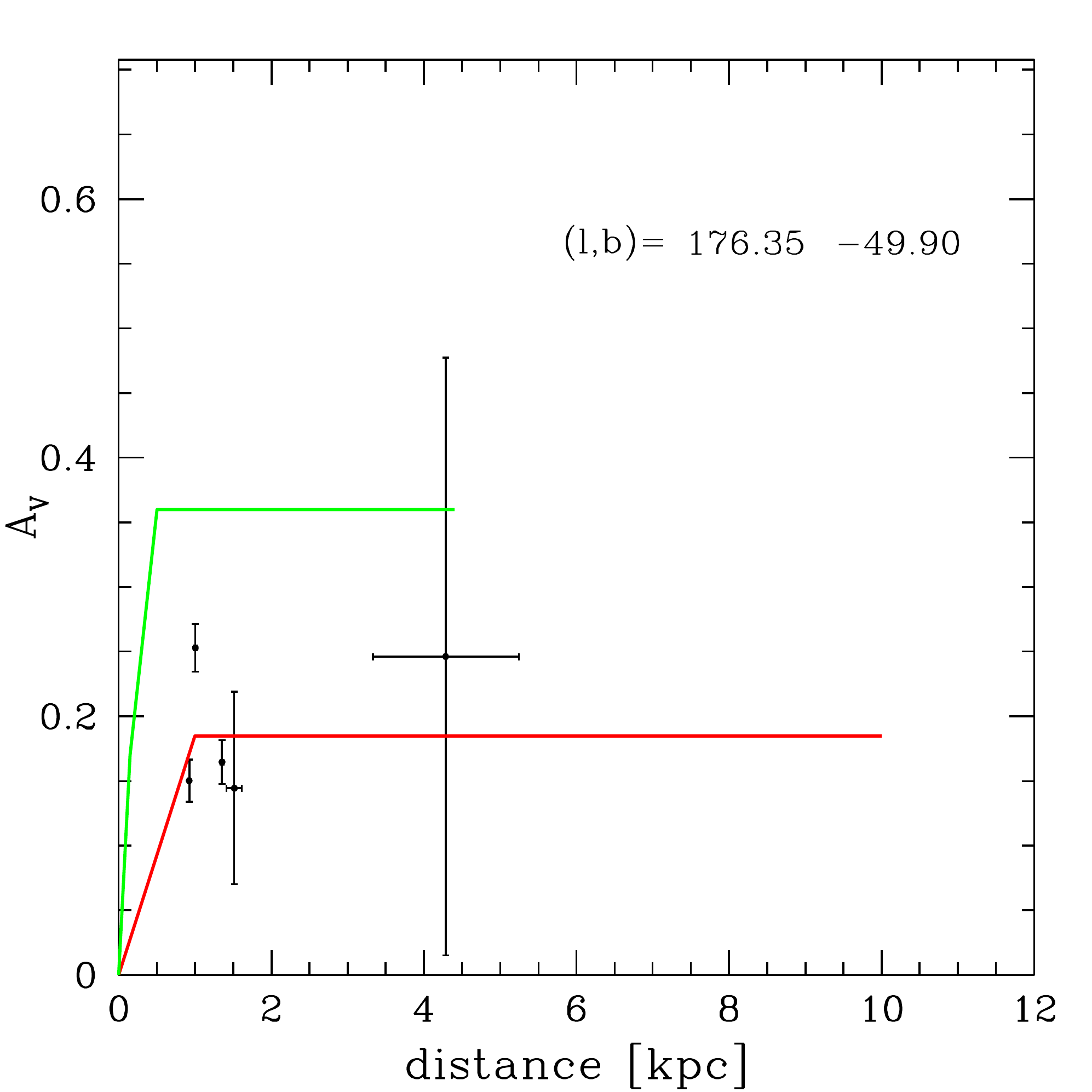} \includegraphics[width=0.33\textwidth]{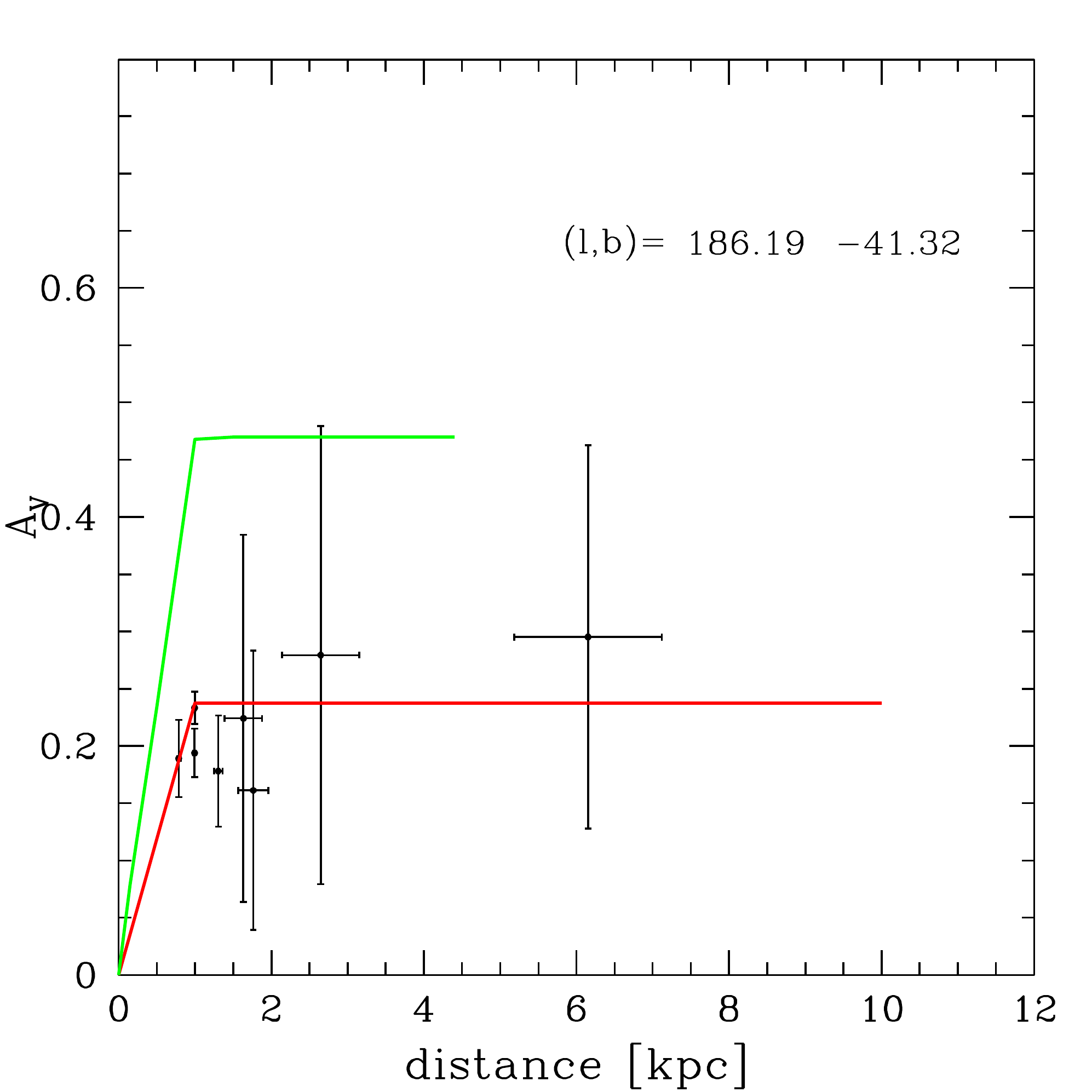}
\includegraphics[width=0.33\textwidth]{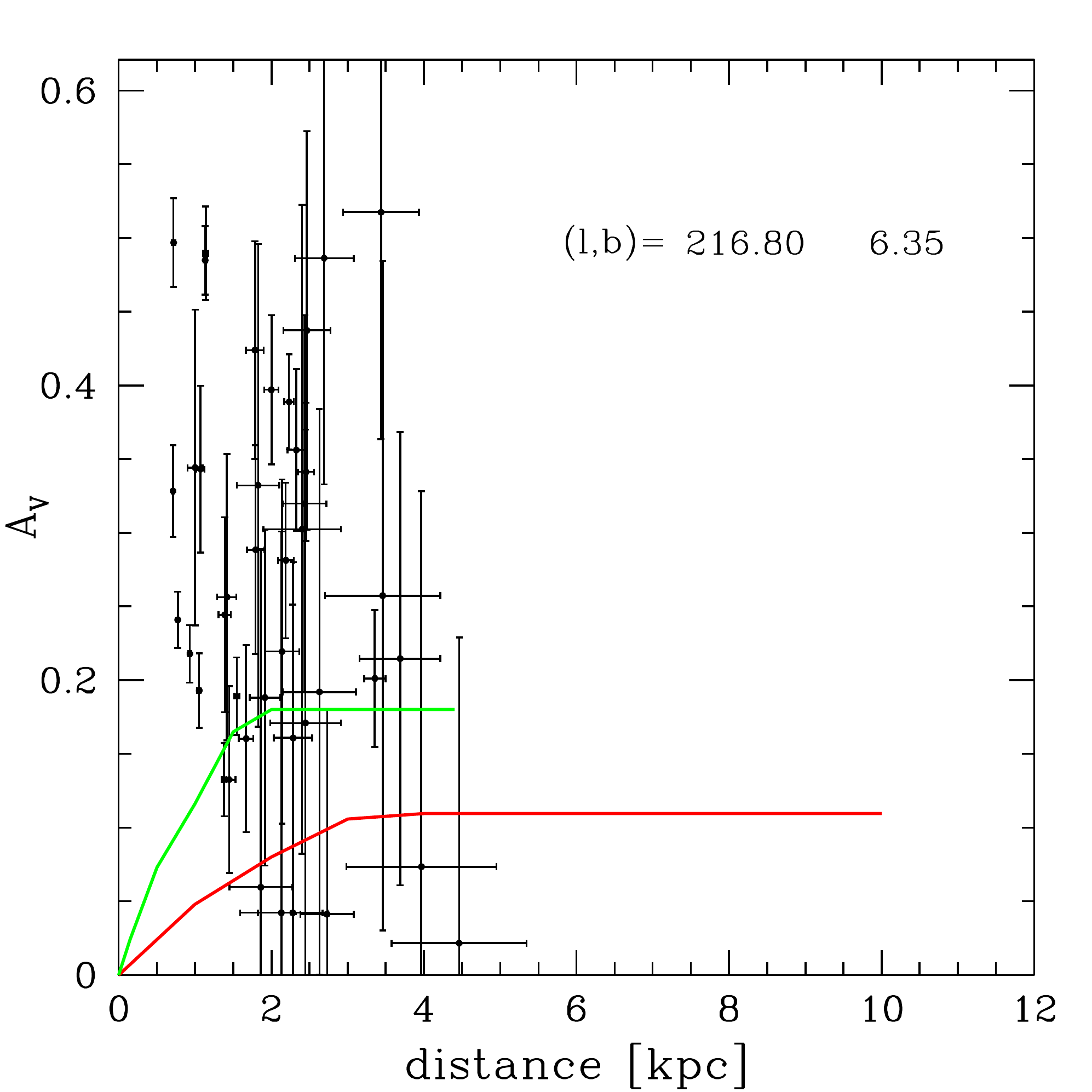} \includegraphics[width=0.33\textwidth]{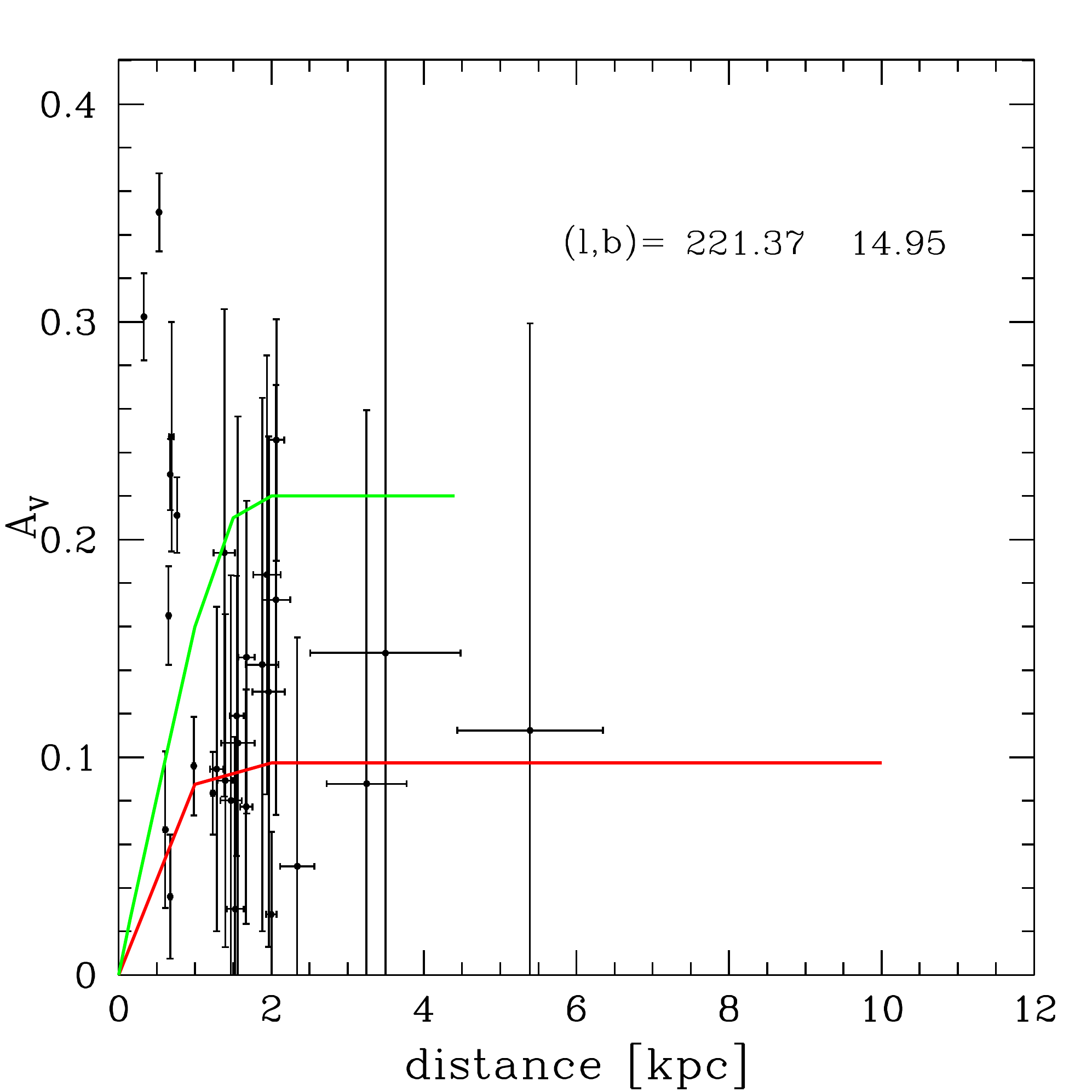} \includegraphics[width=0.33\textwidth]{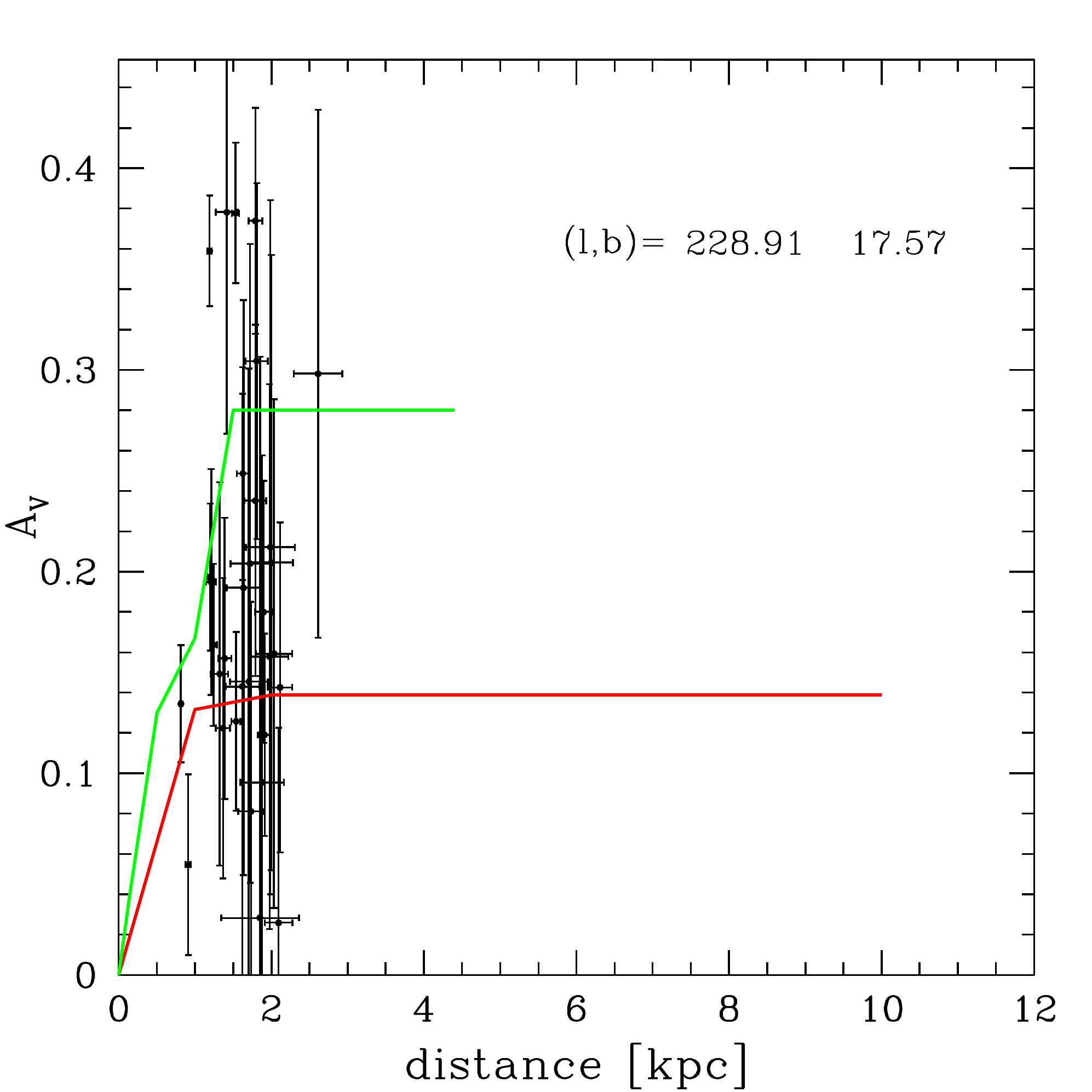}

\caption{Extinction vs distance for different lines of sight where. In red the 3D extinction model of Drimmel et al. (2003)  is superimposed while in green the 3D-model of Chen et al. (2013).} 
\label{chen3Dextinction}
\end{center}
\end{figure*}

\section{Comparison to three-dimensional maps}

In this section, we compare our 3D extinction distributions with the model of \citet{drimmel2003}  and the  map of \citet{chen2014a}. The \citet{drimmel2003}  model is based on the far and near-IR data fits of the dust distribution made by \citet{drimmel2001} from the COBE/DIRBE instrument. The spatial resolution of this map is approximately 21\arcmin $\times$ 21\arcmin, we however recall that the Drimmel et al. model  does not include features related to the Galactic bar nor the nuclear disk, resulting to systematic overestimates of the extinction towards the Galactic Bulge, \citep[see][for further details]{schultheis2014}.

As far as the \citet{chen2014a} 3D map is concerned, we found only a limited spatial overlap with six GES fields. \citet{chen2014a}  combined optical photometry (g, r, i) with 2MASS (J, H, K) and WISE (W1, W2) photometry and used the method of \citet{berry2012} to trace the stellar locus in a multi-dimensional colour space. The constructed  3D map is over roughly 6000 sq. degree towards the Galactic anticenter region, with an angular resolution varying between 3--9\arcmin.

%Figure~\ref{ext3d} displays the face-on view of the  3D map as a function of the scale height, using our GES sample in the Cartesian Galactocentric (X,Y) plane, with the Sun located at 8 kpc from the centre. Superimposed is the illustration of our Galaxy produced by Robert Hurt based on the scientific results of the Spitzer Infrared Space telescope (R. Benjamin).  Compared to the APOGEE targets, where the majority of them are located at distances larger than about 6\,kpc,  the GES targets are concentrated much closer to the Sun with typical distances $\rm d \sim 2-3\,kpc$. Unfortunately, only one direction towards  the direction of the Galactic bar   ($\rm l=28^{0}$, $\rm b=-3^{0}$)  has been observed by GES. However, we see there a higher concentration of dust  which \citet{schultheis2014} identified as a  dust lane in front of the bar. Clearly visible  is also the increased dust amount associated with the Perseus spiral arm, the Sagittarius arm and the Scutum-Centaurus arm. Most of the low extinction is situated in the first few kpc around the
%Sun's position.

Figure~\ref{ext3d} displays for different heights above the Galactic plane, the  $A_K$ extinctions measured from the GES targets. The illustration is made in a Cartesian Galactocentric (X,Y) frame,  with the Sun located at $8\kpc$ from the centre. 
Superimposed in the panel representing the closest distance from the plane,  is the illustration of our Galaxy produced by Robert Hurt based on the  results of the Spitzer Infrared Space telescope (R. Benjamin).  Compared to the APOGEE targets, where the majority of them are located at distances larger than about 6\,kpc,  the GES targets probe a volume much closer to the Sun with typical distances $\rm d \sim 2-3\,kpc$.  From Fig.~\ref{ext3d}, one can see that GES also contains a line-of-sight in the direction of  the Galactic bar  ($\rm l=28^{\circ}$, $\rm b=-3^{\circ}$),    which shows  a higher concentration of dust. According to \citet{schultheis2014} this is associated to a  dust lane in front of the bar. Clearly visible  is also the increased dust amount associated with the Perseus spiral arm, the Sagittarius arm and the Scutum-Centaurus arm. Most of the low extinction is situated in the first few kpc around the Sun's position.

Figures~\ref{3Dextinction} and \ref{chen3Dextinction} show  the 3D extinction for a few selected lines of sight and compare our results with the ones of Drimmel et al. and Chen et al. 
{We present in the online table  (see \ref{online}) additional lines of sights of 3D extinction for different GES fields. Below we describe a few trends:
%The complete map can be found in the online table (see \ref{online}). Below we describe a few trends:
%In contrast to the work of \citet{schultheis2014} using APOGEE data, we lack GES stars at distances 
%larger to the sun that $\sim$ 4\,kpc. However, we can trace extinction for the  first few kpc. Below we describe a few trends. 

\begin{itemize} 
\item Contrary to the APOGEE data,  GES samples  the first few kpc at higher spatial resolution in distance.  In general one sees  a steep rise in $\rm A_{V}$ (see Fig. \ref{3Dextinction}).
\item The Drimmel et al. model underestimates $\rm A_{V}$ systematically   for high Galactic latitudes ($\rm l > |50|$). The Chen et al. map predicts a steeper increase in $\rm A_{V}$ for the first
kpc than Drimmel et al. which seem to be in better agreement with the GES data (see Fig.~\ref{chen3Dextinction}).
\item For most of the lines of sight, we see a steep rise in $\rm A_{V}$ followed by a flattening afterwards which is qualitatively described by the Drimmel model. Due the low absolute extinction values,
 the errors in the derived extinction and distances become significant for distances larger than 4\,kpc.
\item \citet{puspitarini2014} studied the DIBS of 225 GES stars in five fields  and found a good correlation between the DIB strength and the extinction.  We have one field in common which is the ``COROT-ANTICENTER'' field (see upper left panel of Fig.\ref{chen3Dextinction}, located at $\rm (l,b)= (+212.87, -2.04)^{o}$. They found a steep increase in  extinction up to 1\,kpc, a plateau between 1 to 2.5\,kpc and  a second  increase  beyond 2.5 kpc (see their Fig.~6).  We confirm the steep increase in $\rm A_{V}$ between 0 and 1.5\,kpc with a flattening starting at around  2\,kpc. There are too few data points to see their second increase in $\rm A_{V}$. 

\item The GES fields located at $ \rm (l,b)=(+38.30,-6.51)^\circ$, $\rm (l,b)=(+14.60,+21.85)^\circ$, and the field at $\rm (l,b)=(+147.13,-2.04)^\circ$ span the full distance range and follow the  $\rm A_{V}$ vs. distance relation predicted
 by  Drimmel et al.
\item The GES data seem to confirm the general shape of the 3D extinction with a steep rise
in $\rm A_{V}$ for distances up to 4\,kpc and a flattening  which occurs  at shorter distances compared to the  values found with the APOGEE sample of \citet{schultheis2014}.
\end{itemize}

%\begin{figure*}[!htbp]
%   \includegraphics[width=9.0cm]{gallongvsgallat.pdf}
%    \includegraphics[width=8.0cm]{TeffvsdiffEBV_sfd98.pdf}  \includegraphics[width=8.0cm]{loggvsdiffEBV_sfd98.pdf} \includegraphics[width=8.0cm]{FeHvsdiffEBV_sfd98.pdf}
%\caption{(X,Y projection for different |z] of the 3d extinction}
%\label{stell}
%\end{figure*}

With the future data releases of GES in the coming years, we will be able to systematically trace the distance vs. $\rm A_{V}$ behaviour systematically allowing to compare qualitatively spectroscopically  derived extinction with 3D dust models. The combination of both  GES and APOGEE data tracing the low-extinction fields in the Visible as well as the high-obscured fields in the Infrared will  be clearly a  goal for the future for tracing  3D extinction

\section{Interstellar extinction law} \label{law}

Using the extinctions derived by the GES sample, we now investigate the universality of the extinction law, as a function of the position on the sky and the stellar environment. Using APOGEE red clump stars, \citet{wang2014} recently  found a constant power law with $\rm alpha = 1.95$ yielding $\rm A_{J}/A_{Ks} = 2.88$.  They used stars with $\rm 3500 < \teff < 4800\,K$, $\logg < 3$ and
$\rm [Fe/H]  > -1.0\dex$. On the other hand, \citet{yuan2013}  combined SDSS, GALEX, 2MASS and WIDE photometry to determine reddening coefficients from the far UV to the near and mid-IR by using the SDSS spectroscopic archive. They concluded that their newly derived  extinction coefficients differ slightly but favour the R(V)=3.1 Fitzpatrick reddening law (\citealt{fitzpatrick99})  over the  Cardelli et al. (\citealt{cardelli1989})  and  the O'Donnell et al. (\citealt{odonnell94}) reddening laws.

 Here we  perform  a similar study  and aim to push the analysis to the investigation of the systematic dependencies of the extinction law as a function of the stellar parameters ($\teff, \logg, \feh$). Besides the near-IR photometry we use  SDSS photometry  in the filters u,g,r,i,z giving us the the possibility to trace simultaneously 
the extinction law in the optical and the IR. For every band, the colour excess is simply the difference between the intrinsic colour derived
from the isochrone matching method and the observed colour.  Figure~\ref{ce} shows the linear fit results of the different colour excess similar as done by  \citet{yuan2013} and \citet{wang2014}. We forced the intercept to be zero as done as in \citet{wang2014}. Figure~\ref{ce}  displays our best fitting results and Table~\ref{fit} gives the fitting parameters as well as the comparison to \citet{yuan2013} and \citet{fitzpatrick99} and \citet{cardelli1989}.

\begin{figure}[!htbp]

   \includegraphics[width=0.48\textwidth]{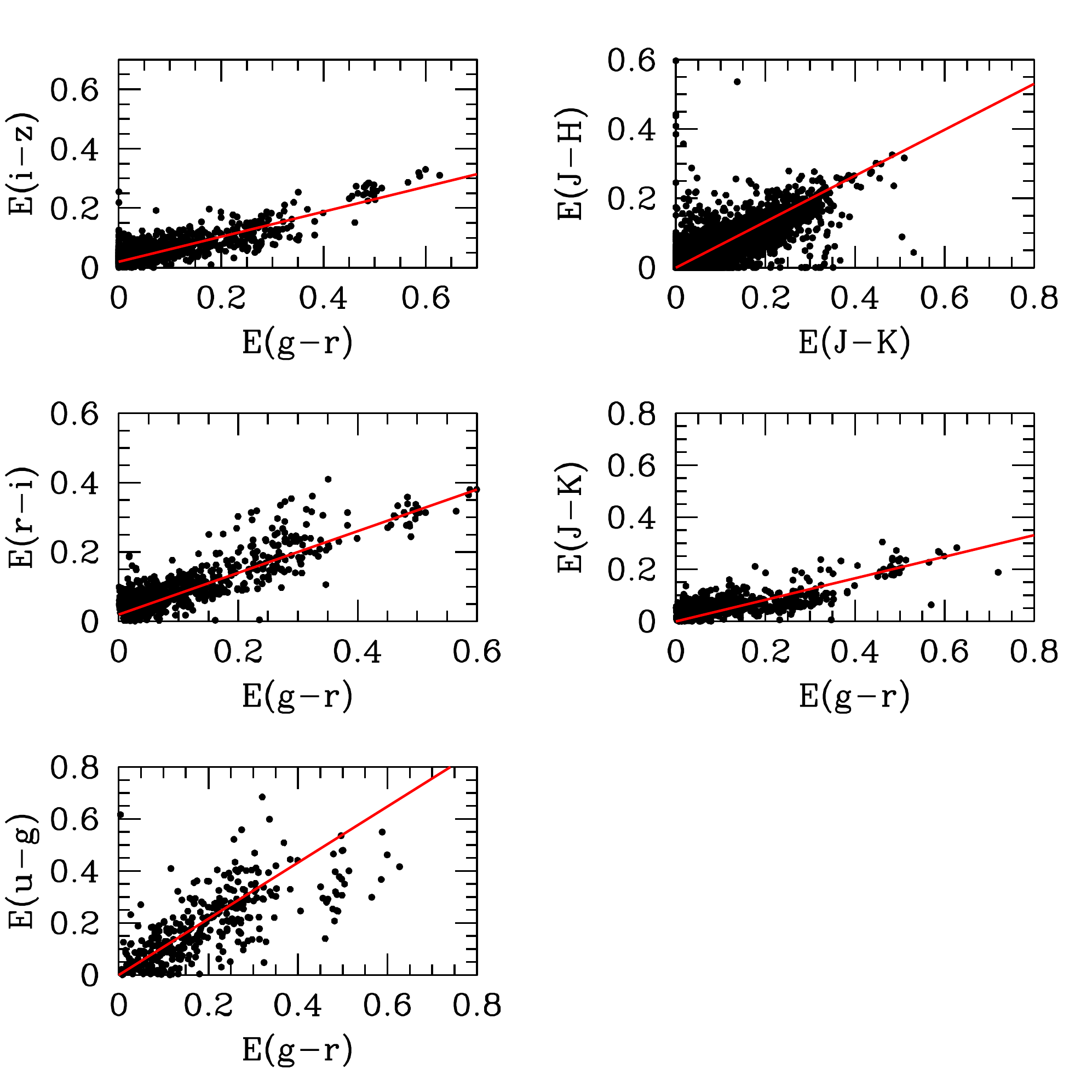}  
\caption{Relation between different colour excesses. The red line denotes the best linear fitting.}
\label{ce}
\end{figure}

\begin{table}
\caption{ Derived slopes of the colour-excesses}
\begin{tabular}{ccccc}
\hline
Colour & This work & Yuan et al. &Fitzpatrick & Cardelli\\
\hline
u -- g$^{a}$&  0.94 $\pm$0.03 &1.08 $\pm$0.010   &  0.945 & 0.984 \\
r -- i$^{a}$&  0.64 $\pm$0.02 &0.60 $\pm$0.010 & 0.582 & 0.557 \\
i -- z$^{a}$& 0.48 $\pm$0.01 &0.43 $\pm$0.004 & 0.426 & 0.496 \\
%z -- J$^{a}$& 0.51 $\pm$0.01 &0.56 $\pm$0.011 &0.544 & 0.544 \\
J -- K$^{a}$& 0.41 $\pm$0.01 &0.414 $\pm$0.01 & 0.411 & 0.466 \\
J -- H$^{b}$& 0.65 $\pm$0.02 & 0.63 $\pm$0.01 & -- & -- \\
\hline
%with $\rm E(g-r)= 0.383 \times E(u-g)$, $\r
\end{tabular}
\begin{description} 
\item[$^a$] derived by fitting versus $E(g-r)$ diagram
\item[$^b$] derived by fitting versus $E(J-K)$ diagram
\end{description}
\label{fit}
\end{table}

The extinction coefficients we derive agree within 10\%  with \citet{yuan2013} except  for u-g where we find  a value  closer to the Fitzpatrick extinction law. The u--g dispersion we measure  is higher than for other colours with some outliers not following  the linear relation between E(u-g) and E(g-r) (see Fig.~\ref{ce}).  The observed r.m.s scatter
is $\sim$ 0.1\,mag in the u-band which is in agreement with  the estimated uncertainties.

%with $\rm E(g-r)= 0.383 \times E(u-g)$, $\rm E(r-i)= 0.729 \times E(g-r)$, $\rm E(r-i)= 0.662 \times E(i-z)$ and  $\rm E(J-H)= 0.662 \times E(J-Ks)$. Our value  $\rm E(J-H)= 0.662 \pm 0.009 \times E(J-Ks)$ is very close to that of \citet{wang2014} with $\rm E(J-H)= 0.641 \pm 0.001 \times E(J-Ks)$.

\begin{figure*}[!htbp]
    \includegraphics[width=6.0cm]{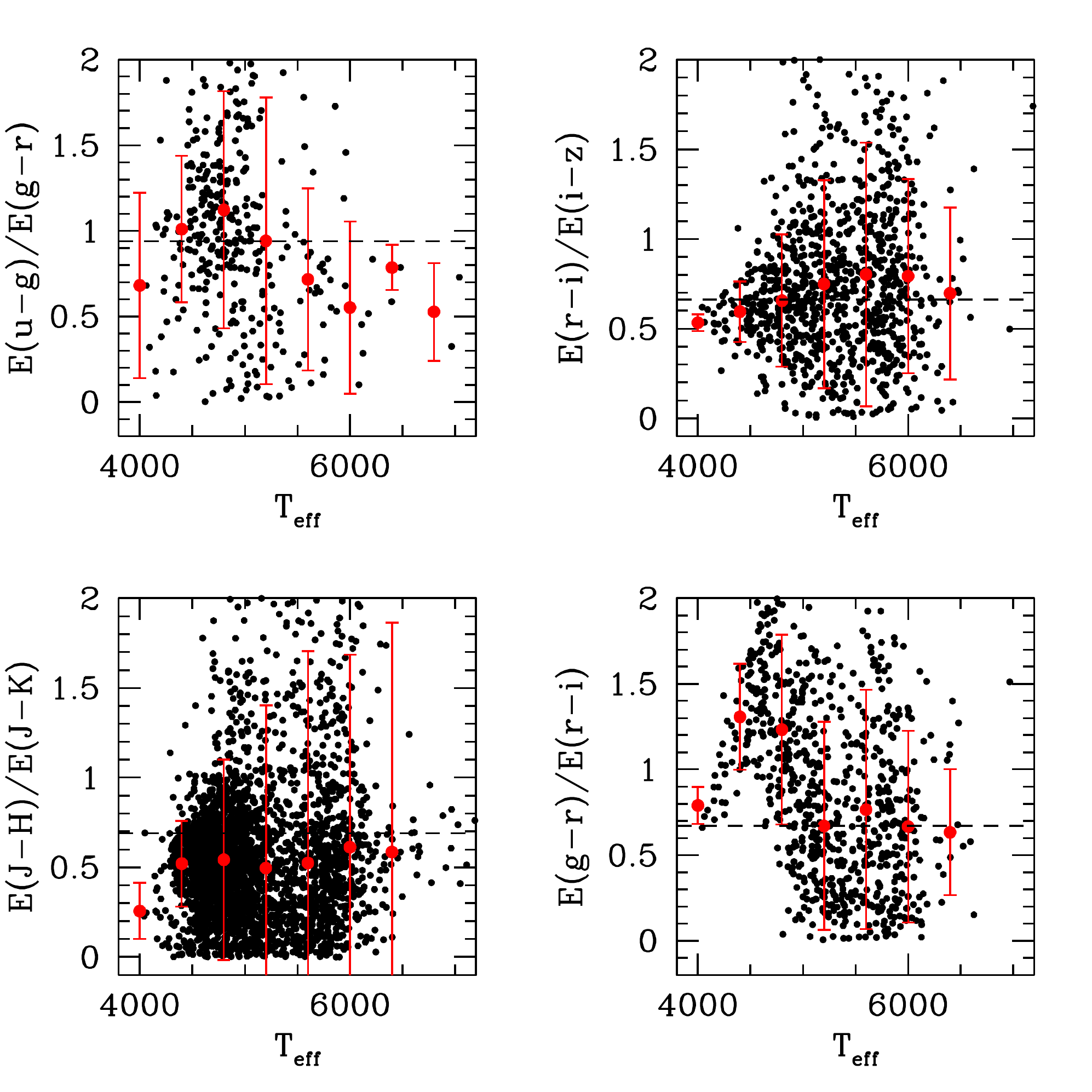}  \includegraphics[width=6.0cm]{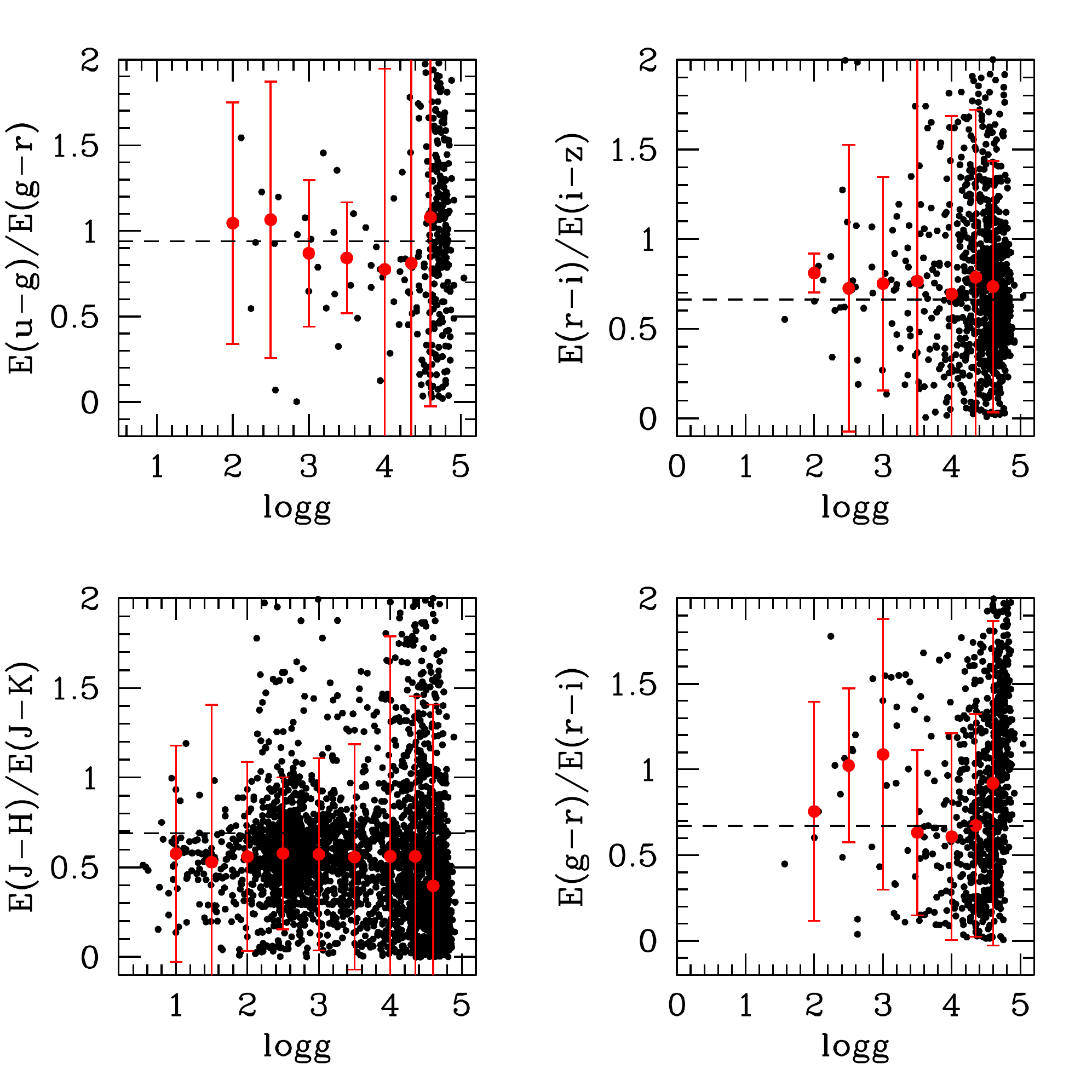}
     \includegraphics[width=6.0cm]{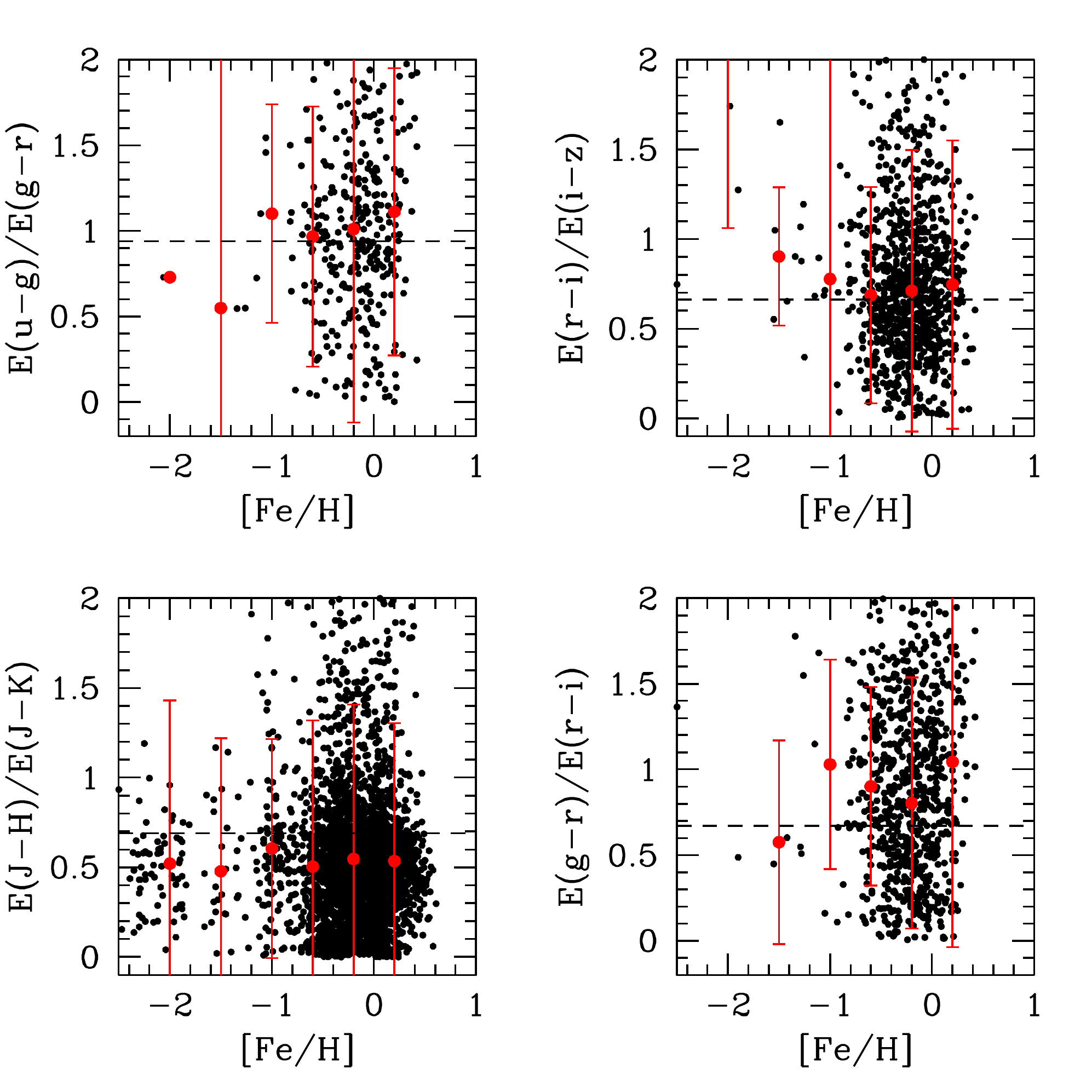}  
\caption{Extinction coefficients as a function of $\teff, \logg$ and [M/H].}
\label{3dextinctionco}
\end{figure*}

Up to now, most of the studies have mainly been concentrating  on deriving interstellar extinction coefficients  using  a specific population or a mixture of stars with different stellar parameters. However, as first suggested by \citet{Wildey1963},  other factors such as the stellar abundance may affect the interstellar extinction. In particular, \citet{Grebel1995} showed that the colour dependence of interstellar extinction is a complicated function of
the temperature, luminosity, and metallicity of the stellar probe. Even  the ratio of total to selective extinction  depends on these parameters  and $\rm R_{V}$ can, according  to their model, vary more than 10\%  between  a  star having $\rm T_{eff} = 3500\,K$ and another one having $\rm T_{eff} = 10\,000\,K$. Figure~\ref{3dextinctionco} shows the extinction coefficient (defined as the ratio of two colour excesses) as a function of $\teff$ and $\logg$. It is clear from Fig.~\ref{3dextinctionco}   that our method does not have the required accuracy to detect  variations at the level of 10\%  as predicted by the theoretical  models of   \citet{Grebel1995}.  However, within the accuracy of our procedure, large scale variations of the different extinction coefficients as a function of  $\teff, \logg$ or $\meta$ are not detected. We note, on the other hand, that the dispersion increases for higher $\logg$.

 Due to the limited spatial overlap between SDSS and the GES, only a
small number of our GES sources have SDSS ugriz photometry.  For the targets for which we have SDSS photometry, the E(g--r)/E(r--i) ratio is  overestimated for cool giant stars with
$\teff < 5000\,K$ and  $\logg < 3.5$.  Within the large dispersion of our method we do not note any hint of a dependence of the extinction coefficient to the Galactic environment, i.e. the metallicity
of the stars. However, we lack of stars with $\rm \meta <  -1\dex$  and small-scale variations cannot  not be excluded by our method. Striking is the  nearly flat  extinction coefficient in the near-IR (e.g. $\rm  E(J-H)/E(J-K)$) plane with no variation
as a function of the stellar parameters.  Within the errors of our method, our work does not show that there is any dependence of the interstellar extinction coefficient on the atmospheric parameters of the stars. This suggests, that extinction maps derived from mean colours of stars such as the RJCE method (\citealt{majewski2011}) or the colour-excesses of stars
(\citealt{lada1994}, \citealt{gonzalez2012}) can be generally used, assuming a constant extinction coefficient which does not depend on the stellar parameter. 
We fitted the E(J--H)  vs. E(J--K) diagram in the same way as \citet{wang2014}. Our value   $\rm E(J-H)= 0.651 \pm 0.009 \times E(J-Ks)$ is slightly higher  than the one of \citet{wang2014} who measured $\rm E(J-H)= 0.641 \pm 0.001 \times E(J-Ks)$, resulting to a power law index of   $\rm \alpha=2.12$  and  $\rm A_{J}/A_{Ks} = 3.15$ (assuming $\rm \lambda_{eff}=1.25\,\mu m, 1.65\,\mu m, 2.15\,\mu m$ for the J,H,$K_{S}$ bands). Our power law index is similar to that of \citet{stead2009} with $\rm \alpha=2.14$ or \citet{fritz2011} with $\rm \alpha=2.11$.

Finally, we investigated the variation of the extinction coefficient along different lines of sight. \citet{zasowski2009} and \citet{gao2009} suggested strong variations in the  extinction law  in the mid-IR as a function of Galactic longitude or
angle from the Galactic Center.  Strong variation of the extinction law as a function of the Galactic latitude was also found by \citet{chen2012} in the
 Galactic Bulge.  \citet{zasowski2009}  suggested the existence of strong longitudinal variations in the infrared extinction law where the slopes increase as the wavelength increase (see their
Fig. 5) resulting in a steeper extinction curve in the outer Galaxy. \citet{gao2009} identified  small variations of the  mid-IR extinction law with the location of the spiral arms.  
Figure~\ref{ceangle}  shows  the extinction coefficient E(J--H)/E(J--K) as a function of the angle  from the Galactic Center (see Fig.~\ref{ceangle}). Note that contrary to APOGEE, GES probes different regions of the Galaxy and
 avoids especially the galactic plane where interstellar extinction is very high. 

Within the  dispersion of our method, we do not find evidence for any trend of the  variation of E(J--H)/E(J--K) with the angle from the Galactic Center nor with Galactocentric distance in agreement with the extinction coefficients in the SDSS bands. This suggests that the extinction law in the SDSS ugriz bands and the near-IR JHKs bands is uniform, confirming  the result of   \citet{wang2014} obtained with APOGEE red clump stars.

%A steady decrease of E(J--H)/E(J--K) with the angle from the Galactic Center is seen which is more pronounced for positive Galactic longitudes. A similar behaviour is also seen in Fig. 5 of \citet{zasowski2009} for the J band getting less steeper going to longer wavelengths. This suggest that the variation of the extinction coefficient with the angle from the GC maybe  due to  variation in thr grain size distribution or the chemical composition of the local ISM.

\begin{figure}[!htbp]

   \includegraphics[width=9.0cm]{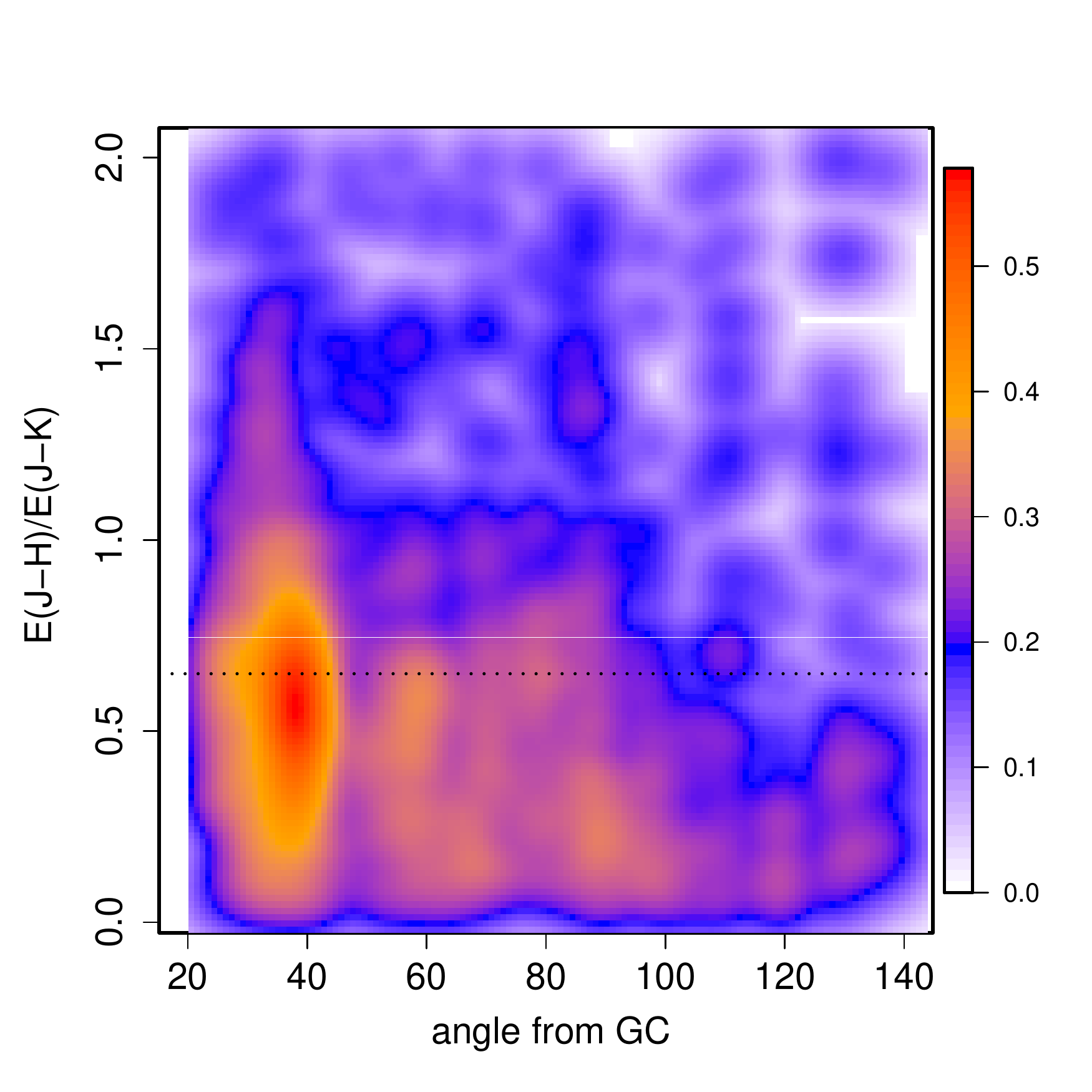}
\caption{Density plot of E(J--H)/E(J--K) vs. angle from the Galactic Center. The dashed blanck line gives the mean of our derived extinction coefficient. }
\label{ceangle}
\end{figure}

%\begin{figure*}[!htbp]

%   \includegraphics[width=9.0cm]{ce_longitude.pdf}   \includegraphics[width=9.0cm]{ce_latitude.pdf}
%\caption{Colour excess  ration as a function of  Galactic longitude (left panel) and Galactic latitude (right panel). The red filled points show the median  value and the error bars the standard deviation.}
%\label{cegal}
%\end{figure*}

\section{Conclusions}
We used data from the GES survey, together with accurate stellar parameters ($\teff, \logg, \meta$) to trace 3D interstellar extinction in intermediate and high-latitude
regions of our Galaxy.  We discuss the influence of different stellar isochrones (Yonsei-Yale and Padova)  on the derived 3D extinction  and compare our results  with the SFD98 dust maps.  We find on average good agreement  with  a mean difference of $\Delta E(B-V) = 0.009\pm 0.075 $, with the dispersion getting larger when  including low galactic latitude regions ($\rm |b| < 10^{o}$). For larger  $\rm E(B-V) > 0.5$  SFD98 gets higher extinction compared to our estimation $\rm E(B-V)_{Padova}$.

We compared our 3D interstellar dust maps with those of \citet{drimmel2003} and \citet{chen2014a}. The GES data
 confirm the steep rise in $\rm A_{V}$ for distances between 0 and 4\,kpc with a flattening of the extinction at larger distances. We studied the influence of the stellar parameters on the extinction coefficients in the  optical (SDSS-bands)  and the  near-IR (JHKs). We do not detect any significant dependence of the extinction coefficient with stellar parameters indicating that a constant extinction coefficient can be assumed. Within the precision of our method, we do not find any  evidence for the variation of the extinction coefficients with the angle from the Galactic centre or Galactocentric distance. We note, however, that our method does not allow to trace small-scale variations of the extinction coefficient. This suggests a uniform extinction-law, as found in \citet{wang2014}.
With the future data releases of GES in the coming years, we will be able to trace the distance vs. $\rm A_{V}$ behaviour, systematically allowing to compare qualitatively spectroscopically  derived extinction with 3D dust models

\begin{acknowledgements}
We want to thank the anonymous referee for her/his extremly useful comments.
Based on data products from observations made with ESO
Telescopes at the La Silla Paranal Observatory under programme ID 188.B-3002.
This work was partly supported by the European Union FP7 programme through ERC grant number 320360 and by the Leverhulme Trust through grant RPG-2012-541. We acknowledge the support from INAF and Ministero dell' Istruzione, dell' Universit\`a' e della Ricerca (MIUR) in the form of the grant "Premiale VLT 2012". The results presented here benefit from discussions held during the Gaia-ESO workshops and conferences supported by the ESF (European Science Foundation) through the GREAT Research Network Programme.  We want to thank C. Bailer-Jones, C. Babusiaux  for extremly useful comments.
M. Schultheis, A. Recio-Blanco, P. de Laverny et V. Hill acknowledge the support of the ``Programme National de Cosmologie et Galaxies'' (PNCG) of CNRS/INSU, France. TB is funded by grant no. 621-2009- 3911 from the Swedish Research Council.

\end{acknowledgements}

\bibliographystyle{aa}
\bibliography{ges_extinction_accepted}

%\end{document}

\appendix
\section{3D extinction along different lines of sights of GES}
\begin{figure*}
\caption{3D extinctions for GES lines of sight for GES field. The red straight line gives the corresponding model of Drimmel et al. (2003)}
\includegraphics[height=0.2\textheight]{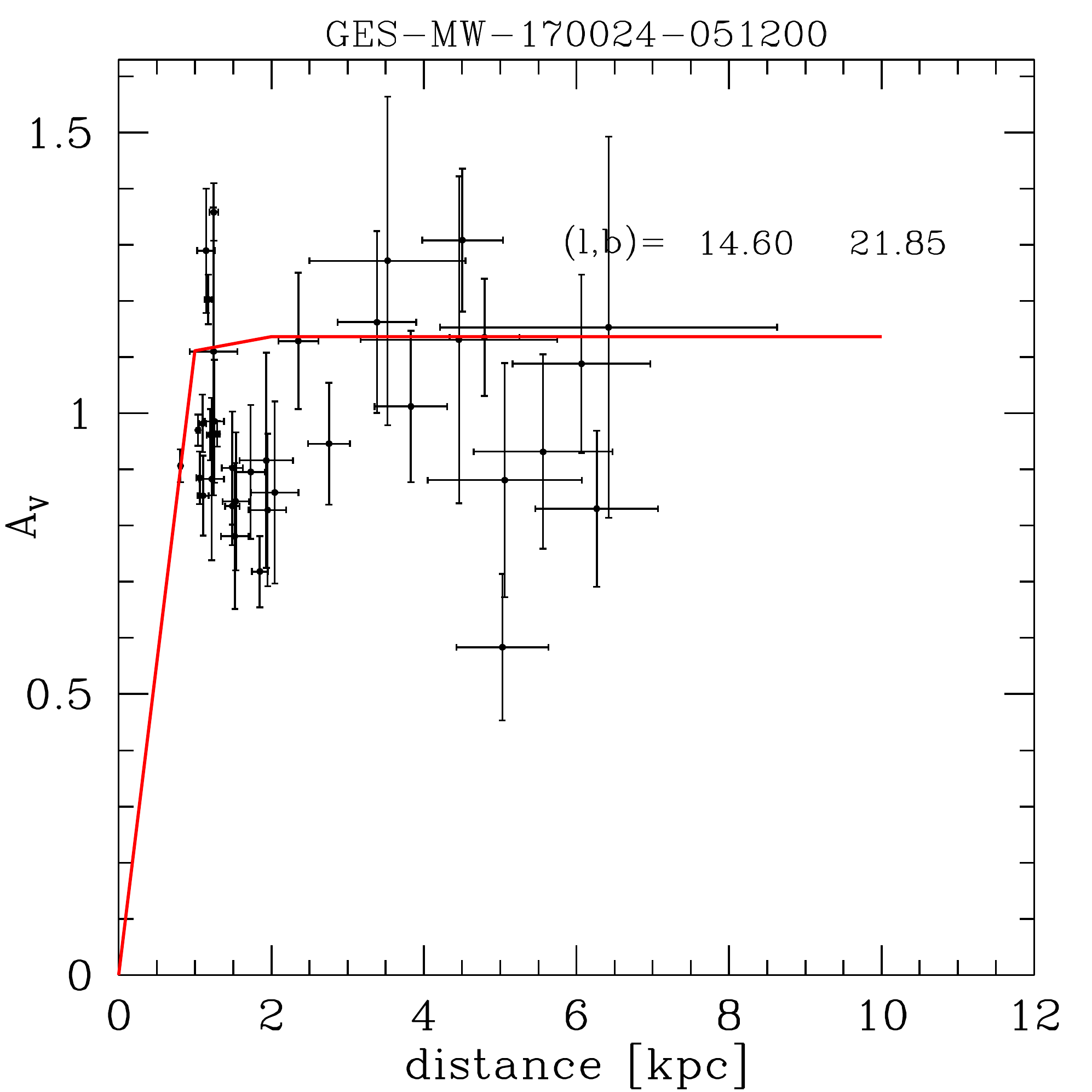}  \includegraphics[height=0.2\textheight]{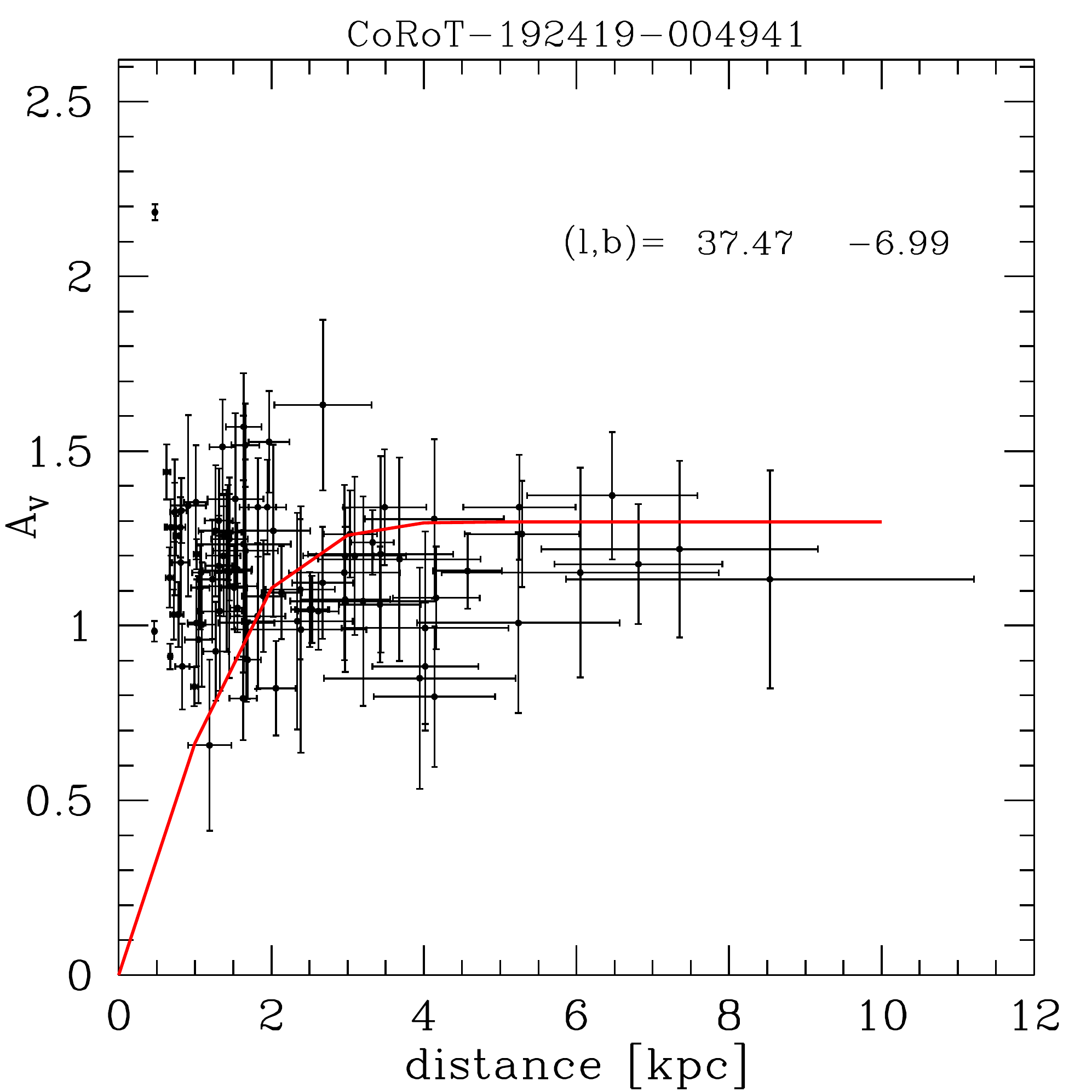} \includegraphics[height=0.2\textheight]{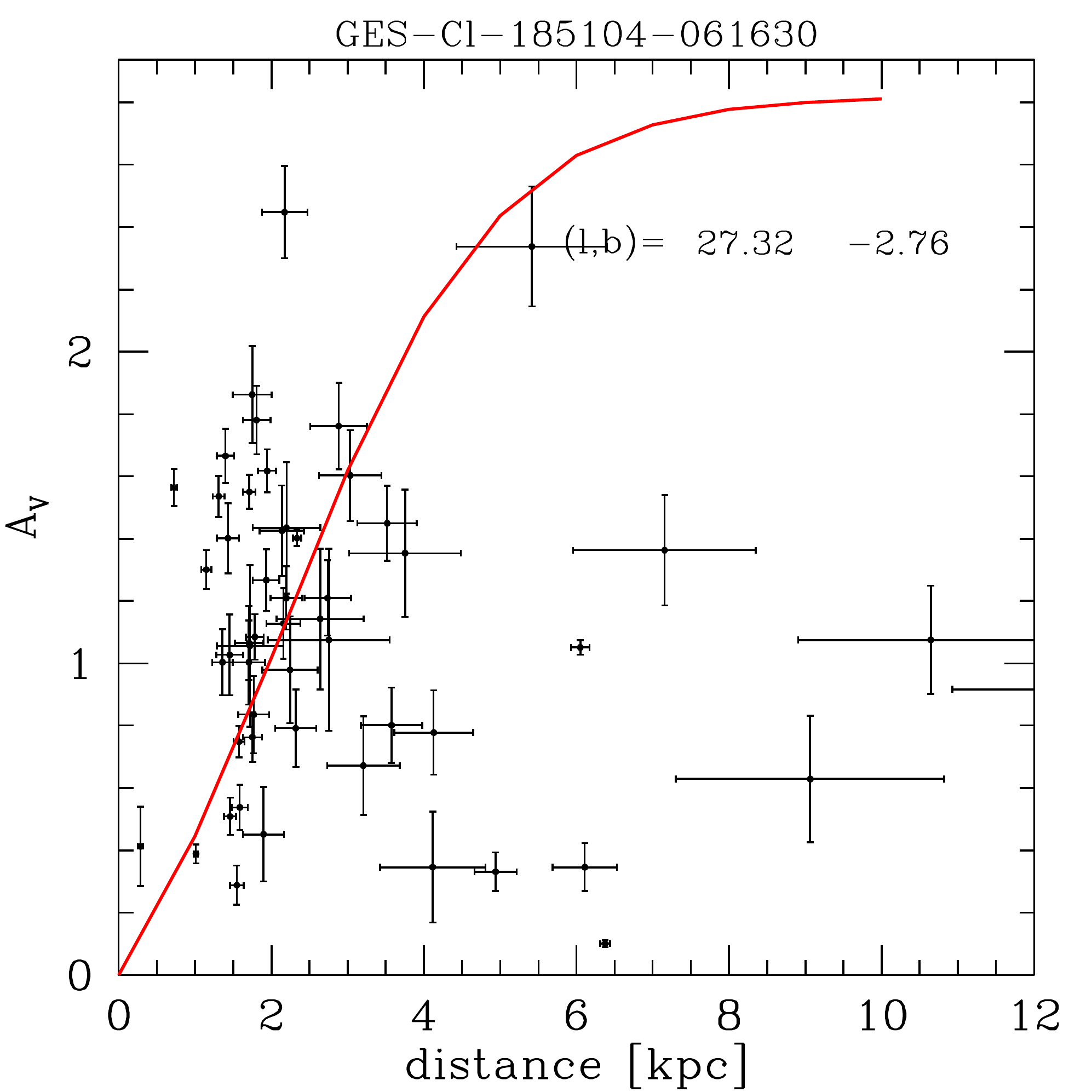} 
\includegraphics[height=0.2\textheight]{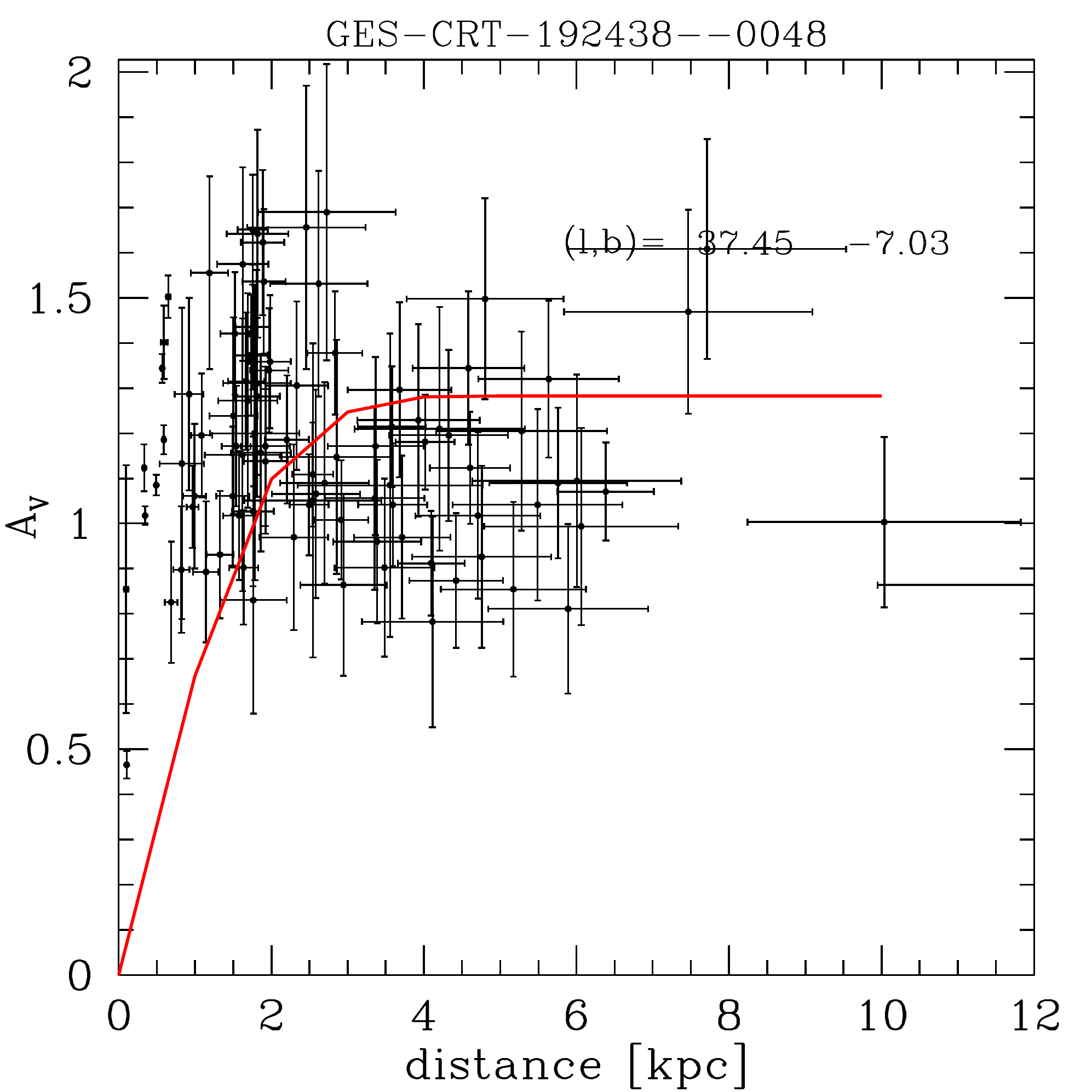}  \includegraphics[height=0.2\textheight]{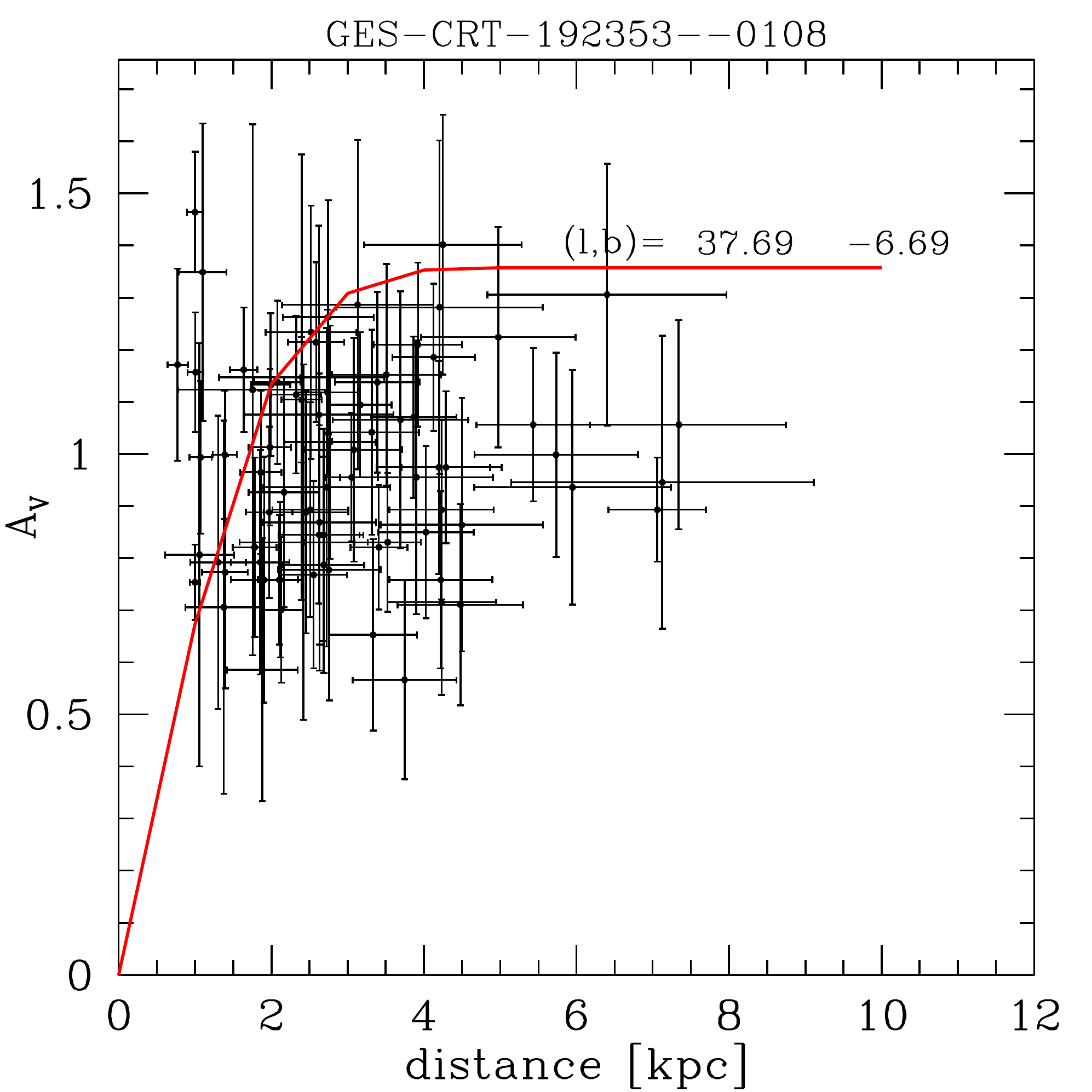} \includegraphics[height=0.2\textheight]{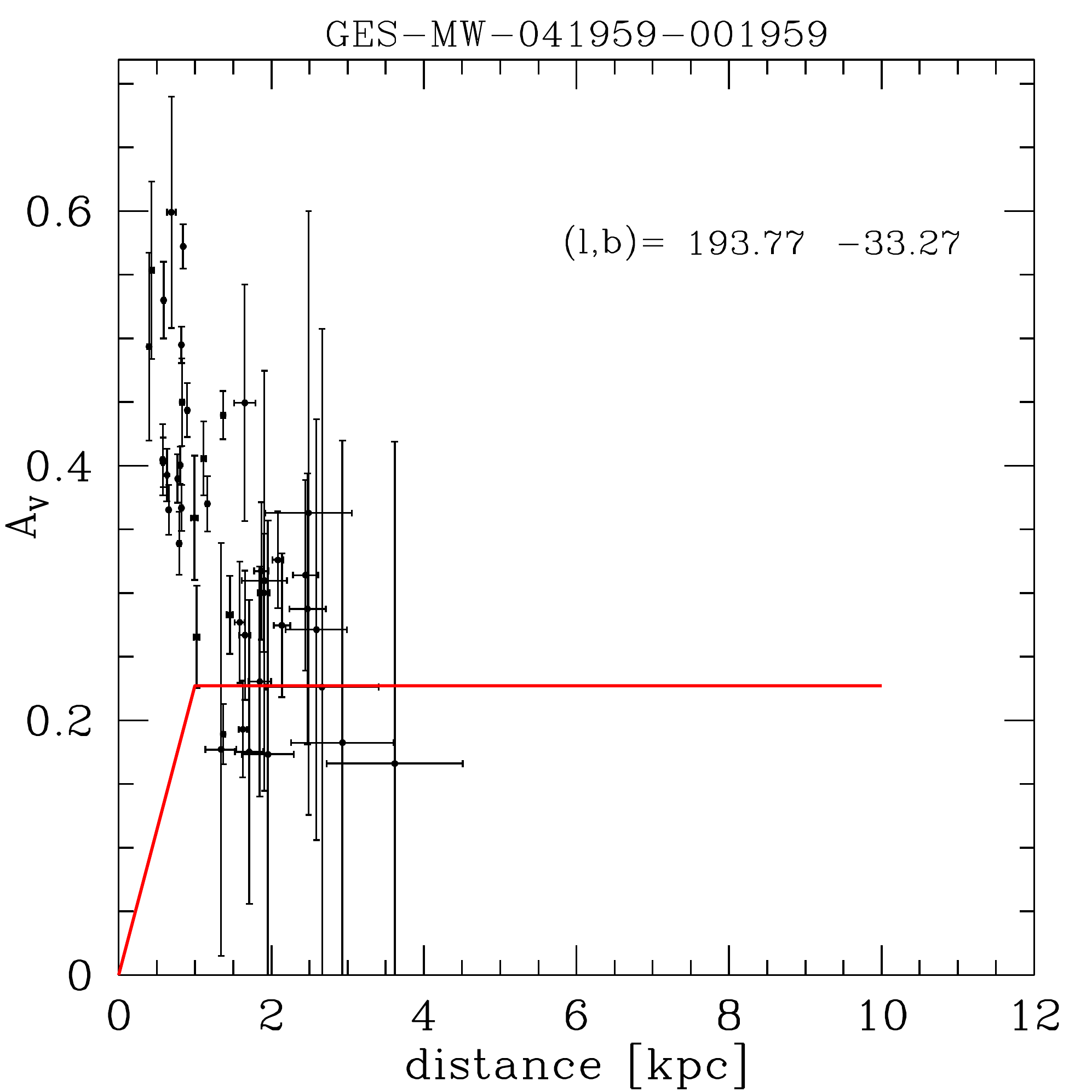}
\includegraphics[height=0.2\textheight]{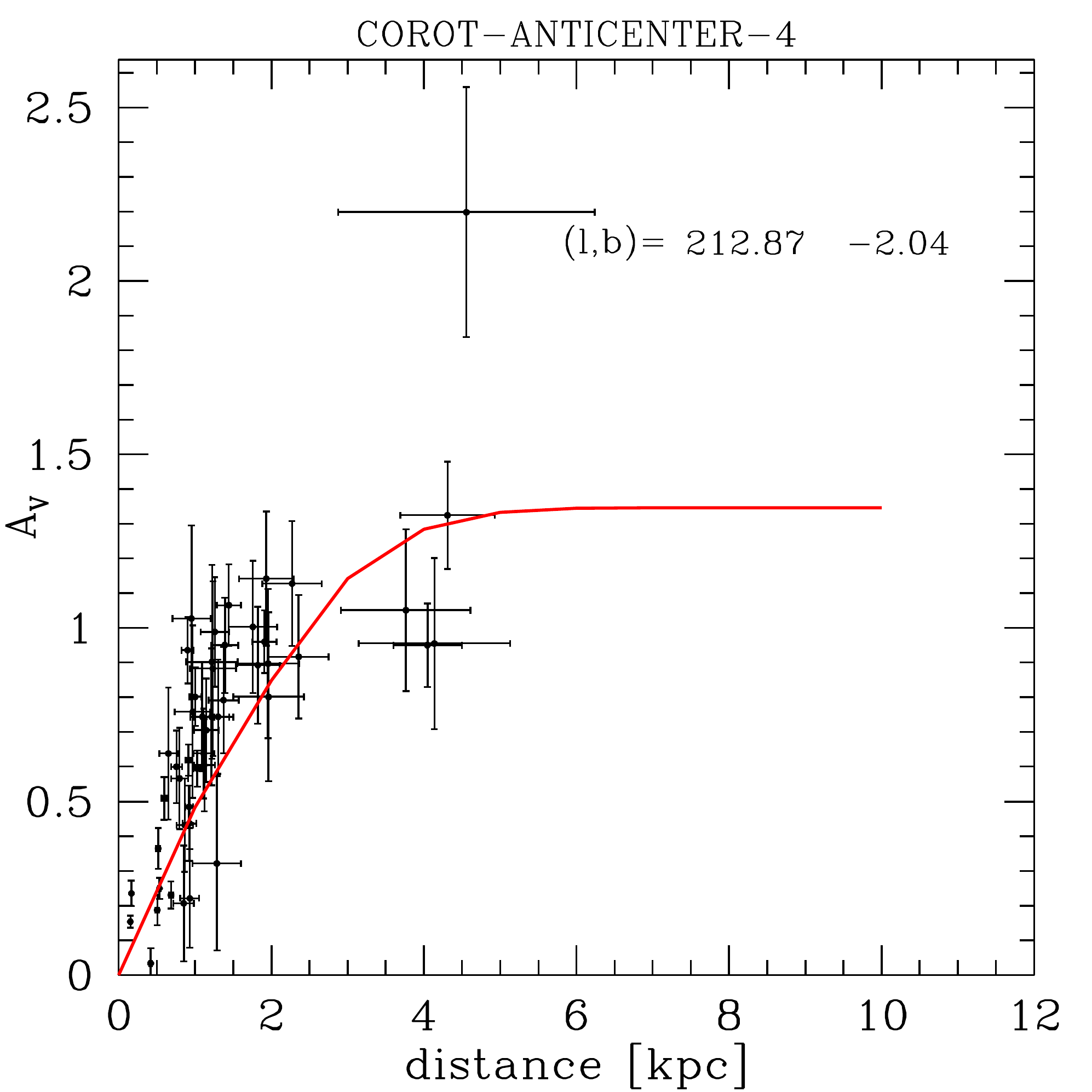}   \includegraphics[height=0.2\textheight]{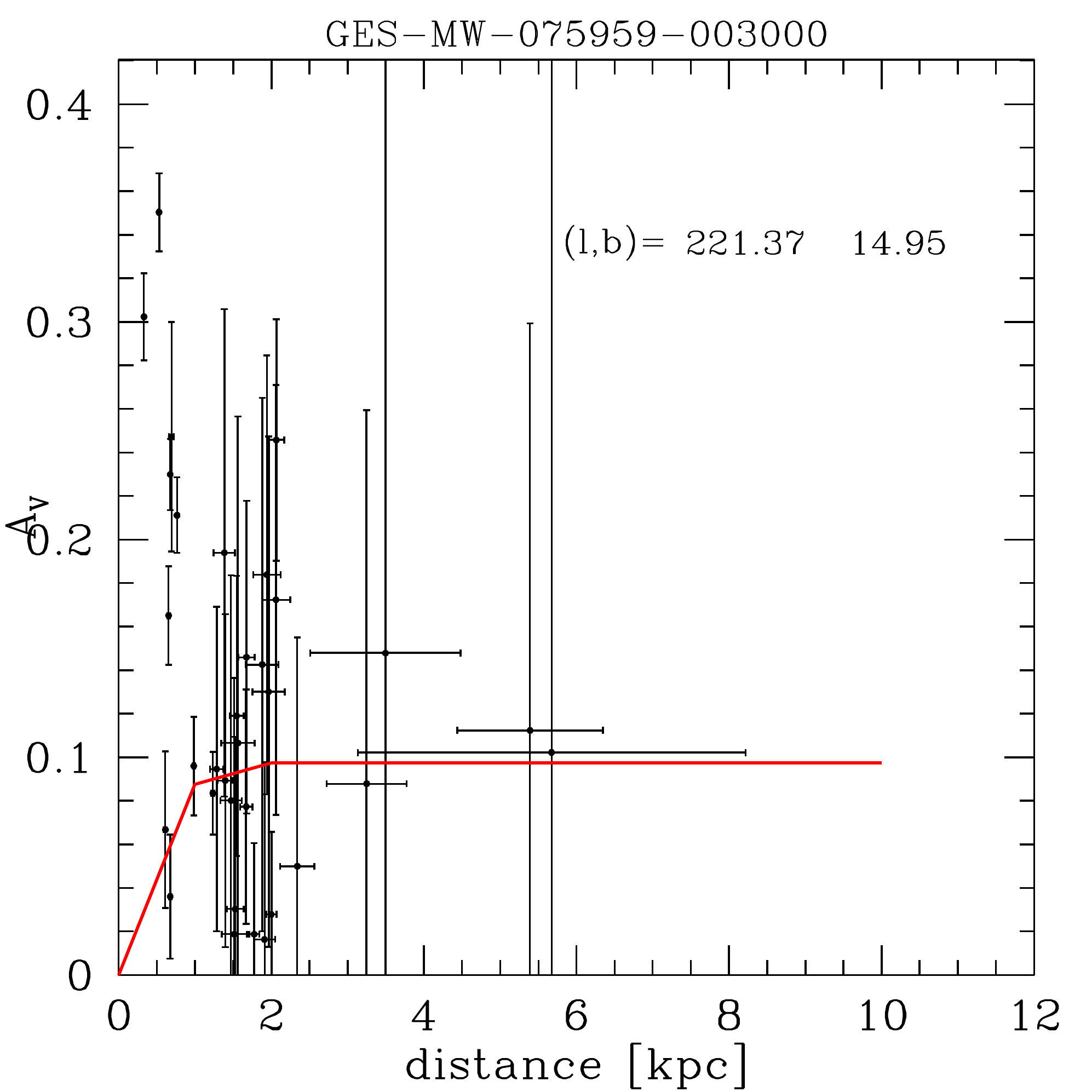}   \includegraphics[height=0.2\textheight]{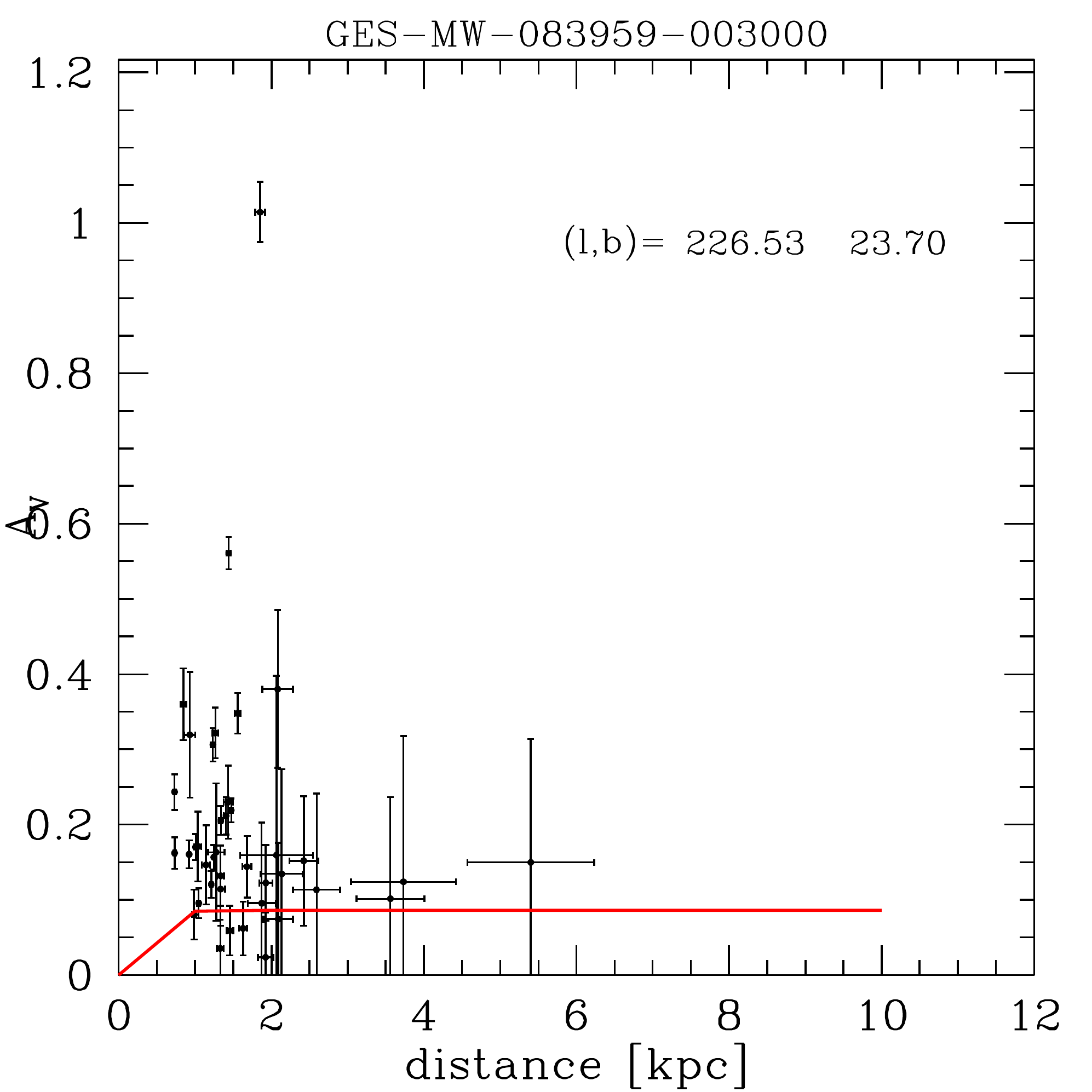} 
\includegraphics[height=0.2\textheight]{GES_MW_083959_003000.pdf}  \includegraphics[height=0.2\textheight]{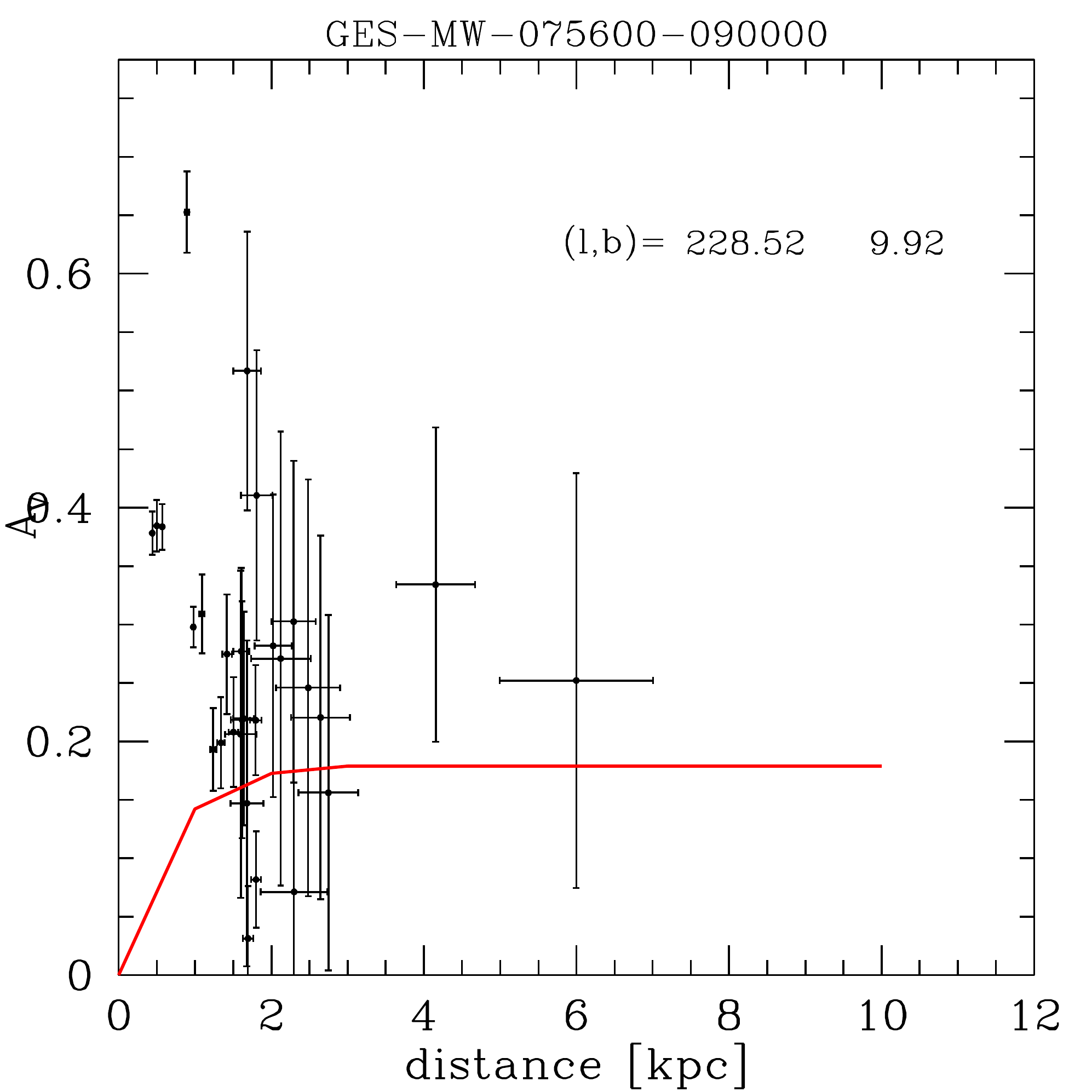} \includegraphics[height=0.2\textheight]{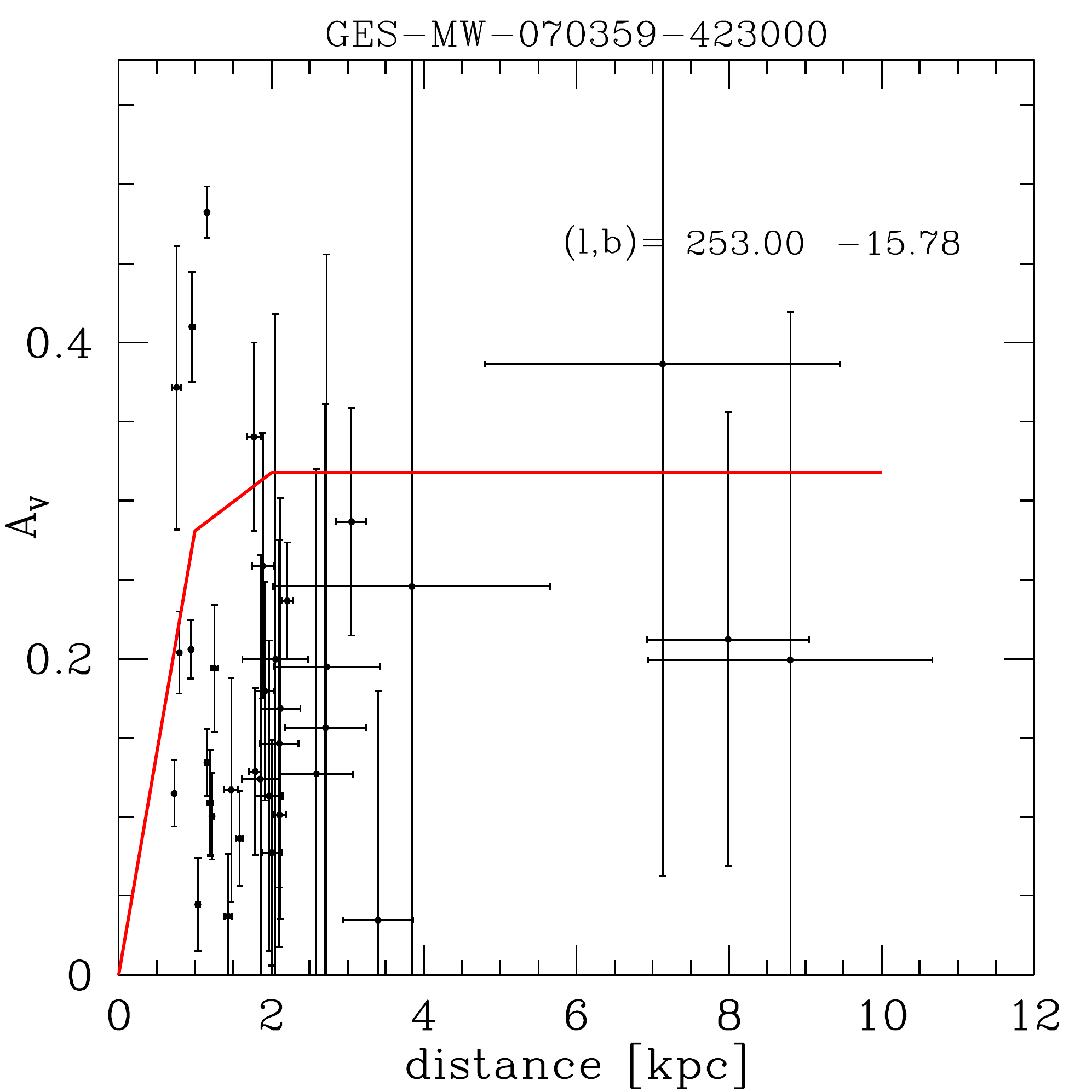}
\includegraphics[height=0.2\textheight]{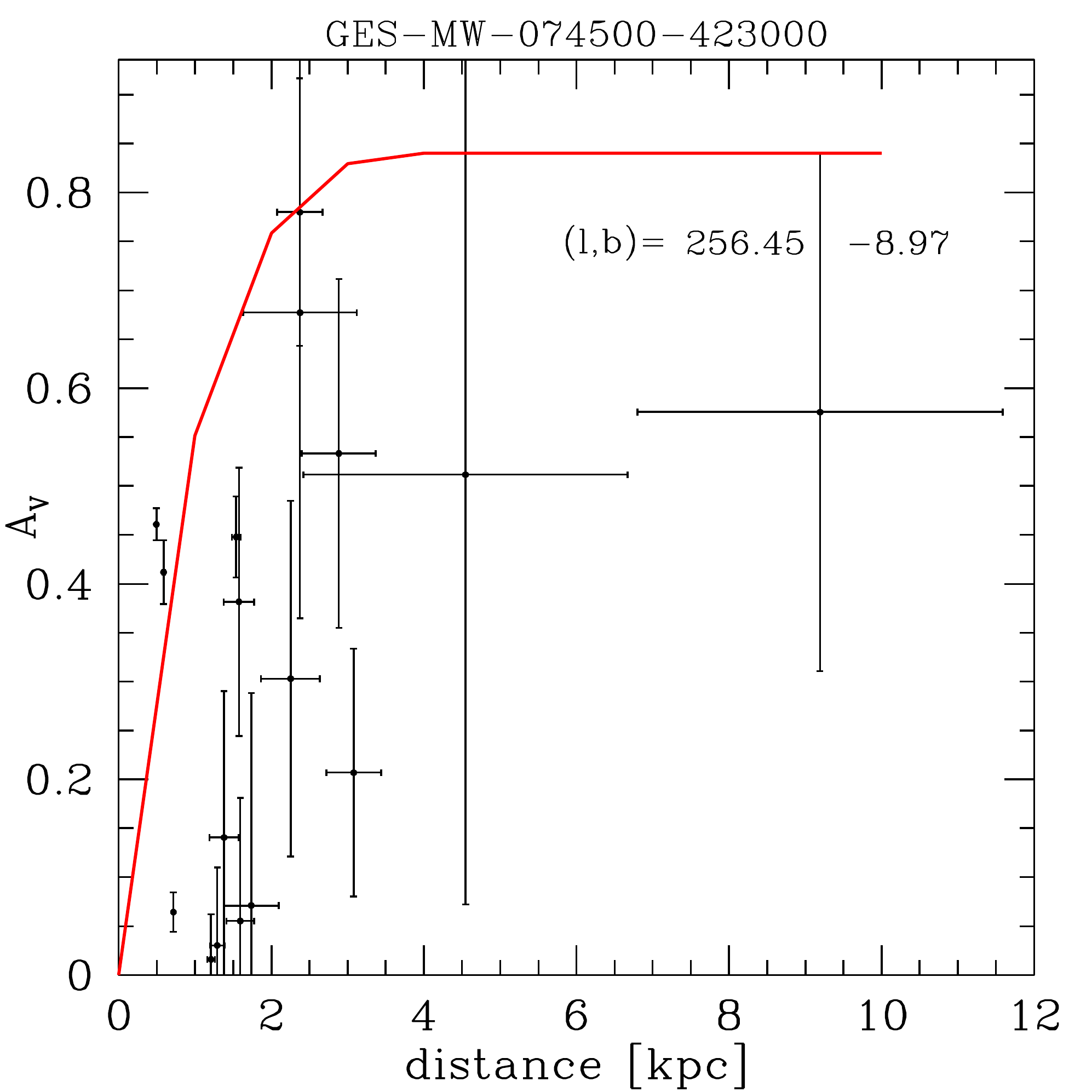}  \hspace{1.6cm}   \includegraphics[height=0.2\textheight]{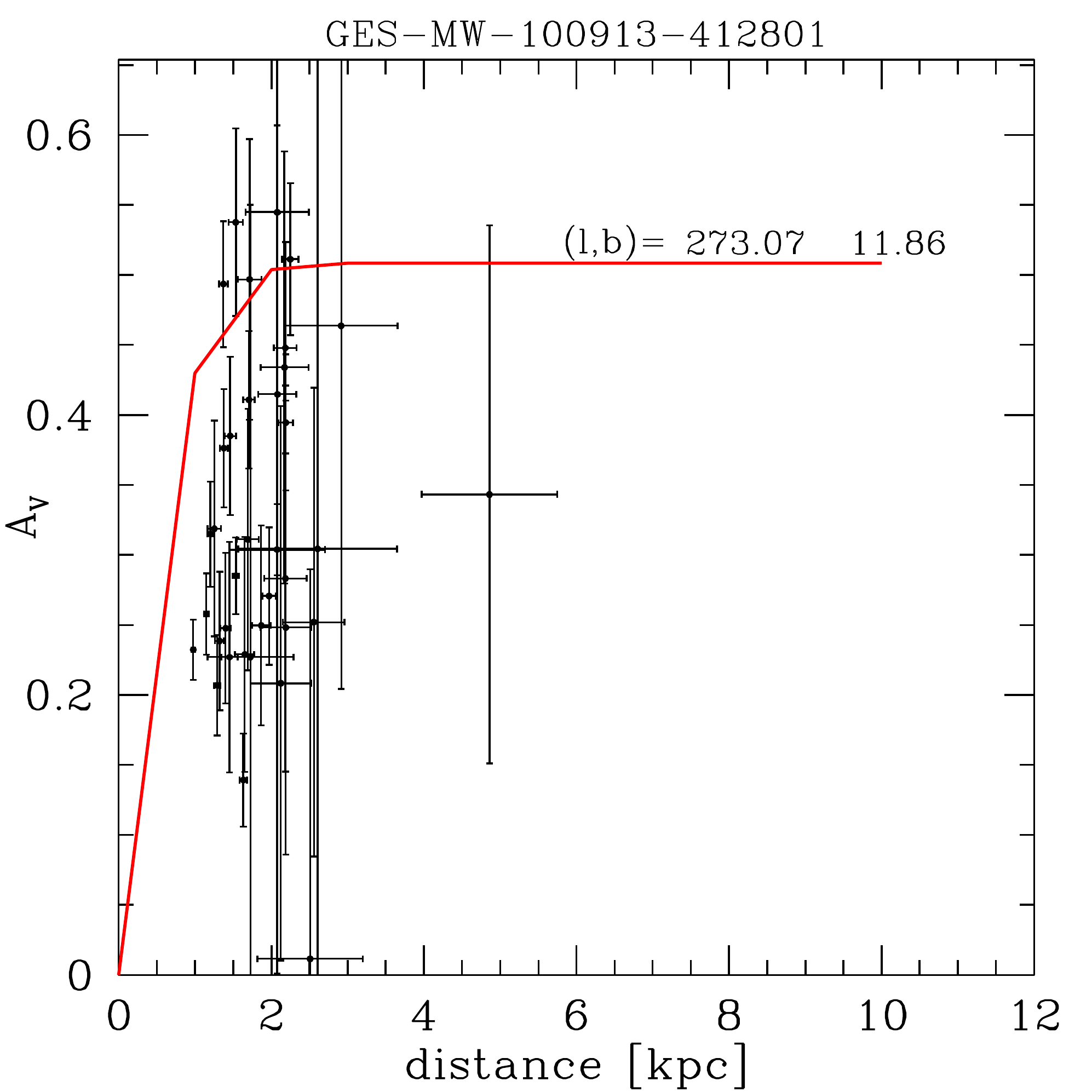}   \hspace{1.6cm} \includegraphics[height=0.2\textheight]{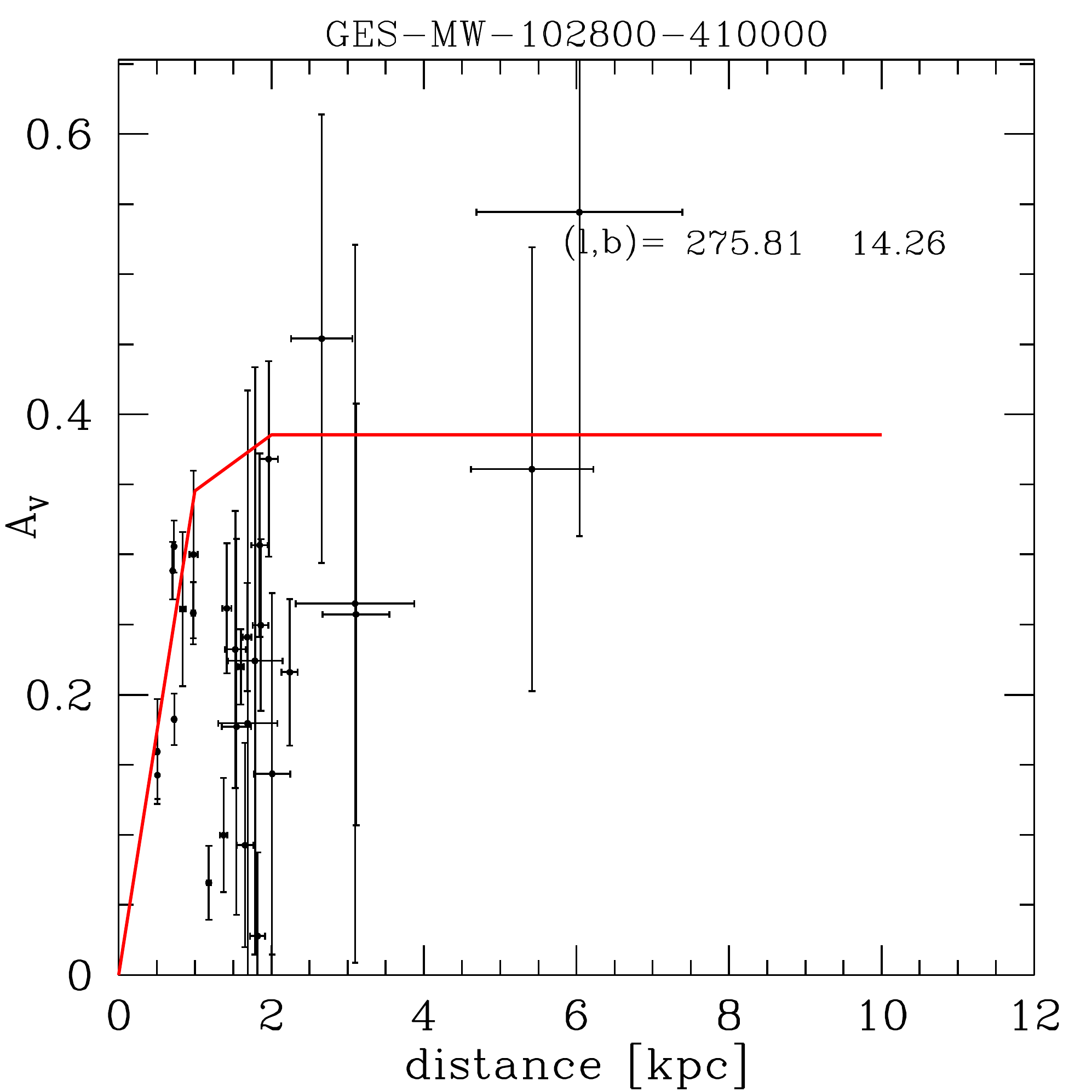}
\label{online}
\end{figure*}

\begin{figure*}
\caption{3D extinctions for GES lines of sight for GES field (continued)}
\includegraphics[height=0.2\textheight]{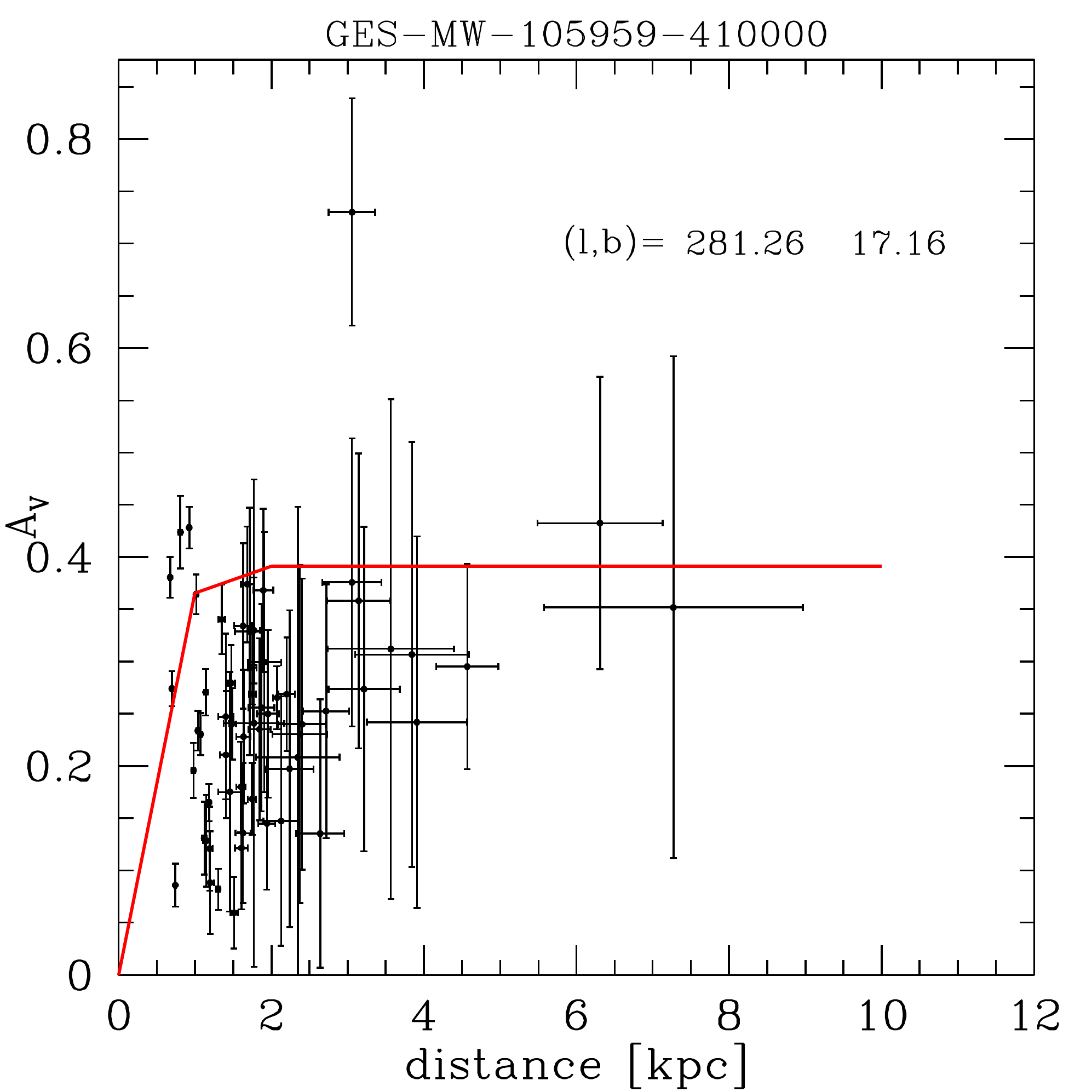} \includegraphics[height=0.2\textheight]{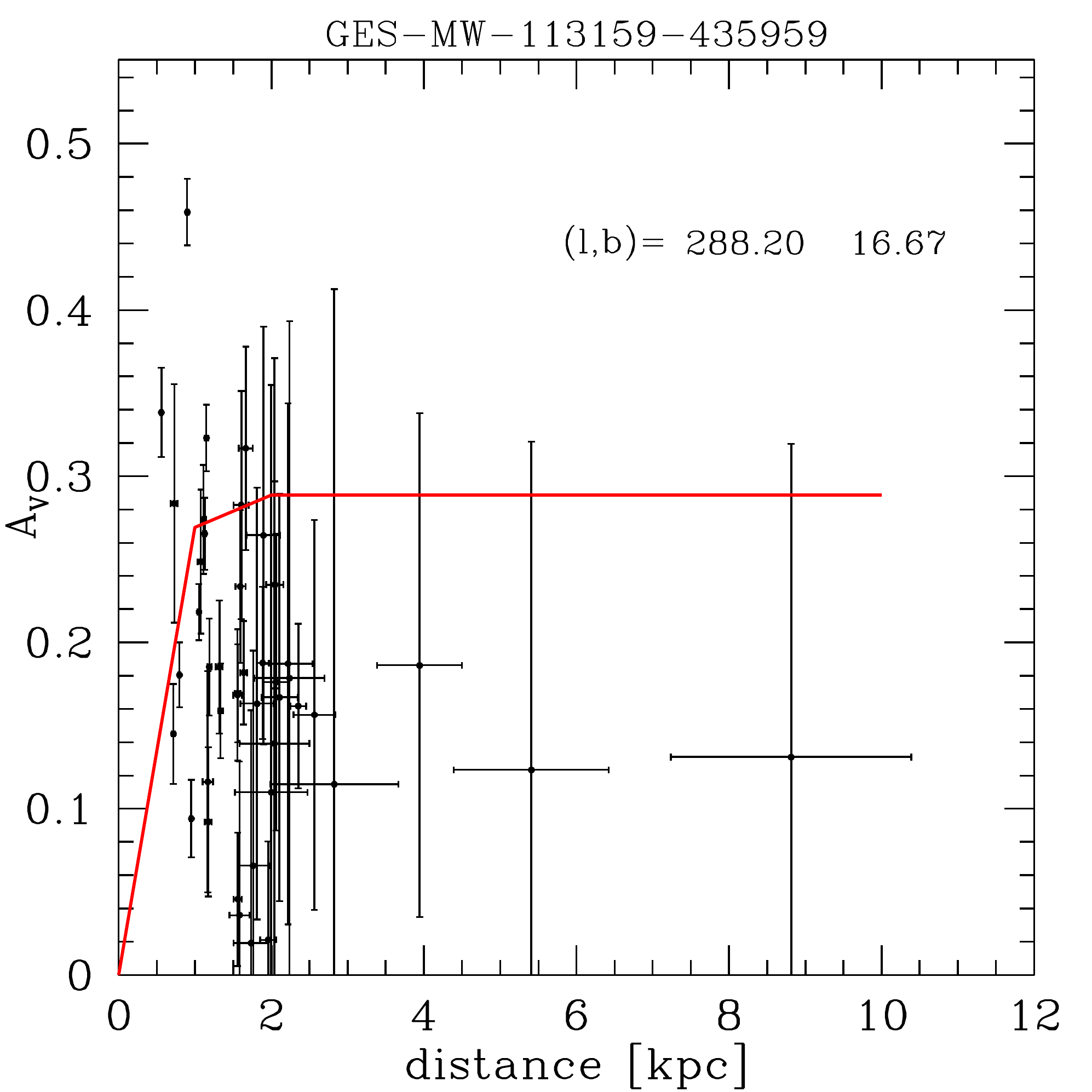} \includegraphics[height=0.2\textheight]{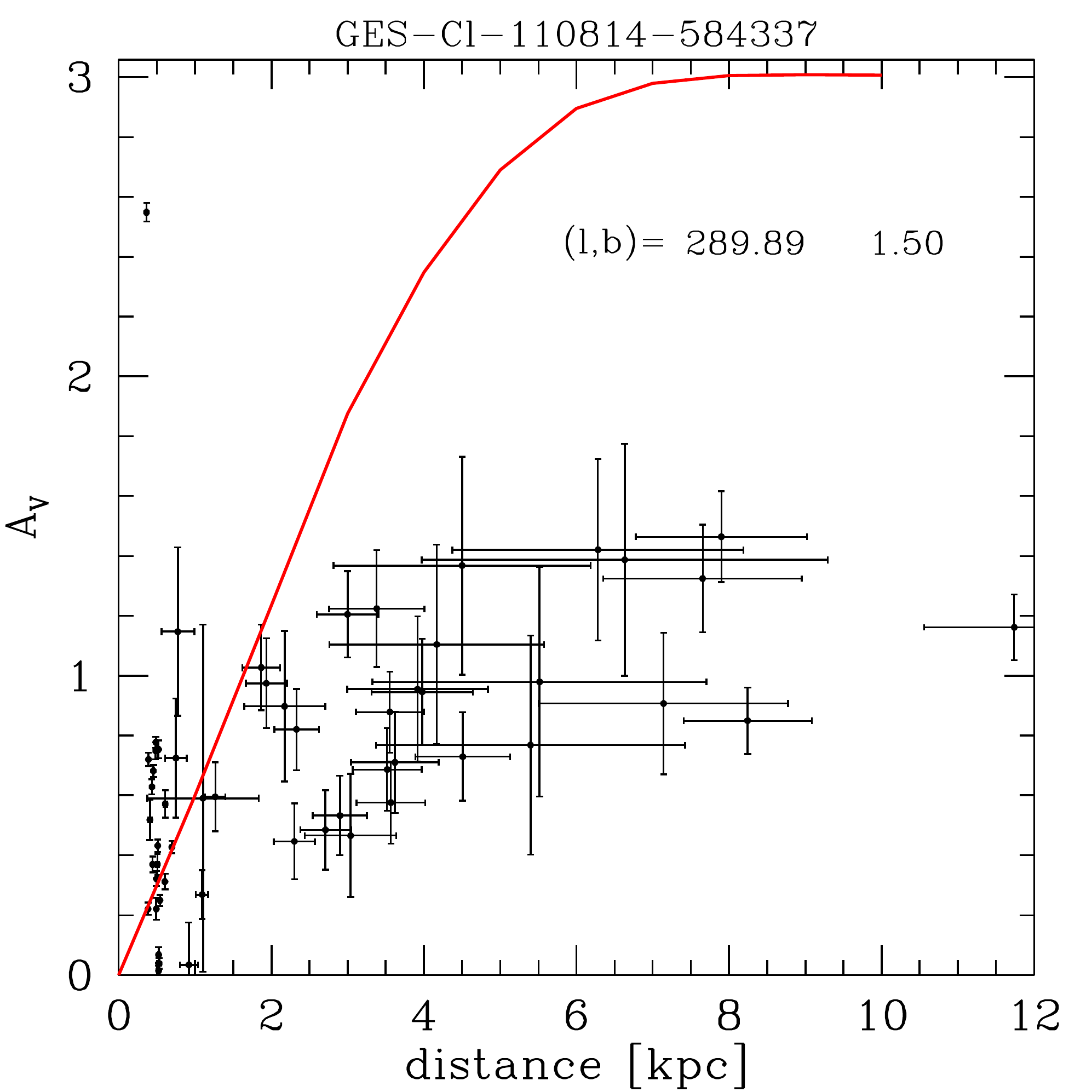}
\includegraphics[height=0.2\textheight]{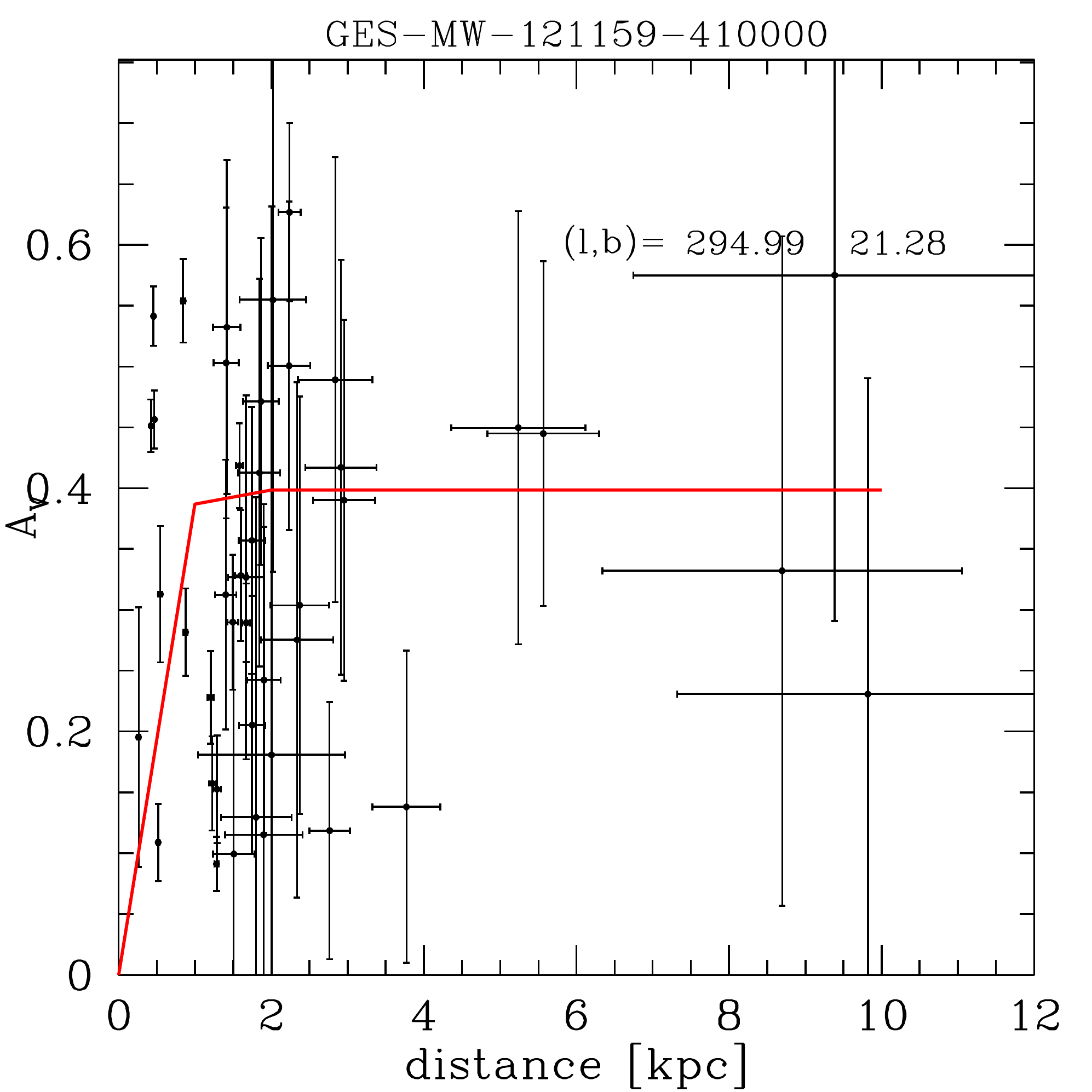}  \includegraphics[height=0.2\textheight]{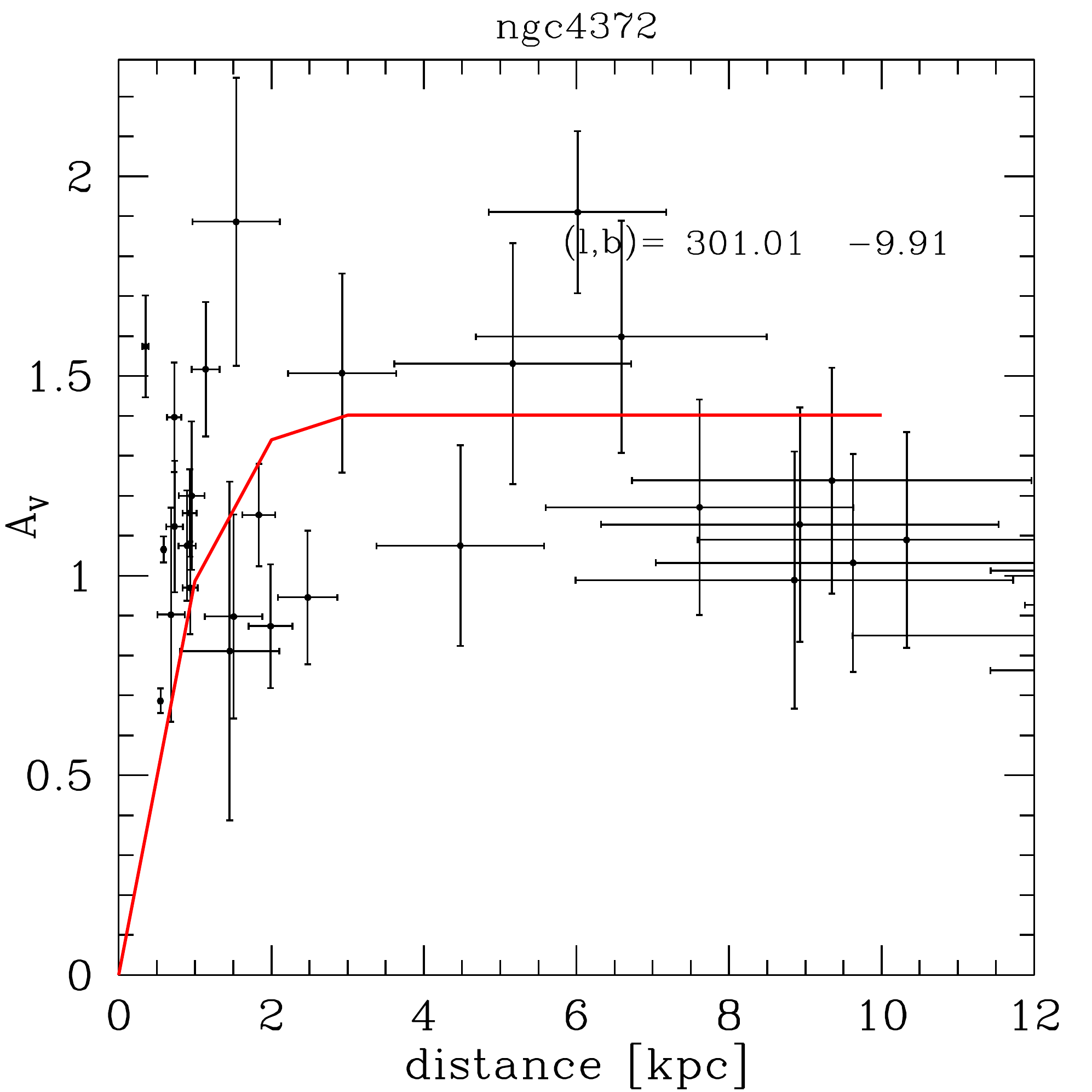} \includegraphics[height=0.2\textheight]{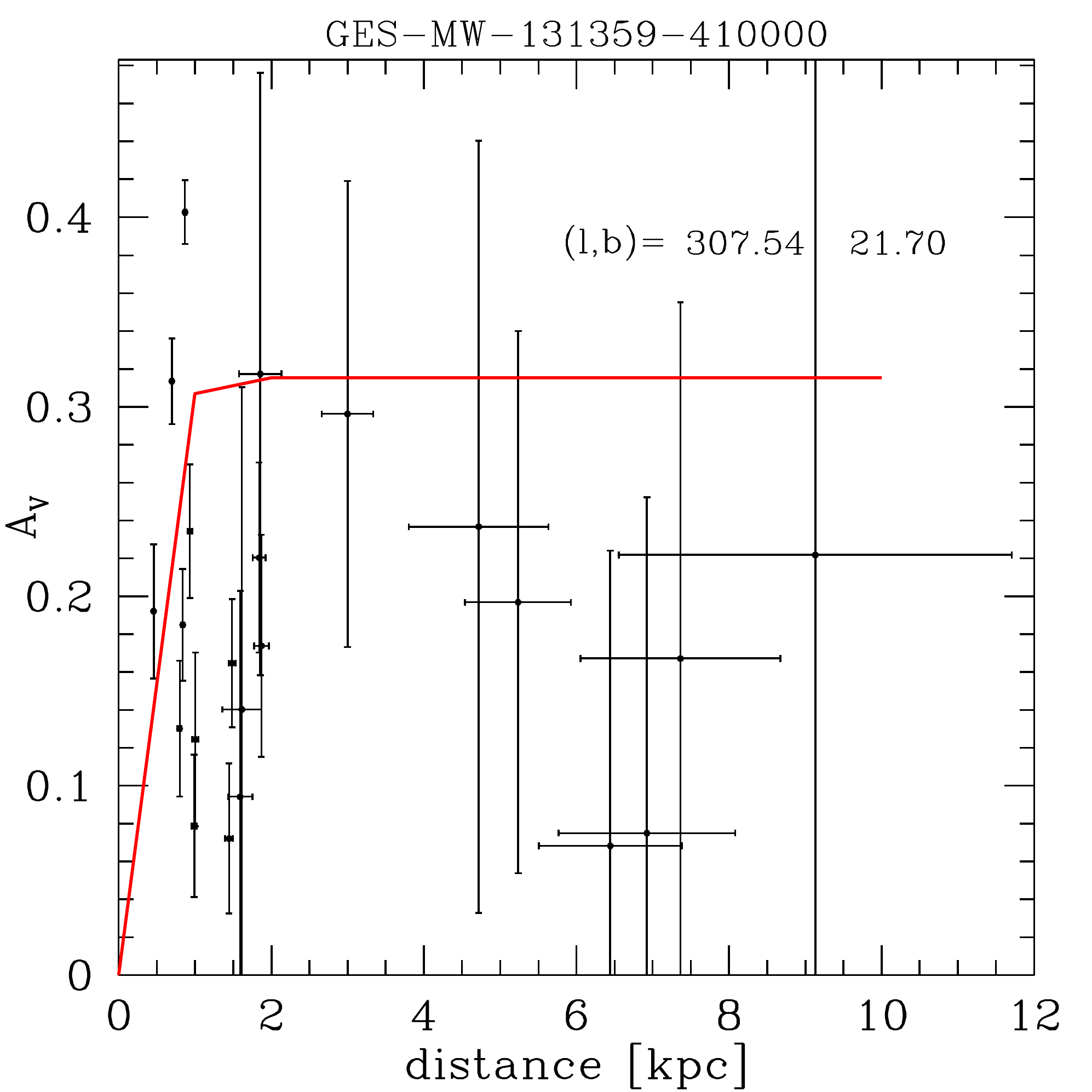}
\includegraphics[height=0.2\textheight]{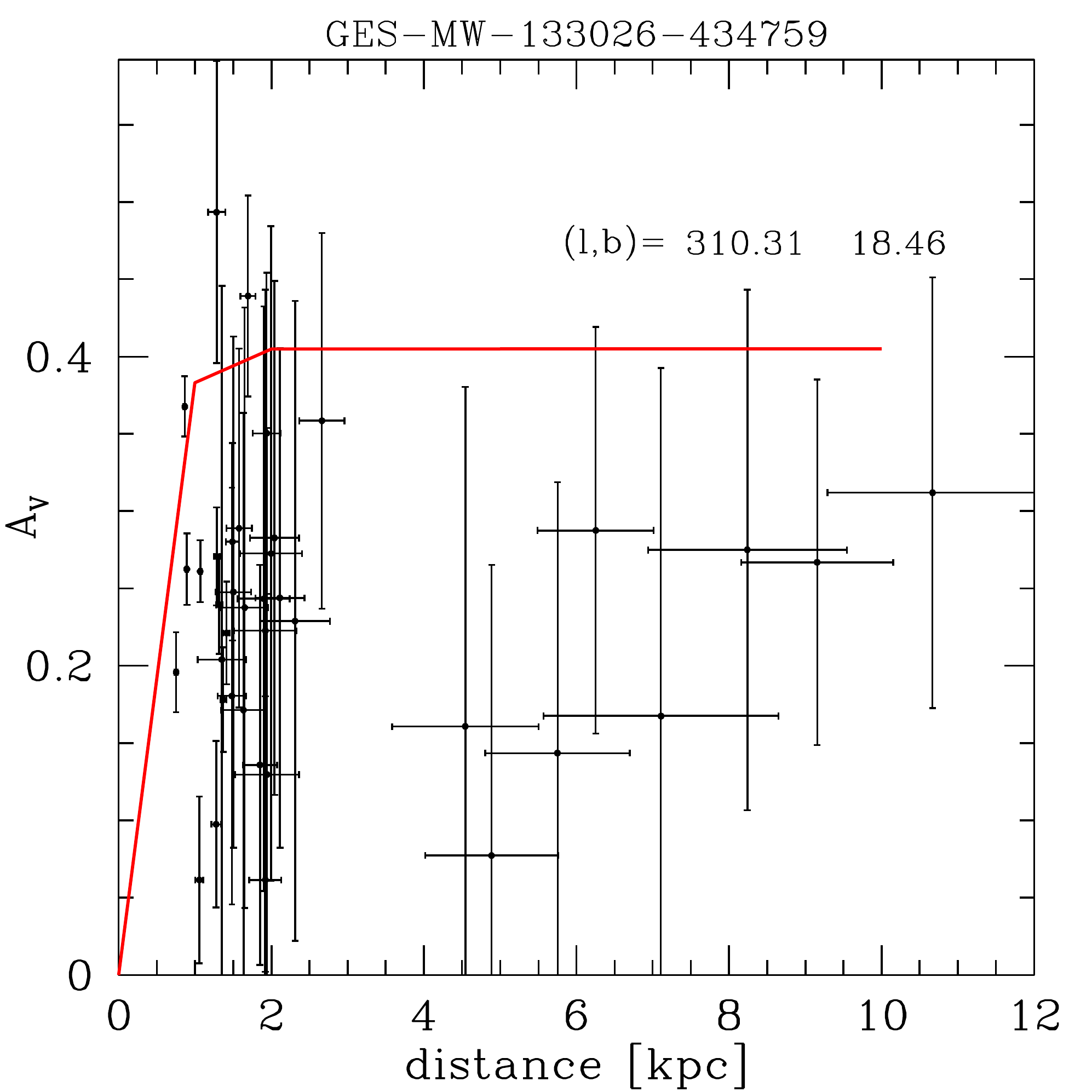}    \includegraphics[height=0.2\textheight]{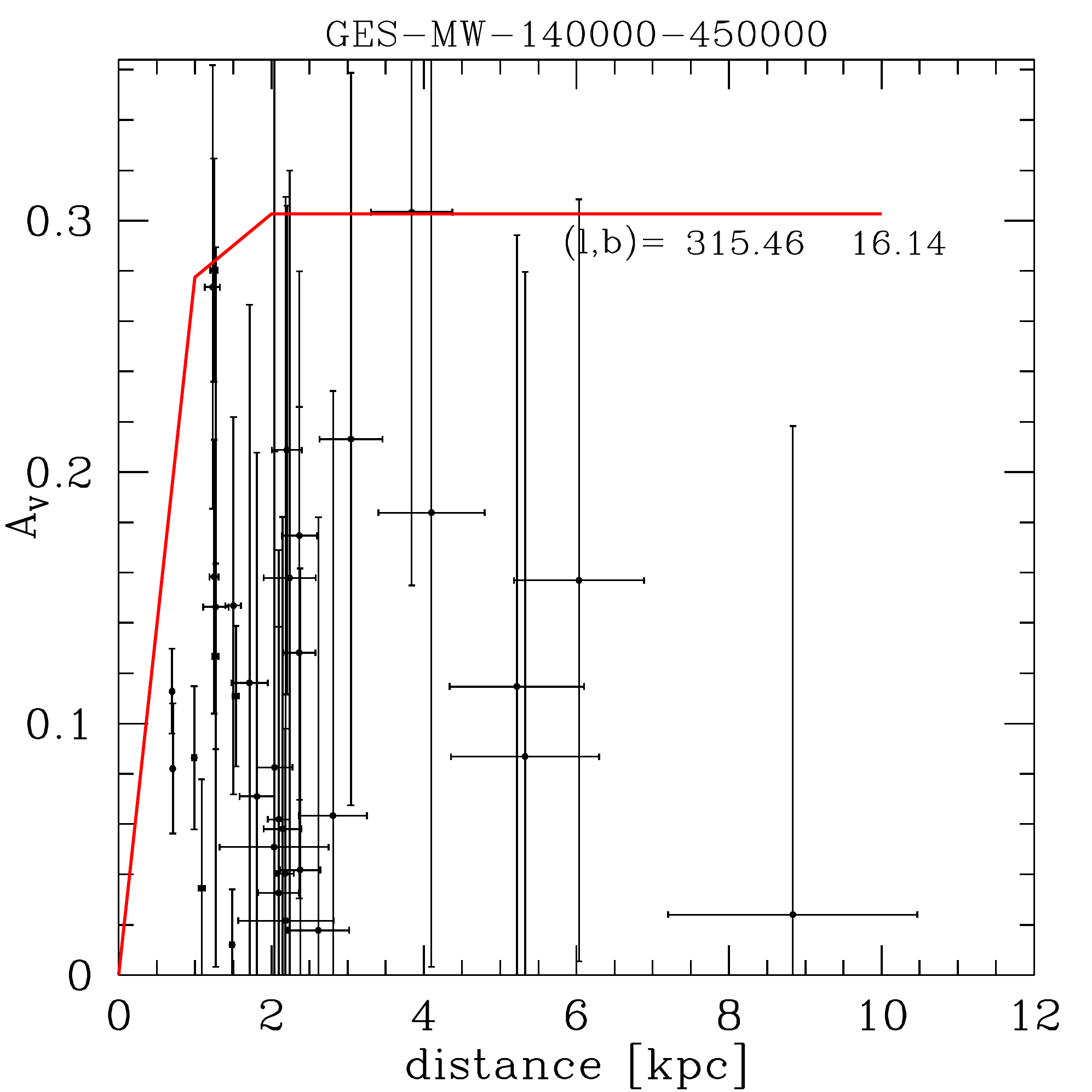}   \includegraphics[height=0.2\textheight]{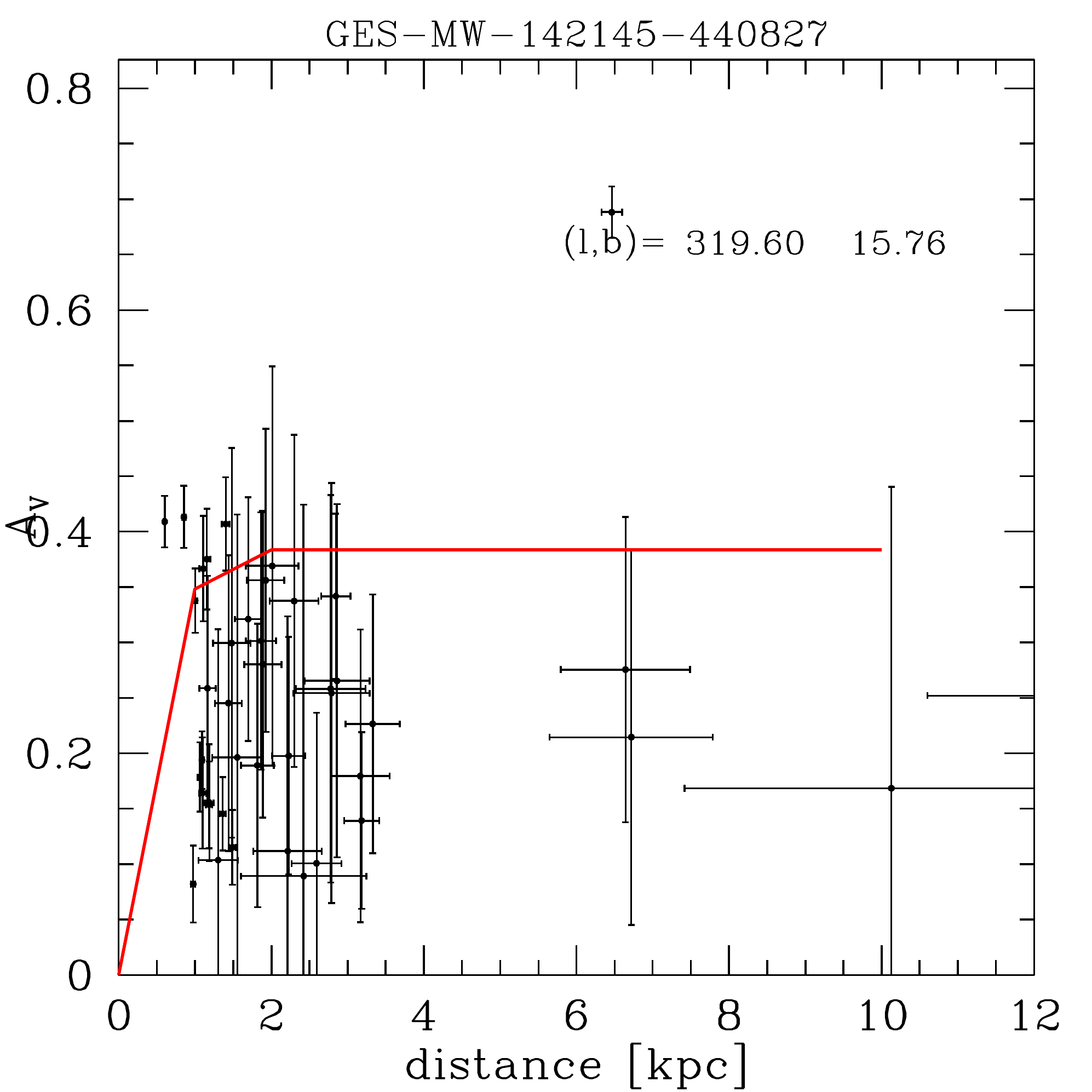}
\includegraphics[height=0.2\textheight]{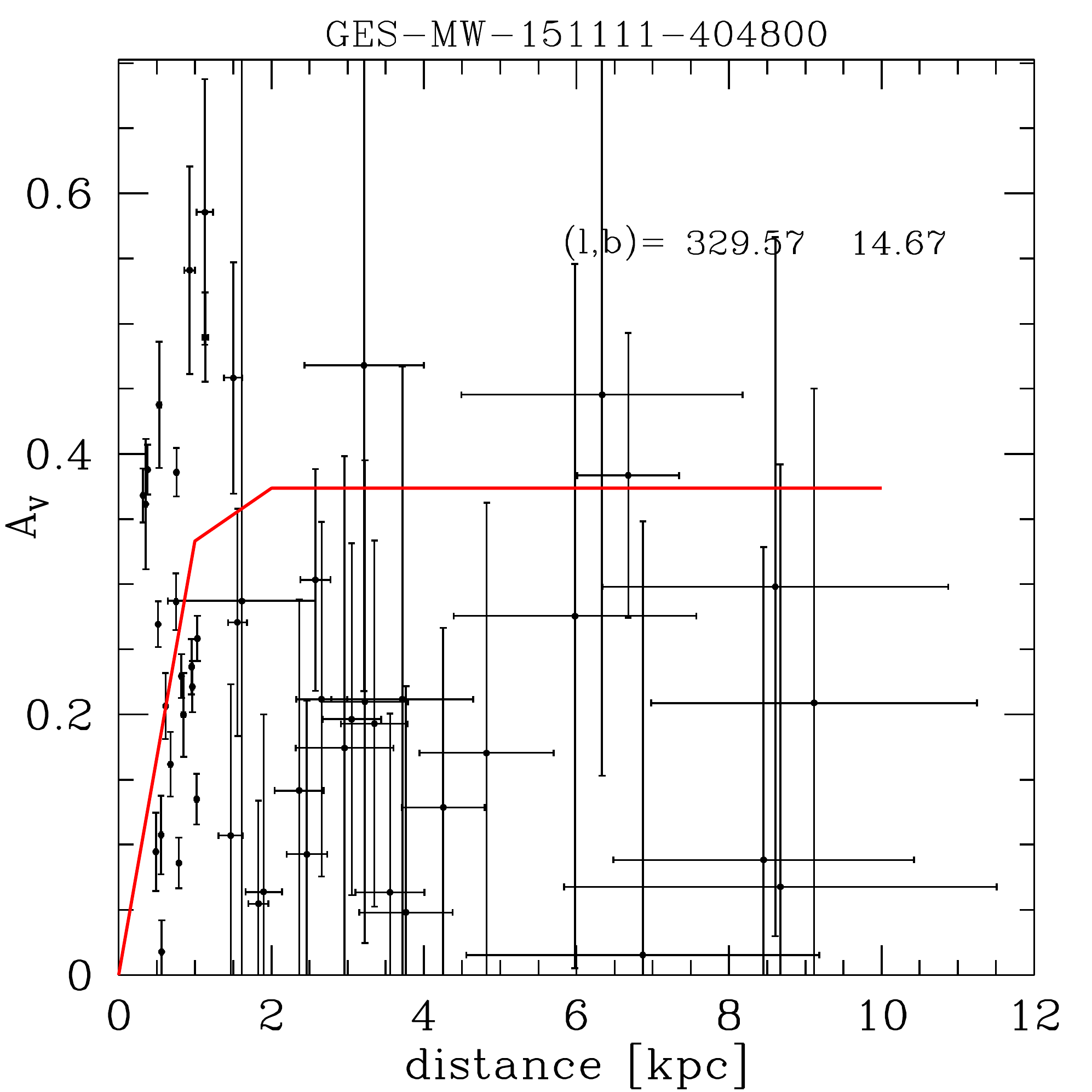}  \includegraphics[height=0.2\textheight]{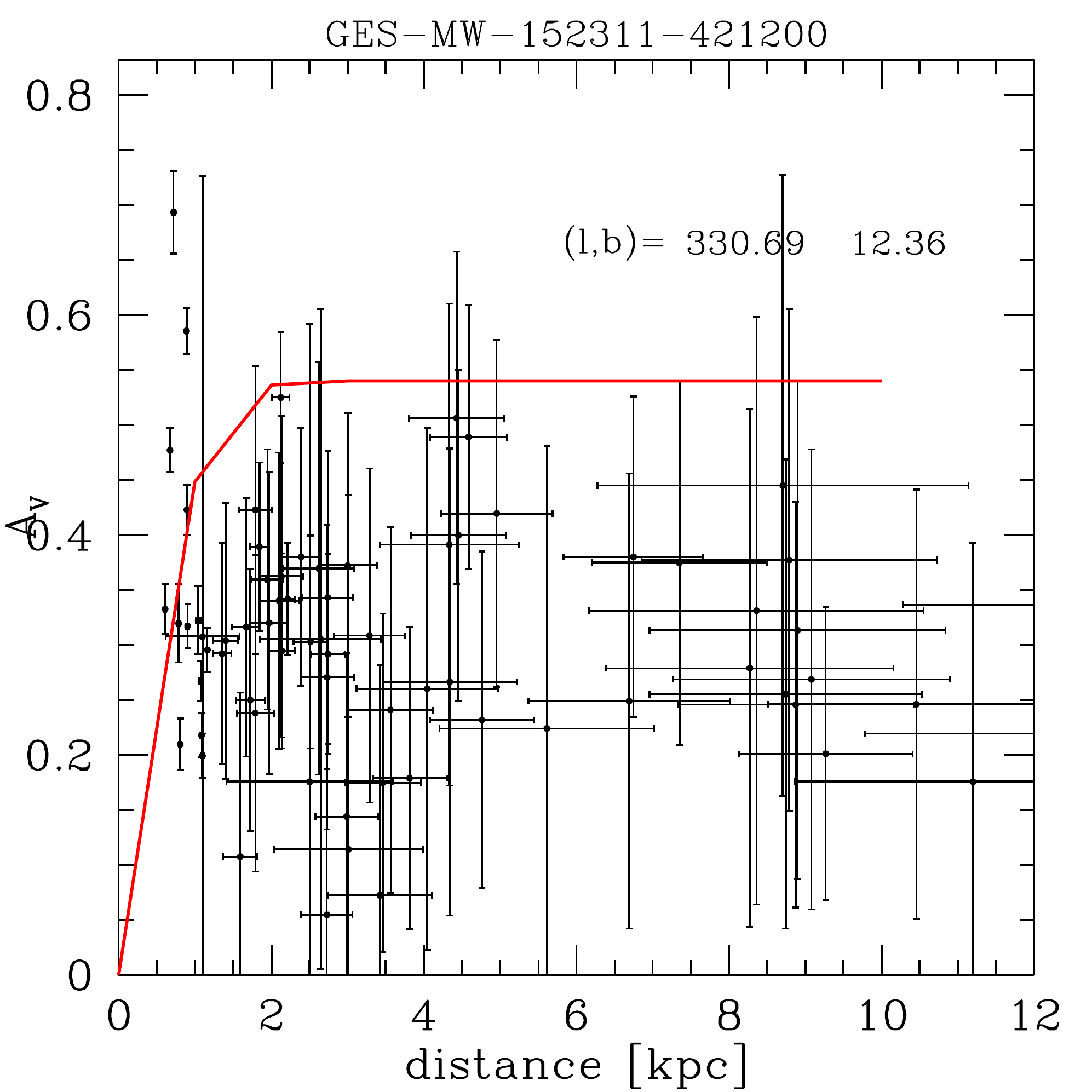} \includegraphics[height=0.2\textheight]{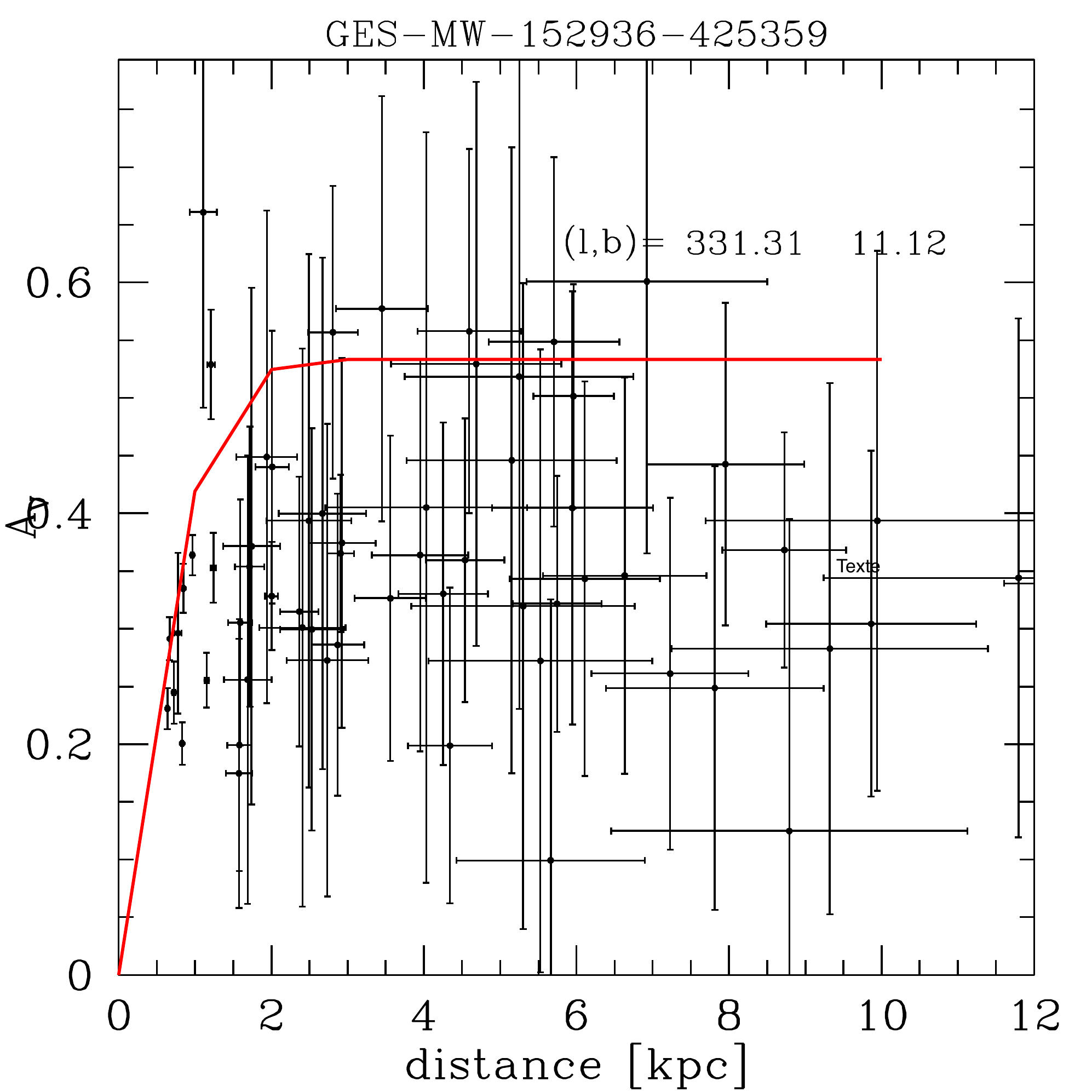}
\includegraphics[height=0.2\textheight]{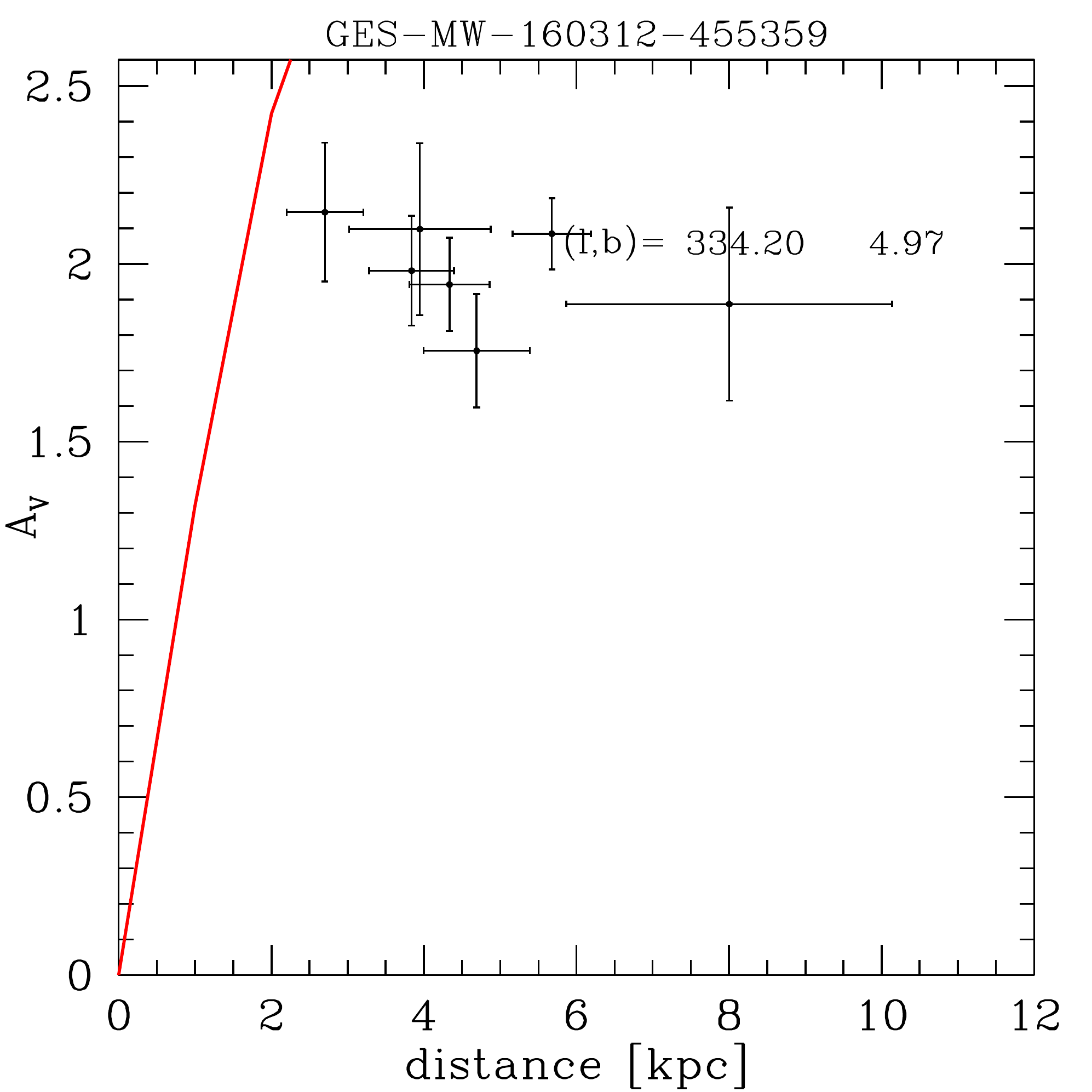} \hspace{1.6cm}   \includegraphics[height=0.2\textheight]{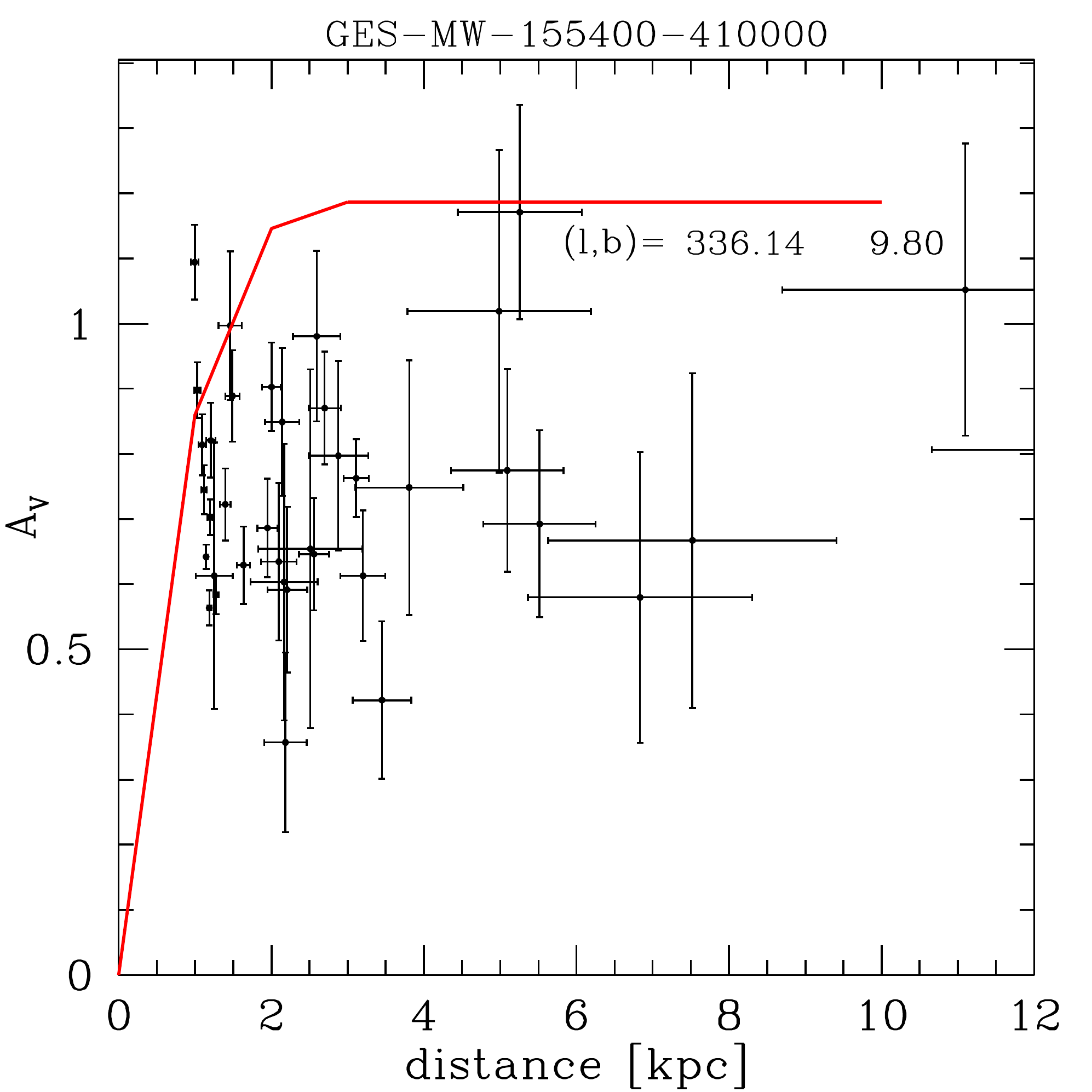}  \hspace{1.6cm} \includegraphics[height=0.2\textheight]{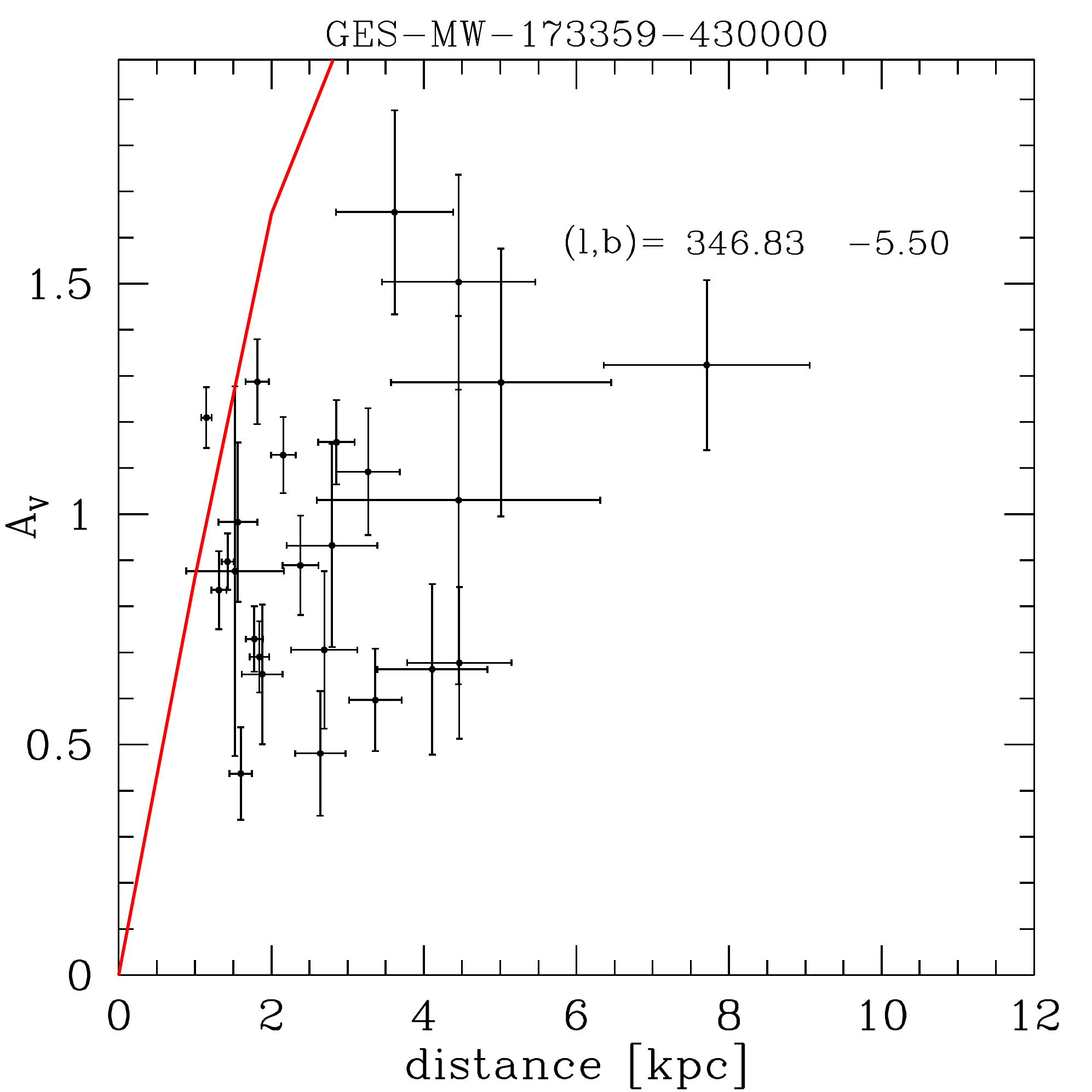}
\end{figure*}

\end{document}